\documentclass[reprint, superscriptaddress, amsmath, amssymb, aps]{revtex4-2}
\usepackage{booktabs}
\usepackage{graphicx}
\usepackage{subcaption}
\usepackage{ragged2e}
\usepackage[usenames,dvipsnames]{xcolor}
\usepackage[most]{tcolorbox}
\usepackage{tabularx}
\usepackage{tikz}
\usepackage[normalem]{ulem}
\usepackage{hyperref}
\hypersetup{
    colorlinks=true,
    linkcolor=blue,
    filecolor=red,
    urlcolor=cyan,
    citecolor=red,
    pdftitle={Proactivity and pinning in the non-reciprocal XY model with vision anisotropy},
    pdfpagemode=FullScreen,
}

\usepackage{booktabs}
\usepackage{tabularx}
\usepackage{array}

\makeatletter
\renewcommand{\@makecaption}[2]{%
  \vskip\abovecaptionskip
  \sbox\@tempboxa{#1: #2}%
  \ifdim \wd\@tempboxa > \hsize
    \justifying #1. #2\par
  \else
    \hbox to\hsize{\hfil\box\@tempboxa\hfil}%
  \fi
  \vskip\belowcaptionskip}
\makeatother

\begin{document}
\preprint{APS/123-QED}

\title{
Proactivity and pinning 
in the non-reciprocal XY model with vision anisotropy}

\author{Gabriele Bandini}
 \email{gbandini@sissa.it}
 \affiliation{SISSA --- International School for Advanced Studies and INFN, via Bonomea 265, 34136 Trieste, Italy}

\author{Asja Jelic}
 \affiliation{ICTP --- The Abdus Salam International Centre for Theoretical Physics, Strada Costiera 11, 34151 Trieste, Italy}

\author{Andrea Gambassi}
 \affiliation{SISSA --- International School for Advanced Studies and INFN, via Bonomea 265, 34136 Trieste, Italy}

\date{\today}

\begin{abstract} 
We study a 
non-reciprocal XY model on a square lattice, in which spins interact with their nearest neighbors through vision-induced anisotropic interaction. 
Such anisotropy 
breaks rotational symmetry and
leads to the pinning of the spin orientation 
along preferred lattice directions. %
We systematically characterize this phenomenon for different interaction kernels, including modulated, sinusoidal, von Mises, and hard vision-cone couplings, and for two classes of microscopic update rules: Glauber and Langevin dynamics. 
A central result of this work is the identification and detailed analysis of two distinct contributions that naturally arise in the Langevin formulation, which we refer to as the reactive and the proactive term. 
We derive the corresponding equations governing both local fluctuations and the global orientation, and use them to characterize the mechanisms responsible for directional pinning. We show that both reactive and proactive contributions can generate global pinning, whereas their role in determining local pinning depends on the specific interaction kernel and may differ qualitatively.
Our analysis clarifies the distinction between local and global pinning, explains the emergence of preferred lattice directions in the different models considered, and reconciles apparent discrepancies reported in previous studies. More generally, it provides a microscopic framework for understanding lattice-induced orientational selection in non-reciprocal XY models.
\end{abstract}

\maketitle
\tableofcontents

\section{Introduction}\label{sec:intro}

Collective motion is one of the most striking manifestations of emergent behavior in active matter.
A paradigmatic example is flocking, which is observed across a wide range of biological systems and length scales,
from bacterial suspensions~\cite{zhang2010} to insect swarms~\cite{kelley2013}, fish schools~\cite{herbert-read2011}, and bird flocks~\cite{ballerini2008, Cavagna2010}.
The Vicsek model and the hydrodynamic theory of Toner and Tu demonstrated
that systems of self-propelled particles can sustain long-range orientational order (LRO), and therefore exhibit a flocking phase, even 
in two spatial dimensions~\cite{Vicsek1995,Toner1995}. While such ordering is forbidden in equilibrium by the Mermin–Wagner theorem, the intrinsically non-equilibrium nature of self-propelled dynamics allows it to emerge. 
Flocking phases with long-range order have also been observed in lattice-gas realizations of active matter, such as the active Ising and active clock models, although with properties that differ from those of Vicsek-type systems~\cite{Solon2022polarflocks}, the dynamics of which is not restricted to a lattice.

In many natural flocking systems, interactions are mediated by visual perception. Incorporating this ingredient at the microscopic level naturally leads to non-reciprocity: particle $i$ may align with particle $j$ if $j$ lies 
within its field of view, while the converse does not necessarily hold. Non-reciprocal interactions are a central theme of many non-equilibrium systems~\cite{Ivlev2015,Fruchart2021, fruchart2026, chakraborty2025, Zhang2023-epr}. They arise whenever the action--reaction symmetry is broken at the microscopic level and appear in a wide variety of contexts, ranging from active matter~\cite{Uchida2010,Shankar2022,Saha_2019,Nagy2010,cavagna2017nonsymmetric,Yllanes_2017,tan2022odd,Besse_2022,marchetti_review,dutta2026,gaur2026, guo2026, parkavousi2026,sandoval2026,welker2025, du2025} to metamaterials~\cite{Scheibner2020,Brandenbourger2019} and neural networks~\cite{Montbri2018,Sompolinsky1986}.

Vision has also been incorporated into lattice spin systems through asymmetric couplings. A notable example is an extension of the classic two-dimensional XY model in which continuous spins align more strongly with neighbors located along the direction of their current orientation~\cite{Dadhichi2020, Loos2023,bandini2025, popli2025, dopierala2025, PhysRevLett.134.167101, rouzaire2026}.
In this setting, the resulting non-reciprocal interactions generate an effective advection of polarizations reminiscent of self-propulsion~\cite{Dadhichi2020}.

Several recent studies~\cite{Loos2023,PhysRevLett.134.167101,bandini2025,solon-journal-club,dopierala2025,Shi2026, popli2025} 
have investigated the non-reciprocal XY model on a two-dimensional lattice,  
using a variety of interaction kernels and dynamical update rules.
Two main classes of microscopic dynamics are typically considered:
(i) Glauber-type Monte Carlo dynamics, in which each spin stochastically relaxes  toward configurations that minimize its ``selfish'' local energy, as explained further below \cite{Loos2023, bandini2025}; and
(ii) Langevin dynamics, where the evolution of each spin is governed by the torque generated through the interactions with neighboring spins \cite{dopierala2025, popli2025, Dadhichi2020, PhysRevLett.134.167101}.
The latter case is amenable to analytical treatment and, in fact, 
it was shown~\cite{dopierala2025} that it has the same mean-field description as constant-density flocks \cite{toner2012birth,Besse_2022}.
This might provide a 
theoretical explanation for the emergence of the LRO observed numerically via Monte Carlo simulations 
in Refs.~\cite{Loos2023,bandini2025}. 
However, in constant-density flocks, the LRO phase is only metastable \cite{Besse_2022}, as topological defects eventually  nucleate and proliferate, ultimately destroying global order. 
This implies that the LRO phase found in the non-reciprocal XY model with Langevin dynamics is also expected to be metastable, as confirmed in Refs.~\cite{dopierala2025,popli2025}.
In addition, these works 
showed that, within this metastable ordered phase, spins tend to align and become pinned along specific directions determined by both the
lattice structure
and the microscopic implementation of the dynamics~\cite{dopierala2025}.

In this paper, we investigate the mechanisms responsible for such a pinning 
in the 
LRO phase of the non-reciprocal XY model on a two-dimensional square lattice. 
Vision anisotropy is introduced through direction-dependent couplings which modulate the interaction strength between a spin and its neighbors according to their relative spatial orientation.
The primary objective here is %
to clarify how the directional pinning emerges and how it depends on the microscopic implementation of the interactions. 
Depending on the specific implementation of the vision anisotropy and the dynamics, the LRO state may be either stable or metastable. Throughout this work, we therefore refer to the aligned state reached by the system as the (quasi)stationary state, keeping in mind that in some cases it is only long-lived on the timescales accessible to our analysis.

At thermal equilibrium, the two distinct ways of implementing the microscopic dynamics mentioned above are usually constructed on the basis of the same energy density, which is locally associated to each single degree of freedom. In particular, the Langevin dynamics of each single spin variable is obtained by adding noise to the deterministic force obtained by differentiating the energy density used in the Glauber dynamics with respect to the spin variable. In extending this construction to the case in which non-reciprocal interactions are present, the energy density has to be replaced by the selfish energy \cite{Loos2023, bandini2025}.
Accordingly, the deterministic part of the Langevin equation contains two distinct contributions. The first, which we refer to as \emph{the reactive term}, corresponds to the usual  interaction torque modulated by the interaction kernel, as in Refs.~\cite{dopierala2025, popli2025}. The second, often overlooked (with few exceptions, see Refs.~\cite{PhysRevLett.134.167101, Shi2026}),
is proportional to the derivative of the interaction kernel with respect to the spin orientation; we denote this contribution as \emph{the proactive term}.
In this work, we show that the second term has a profound impact on the stationary states of the system. 
By systematically comparing different interaction kernels, we clarify the mechanism which is eventually responsible for pinning. 
Specifically, we 
demonstrate that the reactive and proactive contributions generally favor pinning along different spatial directions.
Moreover, we investigate the distinct roles of the reactive and proactive contributions in generating local and global pinning.
We find that both terms always contribute to the global pinning of the mean orientation. At the local level, their contributions depend on the interaction kernel and are analyzed systematically for the various cases considered in this work.
Finally, we compare the analytically predicted pinning coefficients with independent numerical estimates obtained from two observables: 
the autocorrelation time of the mean orientation and 
the effective 
pinning coefficient  extracted from the static structure factor of the stationary configurations.

The remainder of the presentation is organised as follows.  
In Sec.~\ref{sec:glauberVSLangevin} we introduce the Glauber and Langevin formulations of the dynamics of the non-reciprocal XY model. In Sec.~\ref{sec:interactionskernels} we define the various interaction kernels considered in this work.
Section~\ref{sec:proactive} discusses the physical meaning of the proactive term and its relation to the alignment rules which are encountered in models of active matter. 
Section~\ref{sec:meanfield} presents a mean-field argument which explains the emergence of a preferred spin orientation with respect to the axes of the lattice for the various dynamics and interaction kernels considered.
The corresponding pinning term is identified from the equation of motion in Sec.~\ref{sec:pinning}. 
Section~\ref{sec:pinning-local-global} introduces the general framework and discusses the difference between local pinning, associated with the dynamics of individual spins, and the global pinning, associated with the dynamics of the mean spin orientation. 
In Secs.~\ref{sec:pinning-modulated}, \ref{sec:pinning-sinusoidal}, and \ref{sec:pinning-vonmises}, we present and discuss the explicit forms of the various pinning terms for the modulated, sinusoidal, and von Mises kernels, respectively.  A summary of the predictions concerning the presence of the local and global pinning is presented in Sec.~\ref{sec:pin-summary}.
Section~\ref{sec:numerical-mean-orienetation}, instead, illustrates the results of the numerical simulation of the dynamics of the mean orientation. In particular, in Sec.~\ref{sec:transient} we analyze the transient dynamics leading to the selection of the pinned directions, while in Sec.~\ref{sec:stationary-state} we study the stationary pinned state and calculate numerically the global pinning coefficients, showing that they are consistent with the stability of the selected directions.
In Sec.~\ref{sec:comparison} we compare the pinning coefficients obtained analytically and evaluated numerically
with independently measurable quantities. 
In Sec.~\ref{sec:correlation-time}, we relate the global pinning coefficient to the autocorrelation time of the mean orientation, while in Sec.~\ref{sec:correlation-length} we compare the global and local pinning coefficient with the effective pinning parameter extracted from the static structure factor of the fluctuations in the stationary state.
Finally, in Sec.~\ref{sec:conclusions} we draw the conclusions of this work and discuss possible future perspectives.
Appendix~\ref{app:local} is devoted to the derivation of the analytical expressions of the local pinning terms for the various interaction kernels and types of dynamics, while App.~\ref{app:global} contains the corresponding derivations for the global pinning term which governs the dynamics of the mean orientation.

\begin{figure}[t]
    \centering
    \includegraphics[width=0.99\linewidth]{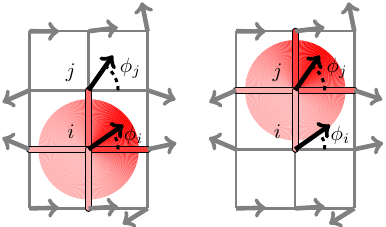}
    \begin{tikzpicture}[remember picture,overlay]
      \node at (-2.4,4.0-2.0) {\reflectbox{\includegraphics[width=0.14\linewidth, angle=0]{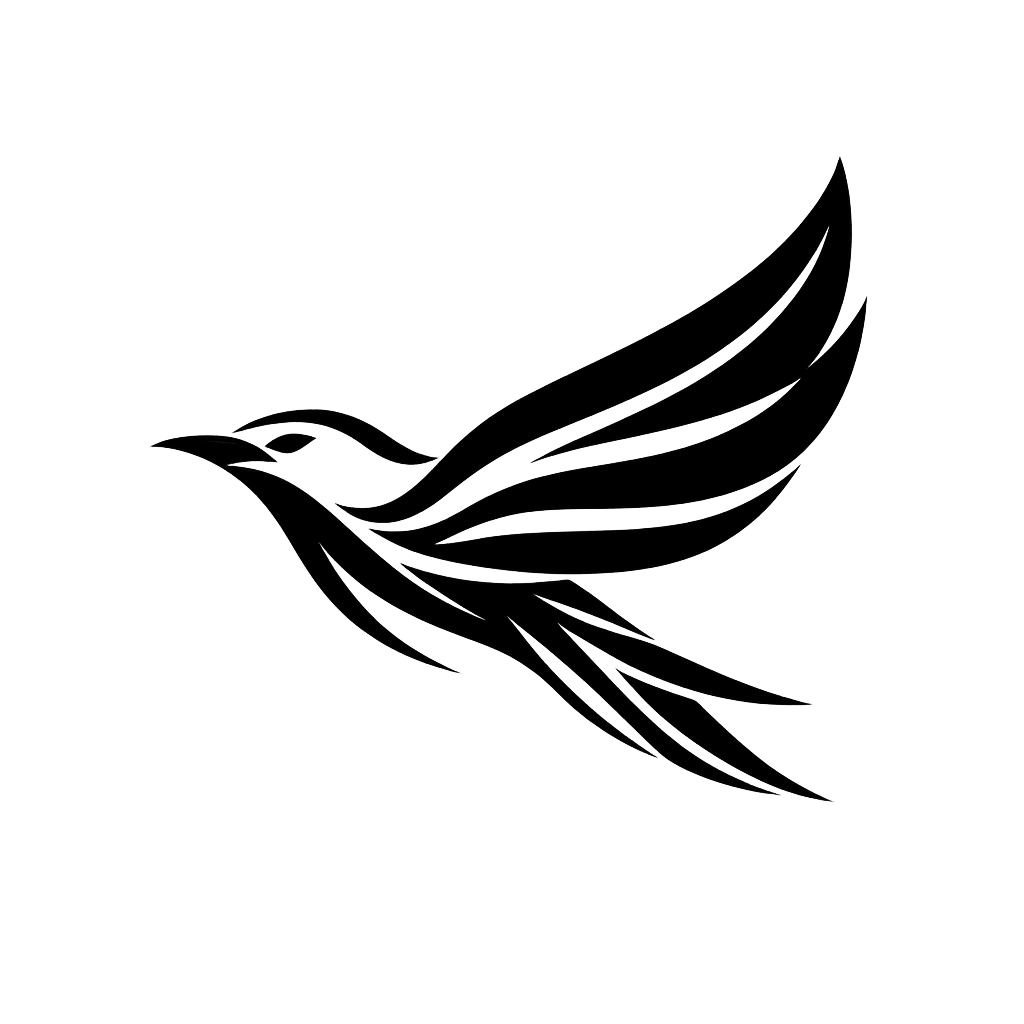}}};
    \end{tikzpicture}
    \begin{tikzpicture}[remember picture,overlay]
      \node at (-2.4,5.39-2.02) {\reflectbox{\includegraphics[width=0.14\linewidth, angle=-20]{figures/bird.png}}};
    \end{tikzpicture}
    \begin{tikzpicture}[remember picture,overlay]
      \node at (2.0,4.0-2.0) {\reflectbox{\includegraphics[width=0.14\linewidth, angle=0]{figures/bird.png}}};
    \end{tikzpicture}
    \begin{tikzpicture}[remember picture,overlay]
      \node at (2.0,5.39-2.02) {\reflectbox{\includegraphics[width=0.14\linewidth, angle=-20]{figures/bird.png}}};
    \end{tikzpicture}
    \caption{\small 
Schematic illustration of the non-reciprocal XY model with vision-induced anisotropic interactions.
The coupling between a spin (black arrow, representing for example a bird in a flock) and one of its nearest neighbors 
depends on the angle formed by the spin orientation and the bond
connecting the two sites (e.g., $\langle i,j\rangle$), 
with stronger interactions occurring at smaller angles, i.e., with the neighbors situated in the direction indicated by the spin. 
The color intensity (red scale) represents the strength of the interaction as a function of 
the relative direction of the bond compared to that of the spin $\phi_i$ on the left panel and $\phi_j$ on the right one. 
In both panels, the resulting coupling strengths along each bond are also indicated.
%
Because the interaction strength associated with a bond depends on the orientation of the spin that generates it, the same bond 
$\langle i,j\rangle$ generally contributes differently to the dynamics of spins $i$ and $j$. This breaks reciprocity, as illustrated by the different color intensities assigned to the bond $\langle i,j\rangle$ in the two panels.
}
\label{fig:visioncone}
\end{figure}


\section{Models}
\label{sec:models}

\subsection{Glauber vs.~Langevin dynamics}\label{sec:glauberVSLangevin}

The non-reciprocal XY model in two spatial dimensions is defined on a square lattice in which spin
$i$, located at position $\mathbf{r}_i$, 
interacts with neighboring spins
depending on its current orientation, represented by the angle $\phi_i$ it forms with a chosen lattice axis.
The microscopic dynamics of each spin can be implemented in two closely related, yet physically distinct, ways, discussed further below. Both cases involve the  ``selfish'' local energy functional,
which for spin \(i\) reads
\begin{equation}
E_i = -J \sum_{j \in \mathcal{N}_i}
g(\phi_i,\vartheta_{ij})\,
\cos(\phi_i-\phi_j),
\label{eq:selfish}
\end{equation}
where \(\mathcal{N}_i\) denotes the set of the four nearest neighbors of spin \(i\),  
and \(\vartheta_{ij}=\arg\left(\mathbf{r}_j-\mathbf{r}_i\right)\) is the angle (integer multiple of $\pi/2$) of the lattice bond connecting sites \(i\) and \(j\).  
The function \(g(\phi_i,\vartheta_{ij})\) defines 
the \emph{interaction kernel}, which modulates the coupling strength between spin $i$ and $j$ according to 
the field of view of spin \(i\).
Typically, spin $i$ aligns more strongly with neighboring spins $j$ whose relative positions $\mathbf{r}_j-\mathbf{r}_i$ lie along its current orientation $\phi_i$, thereby introducing non-reciprocal interactions,
as illustrated in Fig.~\ref{fig:visioncone}.

\subsubsection{Glauber dynamics}
\label{subsec:Glauber}

In the stochastic Glauber formulation of the (discrete-time) dynamics, each spin is updated according to the transition rates $W$ that depend on the change 
of the local energy $E_i$ associated with the attempted update.  
In practice, one runs a Monte Carlo simulation in which, at each step of the dynamics, a spin (say, $i$) is randomly selected for an attempted update of its orientation $\phi_i$ to $\phi_i'$, which is accepted  with probability
\begin{equation}
W(\phi_i\to\phi_i')=
\frac{1}{1+\exp[(E_i'-E_i)/T]}.   
\end{equation}
Here \(T\) plays the role of an effective temperature while $E_i$ and $E'_i$ are the selfish energies of the spin configuration before and after the attempted change of orientation.

\subsubsection{Langevin dynamics}

An alternative continuous-time dynamics of the system is obtained on the basis of a Langevin equation which involves $E_i$ and which drives the system towards its stationary points. In practice, the derivative of Eq.~\eqref{eq:selfish} with respect to \(\phi_i\) provides the deterministic driving force of the evolution of $\phi_i$ to which a noise $\eta_i$ is added. Accordingly, the resulting equation of motion is 
\begin{multline}\label{eq:langevin}
    \partial_t{\phi_i} =
-J \sum_{j\in\mathcal{N}_i}
g(\phi_i,\vartheta_{ij})\,
\sin(\phi_i-\phi_j)\\
+J\sum_{j\in\mathcal{N}_i}
\partial_{\phi_i}g(\phi_i,\vartheta_{ij})\,
\cos(\phi_i-\phi_j)
+\sqrt{2T}\,\eta_i(t),
\end{multline}
where \(\eta_i\) is a zero-mean Gaussian white noise with 
\(\langle \eta_i(t)\eta_j(t')\rangle=\delta_{ij}\delta(t-t')\) and $T$ plays the role of an effective temperature. 
Note that, at thermal equilibrium (which, inter alia, requires \(g\) to be independent of \(\phi_i\)), both dynamics relax to the Boltzmann distribution \(P \propto e^{-E/T}\), where $E = \sum_i E_i/2$.
For non-reciprocal kernels, however, detailed balance is broken, and the two dynamics generally reach distinct stationary states.

The first term on the r.h.s.~of Eq.~\eqref{eq:langevin} corresponds to the familiar torque of the XY model, which tends to align the spins, here weighted by the anisotropic kernel \(g(\phi_i,\vartheta_{ij})\).
The second term, proportional to the derivative \(\partial_{\phi_i}g\), has no counterpart in 
equilibrium 
\cite{PhysRevLett.134.167101, Shi2026} and is discussed in more detail in the following.


\begin{figure*}[t]
    \centering
    \raisebox{0.9cm}{\includegraphics[width=0.49\linewidth]{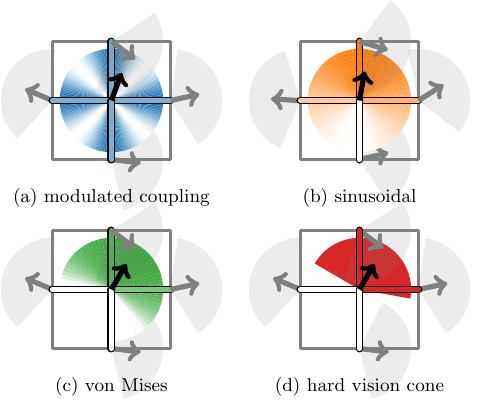}}
    \includegraphics[width=0.5\linewidth]{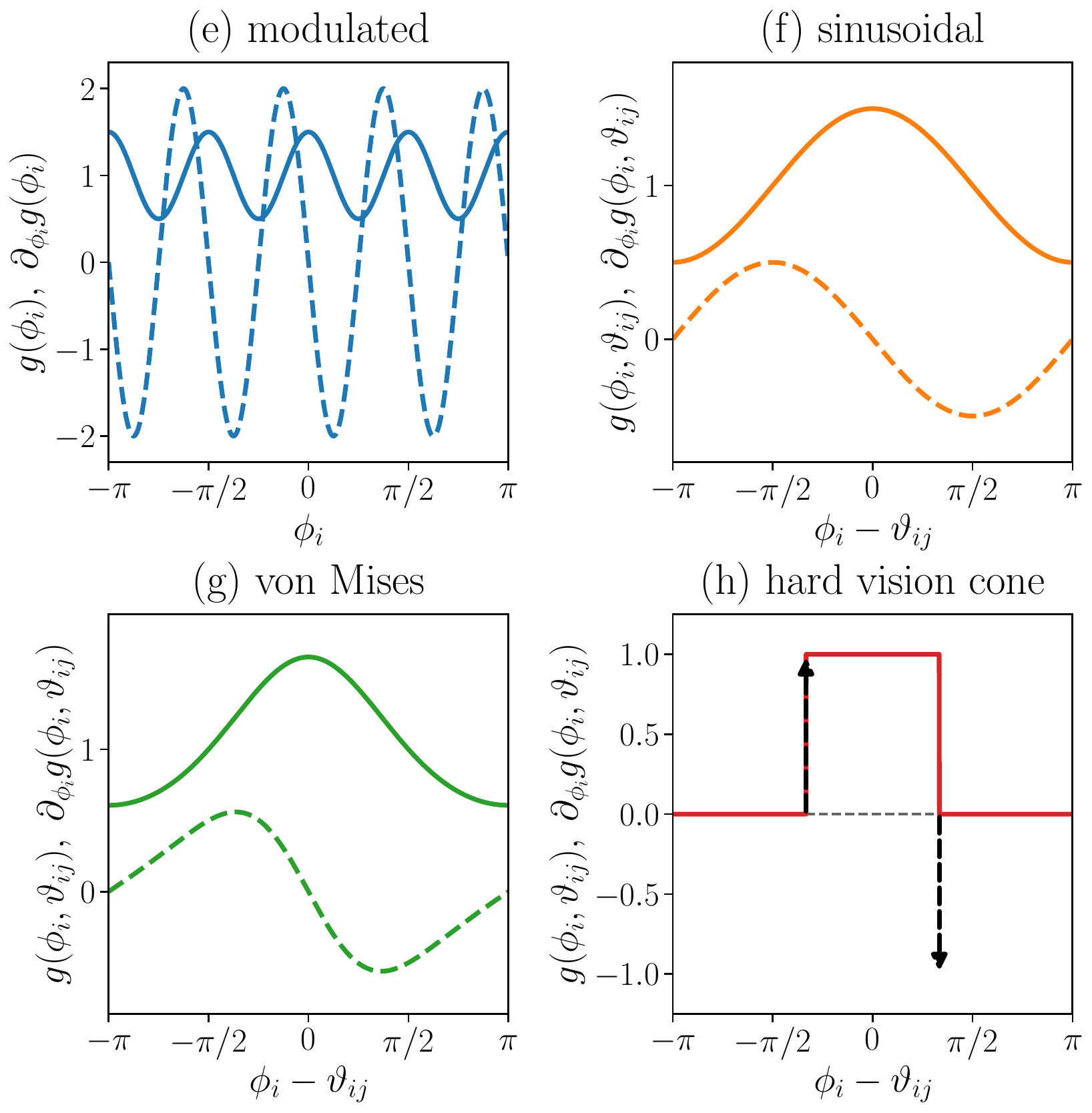}
    \caption{\small Representation of the interaction kernels considered in this work. Panels (a--d) illustrate the interaction strength, while panels (e--h) display the corresponding functional form $g$ of the kernels (solid lines) and of their derivatives $\partial_{\phi_i}g$ (dashed lines). 
(a, e) Modulated coupling (Eq.~\eqref{eq:mod-coup}): all nearest-neighbors bonds have identical strength, with no vision bias or anisotropy ($g$ is not function of $\vartheta_{ij}$), although the interaction is modulated by the spin orientation.
(b, f) Sinusoidal coupling (Eq.~\eqref{eq:sin-kernel}): forward bonds are enhanced through a smooth sinusoidal angular modulation.
(c, g) Von Mises coupling (Eq.~\eqref{eq:vonmises-kernel}): similar to the sinusoidal case, but with a different angular modulation, characterized by a  forward bias which becomes increasingly more pronounced upon increasing the parameter $\sigma$;
(d, h) Hard vision cone (Eq.~\eqref{eq:HVC}): interactions are restricted only to neighbors that lie within a finite vision cone. Accordingly, the kernel derivative exhibits two Dirac delta contributions (dashed arrows) at the boundaries of the vision cone.
}
    \label{fig:kernels}
\end{figure*}

\subsection{Interaction kernels}\label{sec:interactionskernels}
\label{sec:kernels}

The physical behavior of the non-reciprocal XY model is dictated by the choice of the functional form of the interaction kernel \(g(\phi_i,\vartheta_{ij})\).  
For concreteness, we consider here the following four representative cases, illustrated in Fig.~\ref{fig:kernels}:
\begin{enumerate}
    \item \textbf{Modulated coupling (no vision bias)}~\cite{dopierala2025}:  
    \begin{equation}
        g(\phi_i,\vartheta_{ij})=1+\epsilon\cos(4\phi_i),
        \label{eq:mod-coup}
    \end{equation}
    which breaks rotational symmetry via an explicit fourfold modulation of the coupling which is independent of the spatial perception of spin $i$, i.e., $g(\phi_i,\vartheta_{ij})$ is actually independ of $\vartheta_{ij}$. The parameter $\epsilon$ controls the degree of anisotropy. In the present work, we consider couplings which favor the alignment of neighboring spins an thus we assume \(\epsilon \in [0,1]\).
    \item \textbf{Sinusoidal coupling %
    }~\cite{solon-journal-club, dopierala2025, popli2025}:  
    \begin{equation}
        g(\phi_i,\vartheta_{ij})
        = 1+\epsilon\cos(\phi_i-\vartheta_{ij}), 
        \label{eq:sin-kernel}
    \end{equation}  
    where the coupling is maximal when the neighboring spin \(j\) lies along the spatial direction indicated by spin \(i\).  
    \item \textbf{%
    von Mises coupling}~\cite{PhysRevLett.134.167101}:  
    \begin{equation}
        g(\phi_i,\vartheta_{ij})
        = \exp[\sigma\cos(\phi_i-\vartheta_{ij})],
    \label{eq:vonmises-kernel}
    \end{equation}
    where \(\sigma\) is a parameter of the model, 
    which is characterized by a smooth, positive-definite angular dependence of the coupling strenght on the angle formed by the direction of the neighbors and the one of the spin.
    For small values of \(\sigma\), this expression reduces to the sinusoidal kernel above with $\epsilon = \sigma$.
    \item \textbf{Hard vision cone (HVC)}~\cite{Loos2023, bandini2025}:  
   \begin{equation}\label{eq:HVC}
        g(\phi_i,\vartheta_{ij})=
        \begin{cases}
        1 &
        \text{if} \quad %
        \cos(\phi_i-\vartheta_{ij}) \ge \cos(\theta/2),\\[4pt]
        0 &
        \text{otherwise.}
    \end{cases}\
    \end{equation}
    In this case, only neighbors which are within an angular sector of width \(\theta \in [0,2\pi]\) centered around the direction of spin $i$ contribute to the interaction.  
    This kernel introduces discontinuities and sharp geometric constraints.
\end{enumerate}
Note that all these kernels, for either \(\epsilon=\sigma=0\) or \(\theta=2\pi\), render the equilibrium two-dimensional XY model with isotropic couplings. 
%


\subsection{From active to proactive matter}\label{sec:proactive}

\begin{figure*}[t]
    \centering
    \includegraphics[width=0.75\linewidth]{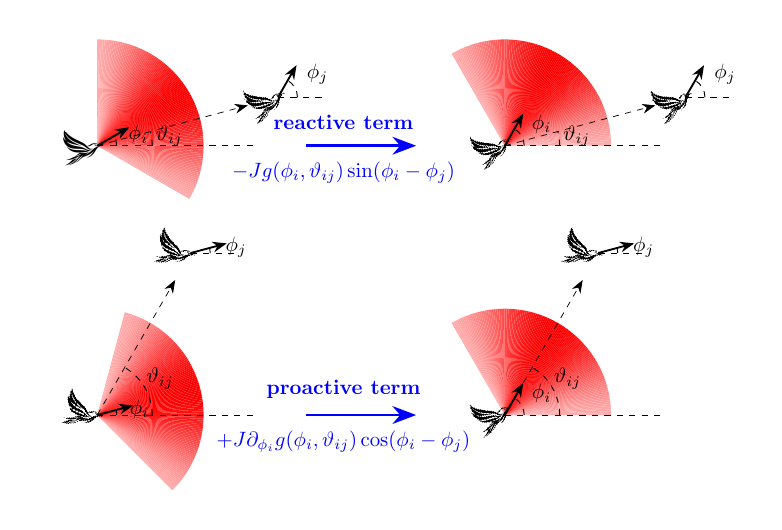}
     \put(-380, 260){\large (a)}
     \put(-380, 110){\large (b)}

    \caption{\small 
Comparison between the reactive and the proactive term in the Langevin dynamics of the non-reciprocal XY model. (a) The reactive term tends to align the orientation of an agent (e.g., a bird at site $i$) with the orientation of the neighbors it perceives (e.g., the one at site $j$), with a strength modulated by the kernel $g(\phi_i, \vartheta_{ij})$, such that $\phi_i\simeq \phi_j$. (Here $\vartheta_{ij}$ is the angle connecting the sites $i$ and $j$, where $\vartheta_{ij} = n\pi/2$ with $n=0$, 1, 2, 3 on the square lattice.)  
In contrast, (b) the proactive term drives the agent to reorient along the direction of the unit vector connecting it to its neighbors, such that $\phi_i\simeq \vartheta_{ij}$. This occurs through an interaction mechanism sensitive to the gradient of the vision coupling, effectively pushing the agent to maximize its sensory input, i.e., the effective number of visible neighbors which are not misaligned with the agent.
As a consequence, the proactive contribution is sensitive to variations in the visual field: when a neighbor is about to exit the field of view, this term induces a reorientation that tends to keep it within sight.
}

    \label{fig:dynamics_comparison}
\end{figure*}

The second term on the r.h.s.~of Eq.~\eqref{eq:langevin}, proportional to 
\(\partial_{\phi_i}g(\phi_i,\vartheta_{ij})\), is often omitted in the Langevin formulation of the non-reciprocal XY model, see, e.g., Refs.~\cite{Dadhichi2020, dopierala2025, popli2025}. 
As we shall demonstrate below, this term actually introduces a qualitatively new ingredient in the microscopic dynamics, the physical interpretation of which can be understood by comparing the torques generated separately by the two contributions on the r.h.s.~of Eq.~\eqref{eq:langevin}:
\begin{itemize}
\item\textbf{Reactive alignment.}  
The first term, i.e.,
\begin{equation}
-J\sum_{j\in\mathcal{N}_i}g(\phi_i,\vartheta_{ij})\sin(\phi_i-\phi_j),
\end{equation}
describes the usual interaction that favors spin alignment. It is the only term present in the equilibrium XY model, where the coupling $g$ is independent of the spin orientation and thus constant. 
Specifically, spin \(i\) tends to rotate towards the orientations of its visible neighbors, with an effective coupling strength modulated by the kernel \(g\) as illustrated in Fig.~\ref{fig:dynamics_comparison}(a).  
This mechanism is analogous to the alignment rule in Vicsek-type models of active matter, where each agent reacts by adjusting its orientation according to what it currently perceives in its local environment. Henceforth we refer to this as the \emph{reactive term}.
Within this interaction term, the kernel $g$ sets the relative weight of the contribution of each neighbor to the alignment dynamics.
%
\item\textbf{Proactive adjustment.}  
The second term on the r.h.s.~of Eq.~\eqref{eq:langevin}, i.e., 
\begin{equation}
+J\sum_{j\in\mathcal{N}_i}
\partial_{\phi_i}g(\phi_i,\vartheta_{ij})
\cos(\phi_i-\phi_j),
\end{equation}
has a qualitatively different effect. It causes spin \(i\) to rotate in the direction that increases (decreases) the value of the coupling kernel $g(\phi_i,\vartheta_{ij})$, whenever $\cos{(\phi_i - \phi_j)}>0 \,(<0)$.
In practice, this contribution generates a torque on spin $i$ which tends to maximize the effective number
of visible neighbors that are at least partially aligned with that spin, as schematically represented in Fig.~\ref{fig:dynamics_comparison}(b). Here, by partial alignment between two spins $i$ and $j$, we mean that 
one has a positive projection onto the other, 
i.e., $\cos(\phi_i-\phi_j)>0$.  
We emphasize that this term is proportional to the gradient of the ``perception'' kernel $g$.
The resulting feedback mechanism therefore induces a tendency to \emph{maximise the perceived alignment} itself, rather than merely responding to it.
\end{itemize}
The behavior of the spin generated by the second term discussed above goes beyond conventional activity.  For this reason, we refer to it as \emph{proactivity}:
each agent modifies its state (i.e., its orientation) in order to enhance its own sensory input (i.e., the effective number of perceived neighbors which are not misaligned), 
rather than merely aligning passively with it.  
At the collective level, proactivity can stabilize orientational order and suppress the nucleation of defects (the so-called asters \cite{Besse_2022, dopierala2025, popli2025})  
in simulations of the Langevin dynamics in Eq.~\eqref{eq:langevin}. 
In fact, the role of this additional proactive contribution in promoting defect annihilation during coarsening starting from a disordered initial state was previously 
highlighted in Ref.~\cite{PhysRevLett.134.167101}.

It is worth mentioning that the Langevin dynamics 
in Eq.~\eqref{eq:langevin}, which includes
both the reactive and proactive terms, can be directly mapped onto the Glauber dynamics associated with the selfish energy in Eq.~\eqref{eq:selfish}. 
By contrast, the Langevin dynamics with the sole reactive term can 
be simulated through Monte Carlo dynamics only by using the auxiliary Hamiltonian introduced in Ref.~\cite{Shi2026}.

To elucidate the differences between the reactive and proactive terms, we proceed in two steps.
First, in Sec.~\ref{sec:meanfield}, we present a mean-field argument showing that the two contributions favor alignment along different spatial directions, and we discuss why this argument is actually inconclusive in the case of the sinusoidal kernel. 
Second, in Sec.~\ref{sec:pinning}, we refine this analysis by deriving the equations of motion at both the local and global levels: locally, first for the discrete variable $\phi_i(t)$ and then for its continuum counterpart $\phi_{\mathbf{x}}(t)$, and globally for the mean orientation $\phi(t)$.
This allows us to show how pinning terms naturally emerge in both cases, either as local or as global contributions. Furthermore, we highlight that the reactive and proactive components enter the effective evolution equations in qualitatively different ways.


\section{Mean field: coupling strength at $T=0$}\label{sec:meanfield}
\begin{figure}
    \includegraphics[width=0.84\linewidth]{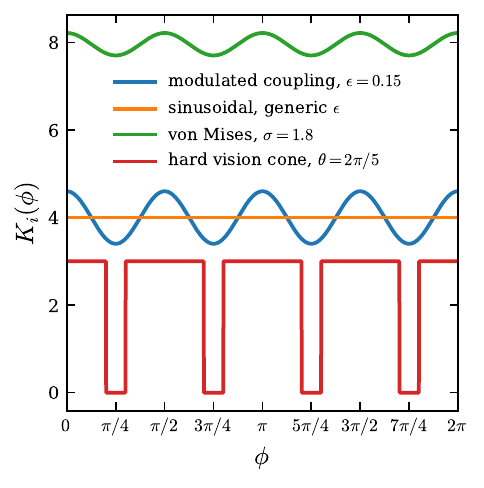}
    \caption{\small
Total coupling $K_i(\phi)$ (see Eq.~\eqref{eq:total-coupling}) for a perfectly aligned spin configuration (corresponding to $T\to 0$) as a function of the global spin orientation $\phi$, for the various interaction kernels considered in this work and listed in Sec.~\ref{sec:kernels}. The qualitative features of these curves are the same for generic choices of the parameters $\epsilon$, $\sigma$, and $\theta$.
}
    \label{fig:total-coupling-rotation}
\end{figure}
We begin our analysis by considering the case in which fluctuations are negligible, i.e., in which all the spins are perfectly aligned. 
Formally, this corresponds to the limit $T\to0$ of the dynamics in Eq.~\eqref{eq:langevin}
with $\phi_i = \phi$ for all $i$, 
such that $\cos(\phi_i - \phi_j) = 1$ and $\sin(\phi_i - \phi_j) = 0$.
Under this assumption, the \emph{total coupling} $K_i(\phi)$ of spin $i$ with \textit{all} its nearest neighbors can be defined, on the basis of Eq.~\eqref{eq:selfish}, as
\begin{equation}\label{eq:total-coupling}
    K_i(\phi) = \sum_{j \in \mathcal{N}_i}  g(\phi,\vartheta_{ij});
\end{equation}
this quantity depends only on the global orientation $\phi$ of the spins and is actually independent of the site index $i$.
Figure~\ref{fig:total-coupling-rotation} shows 
$K_i(\phi)$ as a function of 
$\phi$ for the different kernels considered in this work and summarized in Fig.~\ref{fig:kernels}.

In order to rationalize the emergence of ordering in the system,
a key question to be addressed is whether the dynamics 
drives the system to some stationary point of the total coupling $K_i(\phi)$.
In fact, if this is the case, the spins are expected to be pinned collectively with a value of $\phi$ corresponding to one of these stationary points. 

In the presence of the reactive term alone, $K_i(\phi)$ is indeed \emph{minimized}, as
demonstrated for the modulated coupling kernel and further 
generalized in Ref.~\cite{dopierala2025}.
In fact, when the mean direction $\phi$ corresponds to a minimum of the coupling strength $K_i(\phi)$, a spin that fluctuates away from $\phi$ experiences a stronger interaction with its neighbors than the spins that remain aligned along $\phi$. Consequently, the fluctuating spin is pulled back toward the mean orientation more strongly than the rest of the system is pulled toward the fluctuating spin. This imbalance generates an effective restoring torque that stablizes the configuration around the direction of weakest coupling.
Accordingly, configurations with
weaker effective coupling are dynamically favored.

When the proactive term is added to the reactive one, the situation is reversed. As discussed in Sec.~\ref{sec:proactive}, the additional contribution in the Langevin equation generates a torque that tends to increase the ``sensory input'' of each spin, thereby \emph{maximizing} its coupling $K_i(\phi)$ to the visible neighbors which are not misaligned. This can be understood more clearly by considering the Glauber dynamics associated with the selfish energy in Eq.~\eqref{eq:selfish}, which naturally corresponds to the Langevin dynamics with both the reactive and proactive contributions. 
For an ordered state, the selfish energy in Eq.~\eqref{eq:selfish} becomes $E_i = - J K_i(\phi)$ and thus it is minimized when the coupling $K_i(\phi)$ is maximized, confirming the tendency %
towards configurations with stronger coupling.

Inspecting Fig.~\ref{fig:total-coupling-rotation}, we observe that the total coupling $K_i(\phi)$ is actually independent of $\phi$ for the sinusoidal kernel, while it exhibits a periodic modulation for the modulated coupling, the von Mises kernel, and the hard vision cone (see Sec.~\ref{sec:kernels} for their definitions).
This modulation reflects the $\mathbb{Z}_4$ symmetry of the underlying lattice (explicitly imposed in the case of the modulated coupling). Based on this plot, the argument presented above about the minimization/maximization of $K_i(\phi)$ by the reactive/total dynamics allows one to infer the preferred pinning directions for these three kernels, both for the sole reactive dynamics and for the dynamics that includes the proactive contribution. 
For example, for the modulated coupling and the von Mises kernel,
the reactive dynamics promotes ordering with $\phi = \pi/4 + n\pi/2$, with $n=0$, 1, 2, and 3, while the complete one with $\phi = n\pi/2$.
These directions are summarized in Table~\ref{tab:summary}, which will be illustrated and discussed in, c.f., Sec.~\ref{sec:numerical-mean-orienetation} together with the numerical evidence. 

An important observation is that, for the von Mises coupling kernel, the leading contribution that generates 
a modulation of the total coupling $K_i(\phi)$ arises only at order $\sigma^4$. 
This fact will be taken into account in Sec.~\ref{sec:pinning-vonmises} where the corresponding equations of motion are derived.

The arguments presented above, applicable in the limit of vanishing noise $T\to 0$, also indicate that the sinusoidal kernel alone is not capable of pinning the spins to any specific orientation $\phi$, neither with the reactive dynamics nor with the one which includes the reactive and proactive terms since, in both cases, $K_i(\phi)=0$ independently of $\phi$.
Nevertheless, as shown in Ref.~\cite{dopierala2025}, finite-temperature fluctuations can induce a weak but measurable pinning effect even for the sinusoidal kernel.
In the next section, we systematically analyze how these pinning terms emerge in the presence of fluctuations, both for the sinusoidal and for the other interaction kernels considered in this work.

\section{Search for a pinning term}
\label{sec:pinning}

When the system is in a 
(quasi)stationary
long-range ordered (LRO) state at small but finite temperatures, 
the global orientation of the spins remains pinned along specific lattice directions. 
The selection of these directions depends on both the details of the dynamics and the form of the interaction kernel, as already discussed in Sec.~\ref{sec:meanfield} and as it will be proved numerically in, c.f., Sec.~\ref{sec:numerical-mean-orienetation}. 
In this section, we analyze how pinning emerges at the level of the equations of motion for the modulated, sinusoidal, and von Mises interaction kernels.  The hard vision cone kernel, instead, is not amenable to an analytical treatment because of the discontinuities it generates in the corresponding Langevin equation. This is the reason why we will investigate it only numerically in, c.f., Sec.~\ref{sec:numerical-mean-orienetation}.

\subsection{Local and global equations of motion}
\label{sec:pinning-local-global}

In order to write the equations of motion for both the local  and the global spin orientation, we introduce the following notation. We denote by $\phi_i(t)$ the local spin at lattice site $i$, and by $\phi_{\mathbf{x}}(t)$
its counterpart on the continuum space, which will enter the evolution equations discussed further below.  
The global orientation of the model at time $t$ is defined as the instantaneous spatial average,
\begin{equation}
\phi(t) =\left\langle \phi_i(t) \right\rangle= \frac{1}{N}\sum_{i=1}^N \phi_i(t),
\label{eq:def-phi}
\end{equation}
where $N = L^2$ is the total number of spins in the (finite) system and $\langle \cdots \rangle$ denotes the spatial average over the lattice sites. 
When, in its (quasi)stationary state (i.e.~stable or metastable), the system is pinned along a specific direction, the time average of the global orientation is 
\begin{equation}
\Phi = \overline{\phi(t)} = \overline{\left\langle \phi_i(t) \right\rangle}.
\end{equation}
Here, the overbar indicates a time average in the (quasi)stationary state. 
For the systems considered in this work, the long-time average orientation $\Phi$ is selected from a discrete set of angles determined by the symmetry of the square lattice. In particular,
\begin{equation}
\Phi \in \left\{ n \frac{\pi}{2},   \frac{\pi}{4} + n \frac{\pi}{2} \right\} \quad\mbox{with}\quad n = 0,1,2,3,
\end{equation}
corresponding to alignment along the principal lattice directions and along the lattice diagonals, respectively.

\paragraph*{Fluctuations.}
We now consider the global and local fluctuations. In particular, 
the fluctuation $\delta \phi(t)$ of the global orientation $\phi(t)$ in the stationary state with respect to its average direction $\Phi$ is defined as
\begin{equation}
\delta \phi(t) = \phi(t) - \Phi.
\label{eq:deltaphi}
\end{equation}
At the local level, it is useful to distinguish between fluctuations with respect to the instantaneous global orientation, i.e.,
\begin{equation}
\delta \tilde{\phi}_i(t) = \phi_i(t) - \phi(t),
\label{eq:def-tphii}
\end{equation}
and fluctuations with respect to the long-time average orientation,
\begin{equation}
\delta \phi_i(t) = \phi_i(t) - \Phi.
\label{eq:deltaphii}
\end{equation}
Clearly, $\delta \phi_i(t) = \delta \tilde{\phi}_i(t) + \delta \phi(t)$.
Analogous definitions hold for the fluctuations of the field $\phi_{\mathbf{x}}(t)$ on the continuum.
We point out that in the thermodynamic limit ($N\to\infty$), the mean orientation $\phi(t)$ in the stationary state has vanishing fluctuation $\delta\phi(t)$, i.e., $\phi(t)\to\Phi$ and therefore $\delta\tilde{\phi}_i(t)\to \delta\phi_i(t)$. 

\paragraph*{Evolution of the local fluctuations.}
%
%
%
To determine the pinning terms acting on the local variable $\phi_i$, we consider Eq.~\eqref{eq:langevin}, substitute the explicit expressions of the interaction kernels, and expand the sums over nearest neighbors. 
The final result of this procedure can be conveniently cast in the form
\begin{widetext}
\begin{equation}
\label{eq:local-harmonics-txt}
\partial_t \phi_i
    =
    \sum_{\alpha\in\{\mathrm{R},\mathrm{P}\}}
    \sum_{n=0}^{\infty}
    \Big[
        S_{\alpha,i}^{(n),\mathrm{loc}}(t)\,\sin(n\phi_i)
        +
        C_{\alpha,i}^{(n),\mathrm{loc}}(t)\,\cos(n\phi_i)
    \Big] + \sqrt{2T}\eta_{i}(t),
\end{equation}
\end{widetext}
where the analytic expressions of the coefficients $S_{\alpha,i}^{(n),\mathrm{loc}}(t)$ and $C_{\alpha,i}^{(n),\mathrm{loc}}(t)$  are derived in detail in Apps.~\ref{app:modulated-local}, \ref{app:sin-local}, and \ref{app:vonmises-local}.
These coefficients depend on the instantaneous configuration of the neighbors of spin $i$ and are functions of the local angular differences, i.e.,
\begin{equation}\label{eq:local-coeffs}
    S_{\alpha,i}^{(n),\mathrm{loc}}(t)
    =
    F_{\alpha}^{(n),\mathrm{loc}}
    \Big(
        \{\phi_i(t)-\phi_j(t)\}_{j\in{\cal N}_i}
    \Big),
\end{equation}
and similarly for $C_{\alpha,i}^{(n),\mathrm{loc}}(t)$.
In Eq.~\eqref{eq:local-harmonics-txt} the sum over $\alpha\in\{\mathrm{R},\mathrm{P}\}$ contains separately the contributions due to the reactive (R) and proactive (P) terms in the equation of motion \eqref{eq:langevin}. 

When the spins are pinned around a given direction $\Phi$, Eq.~\eqref{eq:local-harmonics-txt} can be expanded in terms of the fluctuations $\delta\phi_i$, defined in Eq.~\eqref{eq:deltaphii}. The 
resulting equation generally takes the form
\begin{equation}
\partial_t \delta\phi_i
=
\kappa(\delta\phi_i)\,
\nabla_{\rm d}^2
\delta\phi_i
-
r_i^{\mathrm{loc}}(t)\,
\delta\phi_i
+\sqrt{2T}\eta_i(t)
+ \mathcal{A}_i,
\label{eq:local-continuum}
\end{equation}
where $\kappa(\delta\phi_i)$ is a modulated diffusion coefficient and $\nabla_{\rm d}^2$ denotes the discrete Laplacian
\begin{equation}\label{eq:discrete-lap}
\nabla_{\rm d}^2 \delta\phi_i
=\frac{1}{a^2}
\sum_{j\in{\cal N}_i}
(\delta\phi_j-\delta\phi_i).
\end{equation}
Here, $a$ indicates the lattice spacing, %
which sets the unit of length (throughout this work we set $a=1$).
The effective diffusion term $\kappa(\delta\phi_i)\nabla_{\rm d}^2\delta\phi_i$ controls the propagation of local fluctuations, while the term linear in $\delta\phi_i$ acts as a local restoring force with the local pinning coefficient $r_i^{\mathrm{loc}}(t)$ that depends on the harmonic coefficients $S_{\alpha,i}^{(n),\mathrm{loc}}(t)$ and $C_{\alpha,i}^{(n),\mathrm{loc}}(t)$. Instead, $\mathcal A_i$ collects all the contributions arising from the original equation of motion in Eq.~\eqref{eq:local-harmonics-txt}, which are not accounted for by 
the first two contributions.
The stability of the fluctuations $\delta\phi_i(t)$ is controlled by the average value of the local pinning coefficient
\begin{equation}\label{eq:avg-local-pinning}
r^{\mathrm{loc}}
=
\overline{
r_i^{\mathrm{loc}}(t)
},
\end{equation}
%
which needs to be positive, i.e.,
\begin{equation}
r^{\mathrm{loc}}
>
0.
\label{eq:positive-pinning}
\end{equation}
Note that the average $r^{\mathrm{loc}}$ in Eq.~\eqref{eq:avg-local-pinning} does not actually depend on $i$, due to the translational invariance of the system.
In the simplest approximation, $r^{\mathrm{loc}}$ determines the long-time relaxation rate of $\delta\phi_i(t)$ following an initial fluctuation $\delta\phi_i(t=0)$.
The coefficient $r^{\mathrm{loc}}$ might therefore be interpreted as the pinning strength controlling spatial fluctuations.
In general, however, this identification is only approximate. Indeed, the effective pinning governing the long-wavelength dynamics is generally renormalized both by the fluctuations of $r_i^{\mathrm{loc}}(t)$ and by the additional contributions collected in $\mathcal A_i$. A useful measure of the magnitude of the first effect is provided by the standard deviation
\begin{equation}
\label{eq:def-delta-rloc}
\Delta_{r^{\mathrm{loc}}}
=
\sqrt{
\overline{
\left(
r_i^{\mathrm{loc}}(t)
-r^{\mathrm{loc}}
\right)^2
}
},
\end{equation}
which quantifies the fluctuations of the local pinning coefficient around its mean value. When
\begin{equation}
\Delta_{r^{\mathrm{loc}}}
\ll
r^{\mathrm{loc}},
\label{eq:condition}
\end{equation}
the renormalization associated with the fluctuations of $r_i^{\mathrm{loc}}$ is expected to be weak, and $r^{\mathrm{loc}}$ provides a reliable approximation to the effective pinning governing the spatial dynamics. 
Conversely, when $\Delta_{r^{\mathrm{loc}}}$ becomes comparable or larger than $r^{\mathrm{loc}}$, significant renormalization effects are expected and the effective pinning extracted from spatial correlations may differ substantially from  $r^{\mathrm{loc}}$.
As discussed in Sec.~\ref{sec:correlation-length}, this criterion plays an important role when comparing analytical estimates of the pinning coefficient with those extracted from the static structure factor.
More generally, although $r^{\mathrm{loc}}$ provides a natural measure of the local pinning strength, the effective pinning governing spatial fluctuations does not necessarily coincide with it.

Finally, note that the averages appearing in Eqs.~\eqref{eq:avg-local-pinning} and \eqref{eq:def-delta-rloc}, are taken over time. However, in the stationary state, the system is invariant both under time translations and, due to periodic boundary conditions, spatial translations. As a consequence, the quantities $r^{\mathrm{loc}}$ and $\Delta_{r^{\mathrm{loc}}}$ can be estimated equivalently from temporal or spatial averages. In practice, throughout this work we compute them from spatial averages over the lattice (see also App.~\ref{app:local}).%

\paragraph*{Evolution of the global fluctuation.}
To derive the evolution equation for the global fluctuation $\phi$ in Eq.~\eqref{eq:def-phi}, we again start from Eq.~\eqref{eq:langevin}, substitute the explicit form of the interaction kernel, and sum both sides of the equation over all lattice sites in the system . 
This procedure leads to an exact expression of a form analogous to Eq.~\eqref{eq:local-harmonics-txt}, i.e., 
\begin{widetext}
\begin{equation}\label{eq:global-harmonics-txt}
\partial_t \phi
    =
    \sum_{\alpha\in\{\mathrm{R},\mathrm{P}\}}
    \sum_{n=0}^{\infty}
    \Big[
        S_{\alpha}^{(n),\mathrm{glob}}(t)\,\sin(n\phi)
        +
        C_{\alpha}^{(n),\mathrm{glob}}(t)\,\cos(n\phi)
    \Big] + \sqrt{\frac{2T}{N}}\eta(t).
\end{equation}
\end{widetext}
The details of the derivation of the analytic expressions for the various interaction kernels considered in this work are reported in Apps.~\ref{app:modulated-mean}, \ref{app:sin-global}, and \ref{app:vonmises-mean}. 
The coefficients $S_{\alpha}^{(n),\mathrm{glob}}(t)$ and $C_{\alpha}^{(n),\mathrm{glob}}(t)$ 
in Eq.~\eqref{eq:global-harmonics-txt}
involve contributions from all pairs of neighboring spins in the system and they can be expressed as functions of the fluctuations $\delta\tilde{\phi}_i(t)$ as
\begin{equation}
S_{\alpha}^{(n),\mathrm{glob}}(t)
=
\Big\langle
F_{\alpha}^{(n),\mathrm{glob}}
\big(
\delta\tilde{\phi}_i(t), \delta\tilde{\phi}_j(t)_{j\in\mathcal{N}_i}
\big)
\Big\rangle,
\end{equation}
and similarly for $C_{\alpha}^{(n),\mathrm{glob}}(t)$.
Note that these coefficients depend on the fluctuations with respect to the instantaneous mean orientation $\phi(t)$, and not to its temporal average $\Phi$. 

Once the system is pinned, we can expand Eq.~\eqref{eq:global-harmonics-txt} in terms of the fluctuation $\delta\phi$ in Eq.~\eqref{eq:deltaphi}, obtaining an equation that features the pinning term analogous to Eq.~\eqref{eq:local-continuum}:
\begin{equation}
\partial_t \delta\phi = - r^{\mathrm{glob}}(t)\,\delta\phi(t) + \sqrt{\frac{2T}{N}}\eta(t) + {\cal A}(t).
\label{eq:pin-global}
\end{equation}
This coefficient $r^{\mathrm{glob}}(t)$, determined by the harmonic coefficients $S_{\alpha}^{(n),\mathrm{glob}}(t)$ and $C_{\alpha}^{(n),\mathrm{glob}}(t)$ appearing in Eq.~\eqref{eq:global-harmonics-txt}, plays the role of the global pinning coefficient and sets the strength of the restoring force acting on the mean orientation.
In particular, in the stationary state, its time-dependence is due to the fluctuations of these coefficients as functions of time. In the thermodynamic limit, the amplitude of these fluctuations actually vanishes, as they refer to a global quantity, and thus $r^{\mathrm{glob}}(t)$ becomes time-independent and positive, i.e.,
\begin{equation}
r^{\mathrm{glob}}(t) =r^{\mathrm{glob}}   > 0.
\end{equation}
This latter fact ensures the stability of the pinned state against perturbations. Note that, by construction, $r^{\mathrm{glob}}(t)$ is a global quantity and thus it lacks any dependence on space which, instead, is pronounced in its ``local counterpart'' $r_i^{\mathrm{loc}}(t)$.  
In Eq.~\eqref{eq:pin-global}, the quantity ${\mathcal A}(t)$ (which results from a spatial average) 
collects all the contributions arising from the original equation of motion in Eq.~\eqref{eq:global-harmonics-txt}, which are not captured by the linear global pinning term $-r^{\mathrm{glob}}(t)\delta\phi$: it has zero mean and its fluctuations vanish in the thermodynamics limit. 

\paragraph*{Remarks on the pinning terms.}
A key observation is that, in general,
\begin{equation}
{r^{\mathrm{loc}} } \neq r^{\mathrm{glob}}.
\end{equation}
This means that the effective pinning coefficient that governs local fluctuations $\delta\phi_i$ does not necessarily coincide with the one that controls the fluctuations $\delta\phi$ of the global orientation, even after averaging over time.

As an illustrative example, consider first a case in which $r^{\mathrm{loc}}_i = r^{\mathrm{glob}}$, i.e., the reciprocal XY model in the presence of an external field which selects a preferred spin direction $\Phi$. The corresponding dynamics is given by 
\begin{equation}\label{eq:XY-field}
\partial_t \phi_i = -J \sum_{j\in\mathcal{N}_i} \sin(\phi_i - \phi_j) - h \sin\left(\phi_i - \Phi\right).
\end{equation}
Expanding this equation in terms of the local fluctuations $\delta\phi_i$ around the preferred direction, one finds that the local evolution equation for $\delta\phi_i$ contains a linear  term $-h\,\delta\phi_i$ which provides a restoring force toward $\delta\phi_i=0$ and corresponds to $r_i^{\mathrm{loc}} =h$.
Instead, summing Eq.~\eqref{eq:XY-field} over all lattice sites, one obtains the following equation for the spatially averaged orientation  $\phi(t)$:
\begin{equation}
\partial_t \phi(t) = %
- \frac{h}{N} \sum_i \sin(\phi_i - \Phi),
\end{equation}
where the first term in Eq.~\eqref{eq:XY-field} vanishes because the possible contribution of each ordered pair $(i,j)$ is canceled by that of $(j,i)$.
Expanding %
this equation in terms of the global fluctuation $\delta\phi$ of $\phi(t)$ around the preferred direction $\Phi$
one finds %
$-h\,\delta\phi$, corresponding to $r^{\mathrm{glob}} = h$. Consequently, for this reciprocal XY model one obtains
\begin{equation}
r_i^{\mathrm{loc}} =
r^{\mathrm{loc}} = r^{\mathrm{glob}} = h.
\end{equation}
This equality no longer holds in the non-reciprocal XY model, where both local and global effective pinning terms depend explicitly on the instantaneous spin configuration.

The emergence of effective pinning terms in non-reciprocal XY models has been previously investigated in Refs.~\cite{dopierala2025, popli2025}. In particular, Ref.~\cite{popli2025} focuses on \emph{local} equations of motion for $\delta\phi_i(t)$ in the presence of the reactive term, considering both sinusoidal and von~Mises interaction kernels. Reference~\cite{dopierala2025}, instead, investigates the \emph{global} dynamics of the system by deriving an effective equation for the collective variable $\delta\phi(t)$, again in the presence of the reactive term, for both sinusoidal and modulated coupling kernels. 
In the following, we extend these analyses by applying our framework to the modulated, sinusoidal, and von Mises coupling kernels, with particular emphasis on the role of the proactive term.
We report the main findings of this work in the following three subsections, while the details of the numerical simulations are presented in Sec.~\ref{sec:numerical-mean-orienetation}.

\subsection{Pinning terms for the modulated coupling}\label{sec:pinning-modulated}

The case of the modulated coupling was previously studied in Ref.~\cite{dopierala2025}, where only the reactive contribution to the dynamics was considered, with focus on the evolution equation for the global fluctuations $\delta\phi(t)$.
Here, in order to determine the pinning terms, we expand Eq.~\eqref{eq:local-harmonics-txt} for the local orientation $\phi_i$ and Eq.~\eqref{eq:global-harmonics-txt} for the mean orientation $\phi(t)$ as detailed in Apps.~\ref{app:modulated-local} and 
\ref{app:modulated-mean}, respectively.  In the following, we discuss the resulting equations and their physical implications.

In the purely reactive case, the resulting time evolution equations at the local level
\emph{do not} exhibit any explicit pinning term. 
This can be seen from the expansion of $\partial_t\phi_i(t)$ reported in Eq.~\eqref{eq:modulated-reactive-local}: when only the reactive contribution is retained, all coefficients have a vanishing mean value, resulting in a modulated diffusion term but no local restoring force.
At first sight, this conclusion appears to contradict the argument presented in Sec.~\ref{sec:meanfield},
based on which pinning is expected at the minima of $K_i(\phi)$, located at $\phi = \pi/4+n\pi/2$, with $n \in \mathbb{Z}$. Moreover, the numerical simulations discussed in Sec.~\ref{sec:numerical-mean-orienetation} clearly show pinning along the diagonal directions for the purely reactive dynamics.
Our interpretation of this apparent discrepancy is the following. As discussed in Sec.~\ref{sec:meanfield} and in Ref.~\cite{dopierala2025}, the reactive dynamics tends to minimize the total coupling $K_i(\phi)$. 
This happens because the total coupling with the neighbors of a fluctuating spin which departs from the global alignment is stronger than the total coupling of its neighbors with it. As a consequence, the fluctuating spin is pulled back toward the global alignment. 
However, this imbalance cannot be fully captured by the dynamic equations for $\phi_i$ alone: due to non-reciprocity, one must account not only for the bonds activated by $\phi_i$, but also for those activated by its neighboring spins. 
For this reason, pinning does not emerge explicitly at the level of the local equations for $\phi_i$, but 
it appears in the evolution equation for the fluctuations of the global orientation $\delta\phi(t)$. In particular, Ref.~\cite{dopierala2025} expresses the quantity $r^{\text{glob}}$ in terms of the fluctuations $\delta\tilde{\phi}_{\mathbf{x}}$ on the continuum as 
\begin{equation}\label{eq:reactive-mod-pinning-global-cont}
-(16\epsilon/L^2) \int {\rm d}\mathbf{x} \,|\nabla\delta\tilde{\phi}_{\mathbf{x}}|^2,
\end{equation}
which can then be estimated analytically. In our approach, we derive the corresponding  expression on the lattice, i.e., 

\begin{equation}
    r^{\text{glob}} = 4S_{\text{R}}^{(4),\text{glob}}
\end{equation}
which can be evaluated numerically on the basis of the analytic expression in Eq.~\eqref{eq-app:SR4glob}. 
Expanding this expression for small fluctuations and performing a gradient expansion (see App.~\ref{app:modulated-mean}), one recovers Eq.~\eqref{eq:reactive-mod-pinning-global-cont}. Note that, in the absence of an explicit local pinning term, this global pinning coefficient is a quantity
that can be related to the pinning extracted from the static 
structure factor, as discussed in Sec.~\ref{sec:correlation-length}. 

By contrast, when the proactive term is included, explicit pinning terms emerge in the evolution equations for both the local and the global orientations, although with different expressions (see  Apps.~\ref{app:modulated-local} and~\ref{app:modulated-mean} for details). 
The resulting complete dynamics eventually leads to alignment along the lattice directions.

\subsection{Pinning terms for the sinusoidal coupling}\label{sec:pinning-sinusoidal}
The evolution equation for the local orientation $\phi_i$ in the presence of a sinusoidal kernel can be expanded as shown in App.~\ref{app:sin-local}, while its continuum counterpart reproduces Eq.~(5)
of Ref.~\cite{popli2025}.
The corresponding equations for the global orientation $\delta\phi(t)$ are derived in App.~\ref{app:sin-global}; upon expansion for small $\delta\tilde{\phi}_i$, they acquire the same structure as that of the equations reported in Ref.~\cite{dopierala2025}.

Since the system exhibits pinning along one of the four lattice directions, as confirmed by the numerical results presented in Sec.~\ref{sec:numerical-mean-orienetation},
we can set $\Phi = 0$ without loss of generality. 
We find that the evolution equations for both the local fluctuations $\delta\phi_i(t)$ and the global fluctuation $\delta\phi(t)$ contain effective pinning terms with the coefficients $r_i^{\mathrm{loc}}(t)$ and $r^{\mathrm{glob}}$, respectively.
The reactive contribution of the sinusoidal coupling kernel,
yields 
\begin{align}
r_i^{\mathrm{loc}}(t) &=-S_{\mathrm{R},i}^{(1), \mathrm{loc}}(t) \quad \mbox{and} \quad 
r^{\mathrm{glob}} &=-S_{\mathrm{R}}^{(1), \mathrm{glob}},
\end{align}
where the explicit analytic expressions for $S_{\mathrm{R},i}^{(1), \mathrm{loc}}$ and $S_{\mathrm{R}}^{(1), \mathrm{glob}}$ are provided in Eqs.~\eqref{eq-app:SR1loc-SK} and \eqref{eq-app:SR1-glob-SK}, respectively.
This allows us to compare the average of the local pinning strength $r_i^{\mathrm{loc}}$, with the global coefficient $r^{\mathrm{glob}}$. 
For $\epsilon = 1$, $T = 0.2$, and $L = 500$, we obtain numerically 
\begin{equation}
r^{\mathrm{loc}} = 0.0015 \quad \mbox{with} \quad \Delta_{r^{\mathrm{loc}}} = 0.4
\quad\mbox{and}\quad
r^{\mathrm{glob}} = 0.0022.
\label{eq:coeff-pinning-SK}
\end{equation}
The positive value of $r^{\mathrm{glob}}$ 
indicates the presence of a global restoring force acting on the fluctuations of the mean orientation $\delta\phi(t)$. 
Moreover, the values of the global and local pinning coefficients are quantitatively different, i.e., $r^{\mathrm{loc}}\neq r^{\mathrm{glob}}$.

An even more important observation concerns the fluctuations of the local pinning coefficient. Although the average value $r^{\mathrm{loc}}$ is positive, the fluctuations of $r_i^{\mathrm{loc}}$ around this mean, quantified by $\Delta_{r^{\mathrm{loc}}}$, are much larger than the mean itself (see Eq.~\eqref{eq:coeff-pinning-SK}). 
Accordingly, the condition in Eq.~\eqref{eq:condition} is strongly violated. 
In this regime, it is not obvious that the local dynamics can be described in terms of a single effective pinning obtained by renormalizing the average pinning coefficient $r^{\mathrm{loc}}$.
As discussed in Sec.~\ref{sec:correlation-length}, this makes any quantitative comparison between analytical pinning coefficients and those extracted from the stationary fluctuations considerably more subtle.
Note that the numerical estimates reported in Eq.~\eqref{eq:coeff-pinning-SK} were obtained for 
a specific set of simulation parameters.
Nevertheless, the qualitative features $r^{\mathrm{loc}} \neq r^{\mathrm{glob}}$ and that $\Delta_{r^{\mathrm{loc}}} \gg r^{\mathrm{loc}}$ are expected to hold throughout the region of parameter space in which the (quasi)stationary LRO phase exists, as we verified numerically for various other choices.

As far as the proactive term is concerned, its inclusion in the dynamics also results in pinning along the lattice , as it is seen numerically in Sec.~\ref{sec:numerical-mean-orienetation}.
In this case, however, the proactive contribution does not generate any local pinning term (see App.~\ref{app:sin-local}); rather, it produces only a global pinning, which can be shown analytically to coincide with the one arising from the reactive contribution discussed above (see App.~\ref{app:sin-global}).

\subsection{Pinning terms for the von Mises coupling}
\label{sec:pinning-vonmises}

The von Mises kernel (see Eq.~\eqref{eq:vonmises-kernel}) represents a particularly instructive case. 
As discussed in Sec.~\ref{sec:meanfield}, the mean-field analysis already reveals an explicit preference for alignment along specific directions:
when both reactive and proactive terms are included in the Langevin dynamics, the system aligns along the lattice directions, whereas retaining only the reactive term favors alignment along the diagonal directions.
This behavior is similar to that observed for the modulated coupling in Sec.~\ref{sec:pinning-modulated} above, but differs from the case of the 
sinusoidal kernel in Sec.~\ref{sec:pinning-sinusoidal}, 
for which no directional selection emerges at the mean-field level.
At the same time, the von Mises kernel differs fundamentally from the modulated coupling because it intrinsically couples the spin degrees of freedom to the underlying lattice through its explicit dependence on the angles $\vartheta_{ij}$.
In the following, we show how these features manifest themselves at the level of the equations of motion, comparing them with what is reported in the literature.

A convenient starting point is to expand the von Mises kernel in Eq.~\eqref{eq:vonmises-kernel} in powers of $\sigma$ up to the fourth order, i.e.,
\begin{equation}
    e^{\sigma \mathcal{C}}
    =
    1
    + \sigma \mathcal{C}
    + \frac{\sigma^2}{2}\mathcal{C}^2
    + \frac{\sigma^3}{6}\mathcal{C}^3
    + \frac{\sigma^4}{24}\mathcal{C}^4,
\label{eq:vM-exp}
\end{equation}
where $\mathcal{C} = \cos{(\phi_i-\vartheta_{ij}})$.
We first focus on the purely reactive case, i.e., on the Langevin dynamics which includes only the reactive term. 
From the mean-field argument presented in Sec.~\ref{sec:meanfield} and as confirmed by the numerical results presented in Sec.~\ref{sec:numerical-mean-orienetation}, we know that the system features pinning along the diagonal directions, e.g., $\Phi = \pi/4$. 
We therefore seek to identify which term in the expansion in Eq.~\eqref{eq:vM-exp} is responsible for this behavior. 
The term of order $\sigma^0$ generates the reciprocal interaction of the usual XY model and thus it cannot explain the emergence of pinning. 
The contribution of order $\sigma$ to $\partial_t\phi_i$ turns out to coincide, up to an inconsequential prefactor (see App.~\ref{app:vonmises-local}),
with the one due to a sinusoidal coupling and thus, as discussed in Sec.~\ref{sec:pinning-sinusoidal}, it cannot by itself lead to diagonal pinning. 
The contribution of order $\sigma^2$ was analyzed in Ref.~\cite{popli2025}, where it was shown that it corresponds to a term of the form 
\begin{equation}
    \langle \sin\phi_{\mathbf{x}}\,(\partial_x^2\phi_{\mathbf{x}}-\partial_y^2\phi_{\mathbf{x}})\rangle \sin\phi_{\mathbf{x}}
\end{equation}
when the system is described in terms of the variable $\phi_{\mathbf{x}}$ on the continuum.
It was suggested there that this term might be responsible for pinning. 
However, since it is proportional to $\sin\phi_{\mathbf{x}}$, it would favor pinning at $\Phi=0$, in contrast with the diagonal pinning observed here.
Even including the contribution of order $\sigma^3$ of the von Mises kernel does not explain this observation, because the ${\cal O}(\sigma^3)$ contribution to the total coupling $K_i(\phi)$ of Eq.~\eqref{eq:total-coupling} vanishes for a globally aligned configuration and therefore it cannot induce any pinning at the mean-field level. This conclusion is further supported by our numerical simulations: truncating the kernel at order $\sigma^3$ results into a pinning along the lattice directions.

In fact, we identify the term of order $\sigma^4$ in Eq.~\eqref{eq:vM-exp} as
the leading contribution responsible for 
diagonal pinning.
Within the mean-field argument discussed in Sec.~\ref{sec:meanfield}, this is indeed the first term that generates a total coupling $K_i(\phi)$ with the $\mathbb Z_4$ symmetry of the square lattice in a globally aligned state at angle $\phi$. 
More specifically, $K_i(\phi)$ acquires a contribution of the form
\begin{equation}
    \frac{\sigma^4}{12} (\cos^4\phi + \sin^4\phi).
\end{equation}
These observations suggest that the pinning phenomenology of the von Mises coupling can be conveniently captured by 
a simplified model that retains only the contributions of order $\sigma^0$, 
$\sigma$, and $\sigma^4$ from the expansion in Eq.~\eqref{eq:vM-exp}. 
The zeroth-order term is included to recover the reciprocal XY model as $\sigma \to 0$.
The term of order $\sigma$ is also retained because, as shown below, it generates a local pinning contribution.
Interestingly, unlike the purely sinusoidal coupling, this contribution here favors the diagonal rather than the lattice directions.
Within this
truncation scheme, the model effectively reduces to the combination of a sinusoidal coupling and a modulated-coupling term. 

The corresponding equations for the local orientation $\phi_i$ and for the mean orientation $\phi$, obtained within this truncation, are derived in Apps.~\ref{app:vonmises-local} and \ref{app:vonmises-mean}, respectively. In particular, the equation for $\phi_i$, written in the form of Eq.~\eqref{eq:local-harmonics-txt}, contains two leading harmonic contributions $S_{\mathrm{R}}^{(1),\mathrm{loc}}$ and $C_{\mathrm{R}}^{(1),\mathrm{loc}}$ (see Eqs.~\eqref{eq-app:SR1-loc-vM} and \eqref{eq-app:CR1-loc-vM}, respectively) which, by symmetry, satisfy $S_{\mathrm{R}}^{(1),\mathrm{loc}}=-C_{\mathrm{R}}^{(1),\mathrm{loc}}$ (see App.~\ref{app:vonmises-local}),
Moreover, numerical measurements show that $S_{\mathrm{R}}^{(1),\mathrm{loc}}<0$. The corresponding contribution to $\partial_t\phi_i$ can therefore be rewritten as $\sqrt{2}\,S_{\mathrm{R}}^{(1),\mathrm{loc}}\sin(\phi_i-\pi/4)$, 
from which the pinning direction $\phi_i = \Phi =\pi/4$ naturally emerges. 
Upon linearization around $\Phi=\pi/4$, this term generates a local pinning term for $\delta\phi_i$ with coefficient
\begin{align}\label{eq:vonmises-loc-pinning}
    r_i^{\mathrm{loc}}(t)
    =&
    -\sqrt{2} S^{(1),\textrm{loc}}_{{\rm R}}>0.
\end{align}
Similarly, the global pinning term governing the evolution of $\delta\phi$ takes the form of Eq.~\eqref{eq:pin-global}, with (see App.~\ref{app:vonmises-mean}) 
\begin{equation}\label{eq:pin-glob-sin-R-maintext}
    r^{\mathrm{glob}} = -\sqrt{2}S^{(1),\textrm{glob}}_{{\rm R}}+4S_{{\rm R}}^{(4),\textrm{glob}},
\end{equation}
where the analytic expressions of the harmonic coefficients $S^{(1),\textrm{glob}}_{{\rm R}}$ and $S_{{\rm R}}^{(4),\textrm{glob}}$ are provided in Eqs.~\eqref{eq-app:SR1-glo-vM} and \eqref{eq-app:SR4-glob-vM}, respectively.
The numerical estimates of these coefficients, reported in Sec.~\ref{sec:stationary-state}, yield $r^{\mathrm{glob}}>0$, confirming the stability of the pinning. 

It is worth emphasizing that the reactive coefficients $S_{\rm R}^{(1),\mathrm{loc}}$, $C_{\rm R}^{(1),\mathrm{loc}}$, $S_{\rm R}^{(1),\mathrm{glob}}$, and $C_{\rm R}^{(1),\mathrm{glob}}$ for the sinusoidal coupling  have the same analytical expressions as for the von Mises kernel, but with different prefactors. 
In particular, they are proportional to  $\epsilon$ in the former case and to  $\sigma$ in the latter:  compare, e.g., the expressions of $C_R^{(1),\mathrm{glob}}$ in Eqs.~\eqref{eq-app:CR1-glob-SK} and \eqref{eq-app:CR1-glo-vM}. 
(The same applies to the corresponding proactive coefficients, but we consider here the case of reactive dynamics.)
Nevertheless, their physical effect is qualitatively different in the two models, which indeed feature different pinning directions.  
This is due to the fact that coefficients with the same analytical expression actually take different values depending on the pinning direction $\Phi$ selected by the dynamics.

In fact, for the sinusoidal kernel, pinning occurs always along the lattice directions, e.g., $\Phi=0$. For this specific pinning, the  symmetry argument discussed in Apps.~\ref{app:sin-local} and ~\ref{app:sin-global} implies that  $C^{(1),\cdots}_{\cdots} =0$, while the non-vanishing coefficients $S^{(1),\cdots}_{\cdots}$ eventually yield pinning. By contrast, for the von Mises kernel with reactive dynamics, pinning occurs along the diagonal directions, e.g., $\Phi=\pi/4$. 
The relevant symmetry is then the diagonal reflection discussed in App.~\ref{app:vonmises-mean}, which implies $S^{(1),\cdots}_{\cdots}=-C^{(1),\cdots}_{\cdots}$.
However, these first-harmonic terms alone are not sufficient to explain the observed diagonal pinning in the (reactive) von Mises case. 
Indeed, if only these ${\cal O}(\sigma)$ contributions were retained, the resulting dynamics would coincide with that of the sinusoidal kernel, thereby leading to pinning along the lattice directions.
This means that the selection of pinning along the diagonal directions is actually driven by the fourth harmonic term, that however does not appear in the local pinning, but only in the global one with $S_{{\rm R}}^{(4),\textrm{glob}}$ of ${\cal O}(\sigma^4)$, exactly as in the modulated coupling  discussed in Sec.~\ref{sec:pinning-modulated} above.
Summarizing, the fourth-harmonic contribution to the von Mises coupling is ultimately responsible for selecting the diagonal pinning directions. However, this contribution does not appear directly in the dynamics of individual spins. 
This local pinning, instead, originates from the ${\cal O}(\sigma)$ contribution, the analytical form of which is the same as in the sinusoidal kernel. 
The role of the fourth-harmonic term is to modify the stationary state, thereby altering the actual numerical values of the ${\mathcal O}(\sigma)$ coefficients, making them consistent with pinning along the diagonal directions.

In this context, a natural question is whether, for sufficiently small $\sigma$, the pinning direction eventually reverts to the lattice axes, or whether an arbitrarily small fourth-harmonic terms is already sufficient to stabilize the diagonal directions. Since the pinning effects become very weak in this regime, our numerical results (data not shown here) are at present inconclusive on this point. Similarly, the effects on the transient dynamics of adding the terms of order $\sigma^2$ and $\sigma^3$ remains to be investigated numerically.

Finally, when the proactive term is included in the dynamics, the local pinning receives a contribution only from the ${\cal O}(\sigma^4)$ terms (see App.~\ref{app:vonmises-local}), whereas the global pinning receives contributions from both the ${\cal O}(\sigma)$ and ${\cal O}(\sigma^4)$ terms (see App.~\ref{app:vonmises-mean}). In this case, the pinning occurs along the lattice directions, consistently with the mean-field argument presented in Sec.~\ref{sec:meanfield} and with the numerical results reported in Sec.~\ref{sec:numerical-mean-orienetation}.

\subsection{Proactivity and pinning: A summary}
\label{sec:pin-summary}

\begin{table}[t]
\centering
\caption{\small
Summary of the presence of a non-vanishing local pinning coefficients $r_i^{\mathrm{loc}}$ in the effective equation for $\phi_i$, coming from the reactive and proactive part of Eq.~\eqref{eq:local-harmonics-txt} for the various interaction kernels considered in this work. %
}
\vspace{0.2cm}
\label{tab:localr}
\begin{tabular}{l|l|l}
\hline\hline
\textbf{Kernel type}
&
\textbf{Reactive}
&
\textbf{Proactive}
\\
\hline
Modulated
&
no
&
yes
\\
Sinusoidal
&
yes
&
no
\\
von Mises
&
yes at ${\cal O}(\sigma)$
&
yes at ${\cal O}(\sigma^4)$
\\
\hline\hline
\end{tabular}
\end{table}

Before illustrating, in the next Section, the results of our numerical simulations, we summarize the content of this Sec.~\ref{sec:pinning}.
We have systematically investigated here the emergence of local and global pinning terms in the equations governing the dynamics of the individual spins and of the mean orientation for all the interaction kernels considered in this work.

A first important result is that %
pinning does not necessarily appear in the local equations of motion; accordingly, it might be deceptive to focus exclusively on them. Moreover, the possible emergence of pinning and its eventual direction depend on the specific implementation of the dynamics (i.e., Langevin equation with the sole reactive term or with both reactive and proactive terms, this latter case being equivalent to Monte Carlo dynamics).
In particular, for the modulated coupling, the reactive term does not generate any local pinning, whereas the proactive term does. For the sinusoidal kernel, the situation is reversed: the reactive term generates a local pinning contribution, while the proactive term does not. For the von Mises kernel (truncated such as to include terms of order $\sigma^0$, $\sigma$, and $\sigma^4$), instead, the reactive contribution to the local pinning originates from the  term at ${\cal O}(\sigma)$, corresponding to the sinusoidal component of the interaction, whereas the proactive contribution to pinning appears only at ${\cal O}(\sigma^4)$, corresponding to the modulated kernel component. These results are summarized in Tab.~\ref{tab:localr}.

By contrast, \emph{all} the dynamical contributions considered here generate a global pinning term with coefficient $r^{\mathrm{glob}}$, acting on the mean orientation $\phi(t)$. A more detailed analysis of this pinning mechanism is presented in Sec.~\ref{sec:stationary-state}, together with a discussion of its relation with the properties of the system in the stationary state. 

\section{Numerical results for the mean orientation}
\label{sec:numerical-mean-orienetation}

In this section, we present the results of the numerical simulations of the non-reciprocal XY model for the different interaction kernels and dynamical implementations introduced in Sec.~\ref{sec:models}. In particular, we focus on the evolution of the mean orientation during both the transient regime and the subsequent (quasi)stationary state,  allowing a direct comparison between the numerical results and the theoretical predictions developed in the previous sections.  

We performed Langevin simulations for the three smooth kernels (modulated, sinusoidal, and von Mises, see Sec.~\ref{sec:kernels}), considering either the reactive term alone (i.e., only the first term on the r.h.s.~of Eq.~\eqref{eq:langevin}) or the complete dynamics which includes also the proactive term (i.e., both terms on the r.h.s.~of Eq.~\eqref{eq:langevin}).

The hard vision cone requires a separate treatment.
In this case, only the reactive term can be straightforwardly implemented within the Langevin framework. Indeed, the corresponding proactive contribution 
generates Dirac delta functions in the corresponding Langevin equation, 
rendering a direct numerical integration impractical.
To circumvent this difficulty, 
instead of simulating the complete Langevin equation, we resort to Glauber dynamics (see Sec.~\ref{subsec:Glauber}) as a proxy for the 
complete dynamics that includes the proactive term.
The Glauber update rule is constructed from the corresponding selfish energy, i.e., the energy functional whose variation generates the Dirac-delta terms appearing in the Langevin equations.

\subsection{Transient dynamics}\label{sec:transient}

Starting from an initial configuration in which all the spins are aligned in the same direction, i.e., with the same value of $\phi_i$, the system is allowed to evolve until all observables reach stationary values. In particular, the dynamics of the mean orientation $\phi(t)$ defined in Eq.~\eqref{eq:def-phi} provides indication about the eventual pinning direction in phases exhibiting LRO. 
Even in those cases in which such a phase is known to be metastable due to eventual nucleation and proliferation of defects
(for example the sinusoidal and hard vision cone with only the reactive term~\cite{dopierala2025}),
the parameters of the model are chosen such that, under Langevin dynamics, these processes occur on timescales much longer than the  longest time reached in the present work \cite{Besse_2022, dopierala2025}.  
This separation of time scales allows us to gather sufficient statistics for a quantitative analysis of the effective pinning mechanisms. 
%

\begin{figure*}
    \centering
    \includegraphics[width=0.96\linewidth]{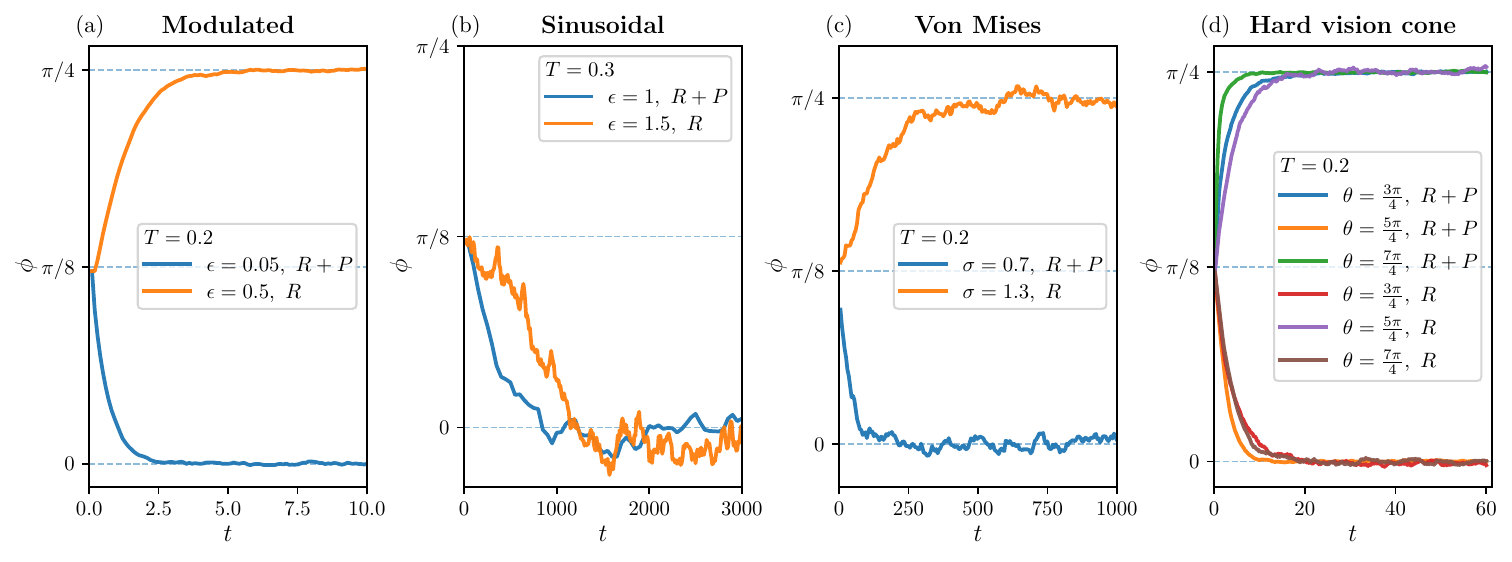}
    \caption{\small 
    Results of numerical simulations of the transient dynamics of the mean orientation $\phi(t)$ (see Eq.~\eqref{eq:def-phi}) for the four interaction kernels considered (system size $L=200$ for all simulations). Starting from an initial condition ${\phi}(0)=\pi/8$, we monitor whether the system relaxes toward ${\Phi}=0$, corresponding to pinning along the lattice directions $n\pi/2$ with integer $n$, or toward ${\Phi}=\pi/4$, corresponding to pinning along the diagonal directions $\pi/4 + n\pi/2$. 
    The numerical results clearly show that, whenever the mean-field argument based on the total coupling $K_i(\phi)$
    and presented in Sec.~\ref{sec:meanfield} 
    applies, 
    the reactive dynamics (curves labeled by R) tends to minimize the coupling (argument of Ref.~\cite{dopierala2025}), whereas the complete Langevin dynamics (curves labeled by R + P) or, equivalently, the Glauber dynamics tend to maximize it.
    For the sinusoidal kernel (panel (b)), the mean-field argument is inconclusive; 
    nevertheless, both these numerical results and the analytical considerations discussed in Sec.~\ref{sec:pinning-sinusoidal} 
    show that pinning occurs along the lattice directions, in agreement with~\cite{dopierala2025}. For the hard vision cone (panel (d)), the orientation that maximizes or minimizes the number of visible neighbors depends explicitly on the vision cone aperture $\theta$, as discussed in Refs.~\cite{Loos2023, bandini2025}: as a consequence, the pinning angle varies accordingly.
    When comparing the evolution under different dynamics (namely, the reactive Langevin, the complete Langevin or, alternatively, the Glauber dynamics)  the parameters  in panels (a), (b), and (c) have been chosen such that the corresponding relaxation timescales are comparable. 
    The observed selection of the pinning direction is robust with respect to choice of the values of the simulation parameters, as we confirmed for a number of instances.
}
    \label{fig:phibar_4panels}
\end{figure*}
%
%
Figure~\ref{fig:phibar_4panels} reports the numerical data corresponding to the transient dynamics. In particular, starting from the initial condition $\phi_i(0)=\phi(0)=\pi/8$ for all $i$,
the system always relaxes towards one of the two possible states: ${\Phi}=0$, corresponding to pinning along the lattice directions, or ${\Phi}=\pi/4$, corresponding to pinning along the diagonal directions. 
Which one is eventually selected depends on the choice of the kernel and of the dynamics, as shown by the various panels in the figure and as summarized in Table~\ref{tab:summary}.

For the modulated (see Fig.~\ref{fig:phibar_4panels}(a)), von Mises (Fig.~\ref{fig:phibar_4panels}(c)), and hard vision cone (Fig.~\ref{fig:phibar_4panels}(d)) kernels, the observed pinning directions are in agreement with the mean-field predictions 
discussed in Sec.~\ref{sec:meanfield}. 
When only the reactive term is included in the dynamics (orange curves in Fig.~\ref{fig:phibar_4panels}), the selected direction corresponds to the minimum of the total coupling $K_i(\phi)$. This yields $\Phi=\pi/4$ for both the modulated and von Mises couplings 
(Figs.~\ref{fig:phibar_4panels}(a) and \ref{fig:phibar_4panels}(c)), while for the hard vision cone kernel one finds (Fig.~\ref{fig:phibar_4panels}(d))
\begin{equation}
    \Phi = \frac{\pi}{4} \,[1-(q\bmod 2)] \quad \text{with}\quad q=\lfloor \theta/(\pi/2)\rfloor \in \{0,1,2,3\}
\end{equation}
where $\theta$ denotes the vision-cone aperture.
%
Equivalently, $\Phi=\pi/4$ for $0<\theta<\pi/2$ and $\pi<\theta<3\pi/2$, whereas $\Phi=0$ otherwise.
Upon including also the proactive term in the Langevin dynamics (or, in the case of the hard vision cone kernel, upon  employing the corresponding Glauber dynamics), the selected direction (indicated by the blue curves in Fig.~\ref{fig:phibar_4panels}) corresponds, instead, to the maximum of the total coupling. Accordingly, one finds $\Phi=0$ for the modulated and von Mises kernels (Figs.~\ref{fig:phibar_4panels}(a) and \ref{fig:phibar_4panels}(c)), while
\begin{equation}
    \Phi = \frac{\pi}{4} \,(q\bmod 2)
\end{equation}
for the hard vision cone (Fig.~\ref{fig:phibar_4panels}(d)), i.e., $\Phi = \pi/4$ for  $\pi/2<\theta<\pi$ and $3\pi/2<\theta<2\pi$, whereas $\Phi=0$ otherwise. 
These results are consistent with the findings reported in Ref.~\cite{dopierala2025} where it was also noted, without further explanation, that the direction of the observed pinning changes when Monte Carlo dynamics is employed instead of  Langevin dynamics. The analysis presented here 
provides an explanation of this behavior by identifying the distinct roles played by the reactive and proactive contributions.

Notably, the sinusoidal coupling is the only case considered here in which the inclusion of the proactive term in the dynamics
does not alter the pinning direction.
In both the purely reactive and the complete dynamics, the system remains pinned along the lattice axes ($\Phi = 0$), as shown by the curves in Fig.~\ref{fig:phibar_4panels}(b). 

Finally, we note that, in the limit $\sigma \ll 1$,
the von Mises coupling becomes asymptotically equivalent to the sinusoidal coupling . It is therefore 
natural to ask
whether there exists a threshold value $\sigma^\ast$ (with $\sigma^\ast\ll 1$ )
at which, in the purely reactive dynamics, the preferred pinning direction changes abruptly from the lattice axes ($\Phi = 0$ for $\sigma < \sigma^\ast$) to the diagonal direction ($\Phi = \pi/4$ for $\sigma > \sigma^\ast$). 
At present, our numerical results at very small values of $\sigma$ 
do not allow us to draw a definite conclusion. 
Resolving this issue would likely require simulations on significantly larger system sizes than those considered here.

\begin{table*}[t]
\centering
\caption{\small  
Summary of the 
pinning directions induced by the presence of the lattice, depending on the specific form of the interaction kernels and dynamics considered.
For the hard vision cone kernel, $q=\lfloor \theta/(\pi/2)\rfloor \in \{0,1,2,3\}$ labels the quadrant of the vision cone aperture $\theta$, and determines the alternation of the $\pi/4$ shift between even and odd quadrants.
}
\vspace{0.2cm}
\label{tab:summary}
\begin{tabular}{l|c|c|c|c}
\hline\hline
\textbf{Kernel type} 
& \textbf{Functional form} 
& \textbf{Vision bias}  
& \multicolumn{2}{c}{\textbf{Pinning direction}}\\
&&
& \textbf{Reactive}
& \textbf{Reactive + Proactive} \\
\hline
Modulated coupling 
& $1+\epsilon\cos(4\phi_i)$ 
& None  
& $\frac{\pi}{4} + n\frac{\pi}{2}$ 
& $n\frac{\pi}{2}$ \\
Sinusoidal 
& $1+\epsilon\cos(\phi_i-\vartheta_{ij})$ 
& Soft  
& $n\frac{\pi}{2}$ 
& $n\frac{\pi}{2}$ \\
von Mises 
& $\exp\{\sigma\cos(\phi_i-\vartheta_{ij})\}$ 
& Soft  
& $\frac{\pi}{4} + n\frac{\pi}{2}$ 
& $n\frac{\pi}{2}$ \\
Hard vision cone 
& step function, aperture $\theta$ 
& Hard  
& $\frac{\pi}{4} \,[1-(q\bmod 2)] +n\frac{\pi}{2}$ 
& $\frac{\pi}{4}\,(q\bmod 2) + n\frac{\pi}{2}$ \\[1mm]
\hline\hline
\end{tabular}
\end{table*}

\subsection{Stationary state}\label{sec:stationary-state}

In this section, we focus on the behavior of the system in the (quasi)stationary state. In particular, we verify that the pinning terms predicted from the analysis of the evolution equations for the global orientation $\phi(t)$ are consistent with the orientation effectively selected by the system and detected numerically.

After the system has reached the stationary state, we compute the harmonic coefficients $S_{\alpha}^{(n),\mathrm{glob}}(t)$ and $C_{\alpha}^{(n),\mathrm{glob}}(t)$ appearing in Eq.~\eqref{eq:global-harmonics-txt}. This is done by numerically averaging over the fluctuations the corresponding expressions derived in the Apps.~\ref{app:modulated-mean}, \ref{app:sin-global}, and \ref{app:vonmises-mean}.
for the different interaction kernels. 
We then perform additional time averages in order to estimate their mean values $\overline{S_{\alpha}^{(n),\mathrm{glob}}}$ and $\overline{C_{\alpha}^{(n),\mathrm{glob}}}$ together with the associated standard deviations. Since these coefficients are global quantities, their fluctuations are expected to vanish in the thermodynamic limit $L\to\infty$.

The resulting averages and standard deviations, along with the values of the parameters used in the simulations, are reported in Table~\ref{tab:pinning_coeffs}.
In the following, we discuss the implications of these numerical results for each interaction kernel separately. 
For notational simplicity, we henceforth omit the superscript ``glob" and the indication of the time average from the harmonic coefficients.
\begin{table*}[t]
\centering
\caption{Numerical estimates of the harmonic coefficients $S_{\mathrm{R},\mathrm{P}}^{(n),\textrm{glob}}$ and  $C_{\mathrm{R},\mathrm{P}}^{(n),\textrm{glob}}$ 
appearing in the equation of the mean orientation $\phi$ (see Eq.~\eqref{eq:global-harmonics-txt}) for the different 
interaction kernels and dynamical implementations.
To simplify the notation, the superscript ``glob'' is omitted in the table.
The reported values are time averages obtained for finite system sizes; uncertainties correspond to the associated standard deviations and are expected to vanish in the thermodynamic limit. 
Entries denoted by ``--'' indicate coefficients that are absent by construction. All the simulations are performed with $L = 200$ and $T= 0.2$.
While the precise numerical values of the various coefficients depend on the chosen parameters, the eventual stability of the pinned state is actually determined only by the sign of their combinations which enter the global pinning coefficient $r^{\mathrm{glob}}$.
These signs are found to be robust upon changing the parameters as long as they give rise to a (quasi)stationary state with pinning.}
\label{tab:pinning_coeffs}
\setlength{\tabcolsep}{4pt}
\renewcommand{\arraystretch}{1.15}

\resizebox{\textwidth}{!}{%
\begin{tabular}{@{} l   cccc  cccc @{}}
\toprule
\textbf{Model} 
& $S_{\mathrm{R}}^{(1)}$ & $C_{\mathrm{R}}^{(1)}$ & $S_{\mathrm{P}}^{(1)}$ & $C_{\mathrm{P}}^{(1)}$
& $S_{\mathrm{R}}^{(4)}$ & $C_{\mathrm{R}}^{(4)}$ & $S_{\mathrm{P}}^{(4)}$ & $C_{\mathrm{P}}^{(4)}$ \\
\midrule
Modulated --  reactive, $\epsilon = 0.5 $
&--  &--  &--  &--
&0.257(1)  &0.0003(10)  &--  &--  \\
Modulated -- total, $\epsilon = 0.05$
&--  &--  &--  &--
&0.0196(1)  &0.00001(200)  & -0.548(2) & 0.0001(10) \\
Sinusoidal --  reactive, $\epsilon= 1.5$
&-0.0057(3)  & -0.0001(3) 
&--  &-- &--  &--  &--  &--  \\
Sinusoidal --  total , $\epsilon = 1.0$
& -0.0065(3) & 0.0001(2) &-0.0065(3) &0.0001(2)
&--  &-- &-- &--  \\
von Mises --  reactive   , $\sigma = 1.5$
& -0.0015(2) & 0.0015(2) &--  &--
&0.00229(4) &0.000004(20)  &--  &--  \\
von Mises --  total , $\sigma = 0.7$ 
& -0.0017(2) & 0.00001(10) &  -0.0017(2)&0.00001(10)
& 0.00120(3) & 0.000001(10) & -0.0234(6) & -0.00006(20) \\
\bottomrule
\end{tabular}%
}
\end{table*}
%

\begin{itemize}
    \item \textbf{Modulated coupling.}
The analytical derivation of the dynamics of the mean orientation $\phi(t)$ under the complete Langevin evolution 
is provided in App.~\ref{app:modulated-mean} and yields the equation:
\begin{equation}
\begin{aligned}
\partial_t \phi =\;&
(S^{(4)}_{\mathrm{R}} + S^{(4)}_{\mathrm{P}})\sin(4\phi) +
 \\
&+
(C^{(4)}_{\mathrm{R}}+
C^{(4)}_{\mathrm{P}})\cos(4\phi) + \sqrt{\frac{2T}{N}}\eta(t),
\end{aligned}
\label{eq:meanphi_final_modulated}
\end{equation}
where the analytic expressions of the harmonic coefficients $S^{(4)}_{\mathrm{R}}$, $S^{(4)}_{\mathrm{P}}$, $C^{(4)}_{\mathrm{R}}$, and $C^{(4)}_{\mathrm{P}}$ are provided by Eqs.~\eqref{eq-app:SR4glob}, \eqref{eq-app:SP4glob}, \eqref{eq-app:CR4glob}, and \eqref{eq-app:CP4glob}, respectively.
As implied by the symmetry arguments presented in App.~\ref{app:modulated-mean},
and confirmed by the numerical averages reported in Table~\ref{tab:pinning_coeffs}, the coefficients $C^{(4)}_{\mathrm{R}}$ and $C^{(4)}_{\mathrm{P}}$ vanish within the numerical precision. This holds both for the complete dynamics and for the one involving only the reactive term, in which case $C^{(4)}_{\mathrm{P}}$ is absent by construction.
When the dynamics is governed solely by the reactive term, Eq.~\eqref{eq:meanphi_final_modulated} reduces to
\begin{equation}
\partial_t \phi =
S^{(4)}_{\mathrm{R}} \sin(4\phi) +\sqrt{\frac{2T}{N}}\eta(t).
\end{equation}
In our simulations, we let the system relax and reach its stationary state, which, as shown in Fig.~\ref{fig:phibar_4panels}(a), is pinned along one of the diagonal directions, $\Phi=\pi/4$, consistently with the mean-field arguments discussed in Sec.~\ref{sec:meanfield}. Evaluating the coefficient $S^{(4)}_{\mathrm{R}}$ in this stationary state, we find $S^{(4)}_{\mathrm{R}}>0$, as reported in Tab.~\ref{tab:pinning_coeffs}.

Expanding around the pinned direction as $\phi=\Phi+\delta\phi$, with $\Phi=\pi/4$, one obtains, at linear order, an equation of the form of Eq.~\eqref{eq:pin-global}, where the relaxation rate is given by
\begin{equation}
r^{\mathrm{glob}}
=
4S^{(4)}_{\mathrm{R}}.
\end{equation}
Since $r^{\mathrm{glob}}>0$, the stationary distribution of $\delta\phi$ is centered around zero with finite variance. This confirms that the pinned direction $\Phi=\pi/4$ is dynamically stable.

By contrast, in the  presence of the additional proactive term, Eq.~\eqref{eq:meanphi_final_modulated} becomes
\begin{equation}\label{eq:pinning-modulated-proactive}
\partial_t \phi =
\left(
S^{(4)}_{\mathrm{R}}
+
S^{(4)}_{\mathrm{P}}
\right)
\sin(4\phi)
+
\sqrt{\frac{2T}{N}}
\eta(t).
\end{equation}
In our simulations, the stationary state is found to be pinned along one of the lattice directions, e.g., $\Phi=0$ (see Fig.~\ref{fig:phibar_4panels}(a)), consistently with the mean-field arguments of Sec.~\ref{sec:meanfield}. In this state, we find numerically
$
S^{(4)}_{\mathrm{R}}
+
S^{(4)}_{\mathrm{P}}
<
0,
$
as reported in Tab.~\ref{tab:pinning_coeffs}.

Expanding around the pinned direction as $\phi=\Phi+\delta\phi$, with $\Phi=0$, one obtains again Eq.~\eqref{eq:pin-global}
with relaxation rate
\begin{equation}
r^{\mathrm{glob}}
=
-4
\left(
S^{(4)}_{\mathrm{R}}
+
S^{(4)}_{\mathrm{P}}
\right).
\end{equation}
Since $r^{\mathrm{glob}}>0$,  the pinned direction $\Phi =0$ is indeed dynamically stable.

These stability arguments, established for the pinned directions $\Phi=\pi/4$ in the purely reactive dynamics and $\Phi=0$ in the full dynamics, extend straightforwardly to all symmetry-equivalent diagonal and lattice directions.

\item \textbf{Sinusoidal coupling.}
The derivation of the evolution equation for the mean orientation is presented in App.~\ref{app:sin-global} and leads to
\begin{equation}
\begin{split}
\partial_t \phi =& 
(S_{\mathrm{R}}^{(1)}+S_{\mathrm{P}}^{(1)})\sin\phi \\
&+ (C_{\mathrm{R}}^{(1)}+C_{\mathrm{P}}^{(1)})\cos\phi+
\sqrt{\frac{2T}{N}}
\eta(t),
\end{split}
\label{eq:meanphi_sinusoidal}
\end{equation}
where the analytic expressions for the harmonic coefficients $S_{\mathrm{R}}^{(1)}$, $S_{\mathrm{P}}^{(1)}$, $C_{\mathrm{R}}^{(1)}$, and $C_{\mathrm{P}}^{(1)}$ are provided in Eqs.~\eqref{eq-app:SR1-glob-SK}, \eqref{eq-app:SP1-glob-SK}, \eqref{eq-app:CR1-glob-SK}, and \eqref{eq-app:CP1-glob-SK}, respectively.
However, under periodic boundary conditions, these coefficients are not independent but they satisfy the relations $S_{\mathrm{R}}^{(1)}=S_{\mathrm{P}}^{(1)}$ and $C_{\mathrm{R}}^{(1)}=C_{\mathrm{P}}^{(1)}$, as shown in Sec.~\ref{app:sin-global}.
In our simulations, the stationary state is found to be pinned along one of the lattice directions, e.g., $\Phi=0$, for both the purely reactive and the complete dynamics (see Fig.~\ref{fig:phibar_4panels}(b)). Unlike the case of the  modulated coupling, here we do not have a simple mean-field argument predicting the selected direction \emph{a priori}. 

Evaluating the coefficients in the stationary state, we find $S_{\mathrm{R}}^{(1)}<0$ and $C_{\mathrm{R}}^{(1)}\simeq0$ 
in the purely reactive case, while for the full dynamics we obtain $S_{\mathrm{R}}^{(1)}=S_{\mathrm{P}}^{(1)}<0$ and $C_{\mathrm{R}}^{(1)}=C_{\mathrm{P}}^{(1)}\simeq0$ (see Tab.~\ref{tab:pinning_coeffs}; the vanishing of the coefficients $C_{\mathrm{R}}^{(1)}$ and $C_{\mathrm{P}}^{(1)}$ is theoretically expected by the symmetry argument reported in App.~\ref{app:sin-global}).
Expanding around the pinned direction as $\phi=\Phi+\delta\phi$, with $\Phi=0$, one obtains Eq.~\eqref{eq:pin-global}
with relaxation rates
$
r^{\mathrm{glob}}
=
-S_{\mathrm{R}}^{(1)} > 0
$
for the purely reactive dynamics, and
$
r^{\mathrm{glob}}
=
-\left(
S_{\mathrm{R}}^{(1)}
+
S_{\mathrm{P}}^{(1)}
\right)
=
-2S_{\mathrm{R}}^{(1)} >0
$
for the full dynamics. In both cases, we find that $r^{\mathrm{glob}}>0$, confirming that the lattice direction $\Phi=0$ is dynamically stable.

In App.~\ref{app:sin-global}, we further show that Eq.~\eqref{eq:meanphi_sinusoidal} preserves the underlying $\mathbb{Z}_4$ symmetry of the lattice. The stability arguments presented above therefore straightforwardly extend to all equivalent lattice directions $\Phi=n\pi/2$, with $n\in\mathbb Z$.

\item \textbf{von Mises coupling.}
In this case, we simulated the dynamics using a simplified version of the von Mises kernel, obtained by truncating its expansion in $\sigma$ and keeping only the terms of order $\sigma^0$, $\sigma^1$, and $\sigma^4$ (see Sec.~\ref{sec:pinning-vonmises}). 
The resulting dynamics is qualitatively equivalent to that generated by the complete von Mises kernel, while allowing for a considerably simpler analytical treatment. 

The analytical derivation of the  dynamics of the mean orientation is provided in App.~\ref{app:vonmises-mean} and yields 
\begin{widetext}    
\begin{equation}
\partial_t \phi
=  
(S_{\mathrm{R}}^{(1)}+S_{\mathrm{P}}^{(1)})\,\sin\phi
+ (C_{\mathrm{R}}^{(1)}+C_{\mathrm{P}}^{(1)})\,\cos\phi
 \\
 \; +
(S^{(4)}_{\mathrm{R}}+ S^{(4)}_{\mathrm{P}})\sin(4\phi) + (C^{(4)}_{\mathrm{R}}+ C^{(4)}_{\mathrm{P}})\cos(4\phi)+\sqrt{\frac{2T}{N}}
\eta(t),
\label{eq:vm-four-terms}
\end{equation}
\end{widetext}
where the analytic expressions of the first harmonic coefficients $S_{\mathrm{R}}^{(1)}$, $S_{\mathrm{P}}^{(1)}$,  $C_{\mathrm{R}}^{(1)}$, and $C_{\mathrm{P}}^{(1)}$ are provided by Eqs.~\eqref{eq-app:SR1-glo-vM}, \eqref{eq-app:SP1-glo-vM}, \eqref{eq-app:CR1-glo-vM}, and \eqref{eq-app:CP1-glo-vM}, respectively; 
those of the forth harmonic $S^{(4)}_{\mathrm{R}}$, $S^{(4)}_{\mathrm{P}}$, $C^{(4)}_{\mathrm{R}}$, and $C^{(4)}_{\mathrm{P}}$ are provided by Eqs.~\eqref{eq-app:SR4-glob-vM}, \eqref{eq-app:SP4-glob-vM}, \eqref{eq-app:CR4-glob-vM}, and \eqref{eq-app:CP4-glob-vM}, respectively. 
In Eq.~\eqref{eq:vm-four-terms} we do not report the terms proportional to the second harmonics $\sin(2\phi)$ and $\cos(2\phi)$, which are discussed in detail in App.~\ref{app:vonmises-mean}. These terms are also of order ${\cal O}(\sigma^4)$, like the fourth-harmonic contributions. However, we find numerically that their contribution to the pinning is significantly smaller than that of the corresponding fourth-harmonic terms and therefore they do not affect the conclusions drawn here.

Symmetry arguments analogous to those used for the modulated coupling and discussed in App.~\ref{app:vonmises-mean} imply that %
$C_{\mathrm{R}}^{(4)} = C_{\mathrm{P}}^{(4)} = 0$. In addition, by the same reasoning employed for the sinusoidal coupling, one finds that, under periodic boundary conditions and in the presence of both reactive and proactive contributions, 
$S_{\mathrm{R}}^{(1)}=S_{\mathrm{P}}^{(1)}$ and $C_{\mathrm{R}}^{(1)}=C_{\mathrm{P}}^{(1)}$.
All these relations are confirmed by the numerical values reported in Tab.~\ref{tab:pinning_coeffs}. 

In the purely reactive case, the stationary state observed in simulations is pinned along one of the diagonal directions, $\Phi=\pi/4$ (see Fig.~\ref{fig:phibar_4panels}(c)), consistently with the mean-field arguments discussed in Sec.~\ref{sec:meanfield}. Evaluating the coefficients in this stationary state, we find
$
S_{\mathrm{R}}^{(1)}
=
-
C_{\mathrm{R}}^{(1)}
<
0,\,\,
S_{\mathrm{R}}^{(4)}
>
0,
$
as reported in Tab.~\ref{tab:pinning_coeffs}. 
Note that the relation $S_{\mathrm{R}}^{(1)} = - C_{\mathrm{R}}^{(1)}$ is theoretically expected based on the symmetries of the problem, as discussed in App.~\ref{app:vonmises-mean}. 
The resulting equation for the mean orientation therefore reduces to
\begin{equation}
\partial_t\phi
=
S_{\mathrm{R}}^{(1)}
\big(
\sin\phi
-
\cos\phi
\big)
+
S_{\mathrm{R}}^{(4)}
\sin(4\phi)
+
\sqrt{\frac{2T}{N}}
\eta(t).
\end{equation}
Expanding around the pinned direction as $\phi=\pi/4+\delta\phi$, one obtains Eq.~\eqref{eq:pin-global}
with
\begin{equation}
r^{\mathrm{glob}}
=
-\sqrt{2}\,
S_{\mathrm{R}}^{(1)}
+
4S_{\mathrm{R}}^{(4)}.
\end{equation}
Since the numerical values of the coefficients in Tab.~\ref{tab:pinning_coeffs} gives $r^{\mathrm{glob}}>0$, 
the diagonally pinned direction is dynamically stable. 

By contrast, when the proactive contribution is included, the stationary state observed in our simulation becomes pinned along one of the lattice directions, $\Phi=0$ (see Fig.~\ref{fig:phibar_4panels}(c)), again in agreement with the mean-field arguments. In this case, we find numerically that 
$
S_{\mathrm{R}}^{(1)}
=
S_{\mathrm{P}}^{(1)}
<
0,
\,\,
C_{\mathrm{R}}^{(1)}
=
C_{\mathrm{P}}^{(1)}
\simeq
0
$
and $S_{\mathrm{R}}^{(4)} + S_{\mathrm{P}}^{(4)}<0$ (with $S_{\mathrm{R}}^{(4)}$ and $S_{\mathrm{P}}^{(4)}$ having opposite signs),  as reported in Tab.~\ref{tab:pinning_coeffs}. 
The vanishing of the coefficients $C_{\mathrm{R}}^{(1)}$ and $C_{\mathrm{P}}^{(1)}$ observed numerically is also theoretically expected from the same symmetry arguments as in the sinusoidal case. 
Expanding around the pinned direction $\Phi=0$ as $\phi=\delta\phi$, we obtain Eq.~\eqref{eq:pin-global} with
\begin{equation}\label{eq:pin-glob-vonmises}
    r^{\mathrm{glob}} = -2S_{\mathrm{R}}^{(1)} - 4(S_{\mathrm{R}}^{(4)}+S_{\mathrm{P}}^{(4)}) 
\end{equation}
Since the numerical values of the coefficients yield $r^{\mathrm{glob}}>0$,  
the pinned direction is dynamically stable.

The arguments above straightforwardly extend to all equivalent diagonal and lattice directions selected by the $\mathbb Z_4$ symmetry of the underlying lattice.

\item \textbf{Hard vision cone.}
This case does not 
admit a straightforward 
analytical treatment due to the singularities introduced by the discontinuous interaction kernel. 
We therefore restrict the discussion to the numerical results which show how the selected orientation depends on the aperture angle $\theta$ 
and of the type of dynamics, as already discussed in Sec.~\ref{sec:transient}.

\end{itemize}

\section{Comparison with correlation time and pinning from the structure factor}\label{sec:comparison}

In this section, we consider two alternative approaches to determine quantities which may provide estimates of the strengths of the local and global pinning terms discussed above. 
In particular, we focus on (i) the correlation time associated with the Ornstein--Uhlenbeck (OU) process 
approximately 
obeyed by the mean orientation $\phi(t)$ and (ii) the pinning term extracted from the structure factor of the fluctuations in the stationary state, which was considered in Refs.~\cite{dopierala2025, popli2025}.

\subsection{Correlation time}
\label{sec:correlation-time}

Once the mean orientation $\phi(t)$ is pinned along the direction $\Phi$, its fluctuations 
$\delta\phi(t)$ around this value (see Eq.~\eqref{eq:deltaphi}) evolve according to Eq.~\eqref{eq:pin-global}. 
In the stationary state, $r^{\rm glob}(t)$ becomes time-independent as discussed after Eq.~\eqref{eq:pin-global}.
While $\mathcal{A}(t)$ has zero mean, i.e., $\overline{\mathcal{A}}=0$,
in systems of finite size it exhibits fluctuations that vanish in the thermodynamic limit, with standard deviation scaling as $\sigma_{\mathcal{A}}\sim N^{-1/2}$, where $N$ is the total number of spins in the system. 
Likewise, the fluctuations of the mean orientation satisfy $\overline{\delta\phi}=0$ and $\sigma_{\delta\phi}\sim N^{-1/2}$. 
Since both $\delta\phi$ and $\mathcal{A}$ vanish as $N^{-1/2}$ in the thermodynamic limit $N\to\infty$, 
we expect Eq.~\eqref{eq:pin-global} to reduce to the linear evolution equation of a standard OU process whenever the component of $\mathcal A(t)$ correlated with the fluctuations $\delta\phi(t)$ is small compared to the deterministic pinning term $r^{\mathrm{glob}}\delta\phi(t)$. Indeed, any component of $\mathcal A(t)$ that is 
uncorrelated with $\delta\phi(t)$ acts effectively as an additional noise source: it modifies the amplitude of the stationary fluctuations, but not the relaxation rate extracted from the autocorrelation function. 
This leads to the 
condition for the validity of the OU description: 
\begin{equation}
\label{eq:OU-validity}
\sigma_{\mathcal A}\,
\left|
\mathrm{Corr}(\mathcal A,\delta\phi)
\right|
\ll
r^{\mathrm{glob}}\,
\sigma_{\delta\phi},
\end{equation}
where $\mathrm{Corr}(\mathcal A,\delta\phi)$ denotes the normalized correlation coefficient between $\mathcal A(t)$ and $\delta\phi(t)$ in the stationary state.
Under this condition, the contribution of $\mathcal{A}$ in Eq.~\eqref{eq:pin-global} becomes negligible, and the dynamics of $\delta\phi$ is effectively described by an OU process characterized by the relaxation rate $r^{\rm OU} \equiv r^{\mathrm{glob}}$ and noise amplitude $\sqrt{2T/N}$.
%
%
The autocorrelation function of this process is given by 
\begin{equation}
    C_\phi(t) = \langle \delta\phi(0)\,\delta\phi(t)\rangle \sim \exp\left( -r^{\mathrm{OU}} |t|\right),
\end{equation}
which decays exponentially with increasing time, independently of the noise amplitude.
The relaxation rate $r^{\mathrm{OU}}$ can therefore be extracted numerically by fitting  $C_{\phi}(t)$ 
with an exponential law.
This allows a direct comparison between the measured relaxation rate $r^{\rm OU}$ and the pinning coefficient predicted by the analytical theory $r^{\rm glob}$ and calculated from the simulations. 
As we show below, the agreement between the two quantities depends crucially on the validity of the condition in Eq.~\eqref{eq:OU-validity}.
In the following, we discuss two representative cases: the modulated coupling, for which the condition in Eq.~\eqref{eq:OU-validity} is satisfied and the OU description is accurate, and the sinusoidal coupling, for which it is not.

For the modulated coupling (see Sec.~\ref{sec:kernels}) the analytical expression for the dynamics of the mean orientation $\phi(t)$ in the purely reactive case is obtained from Eq.~\eqref{eq:meanphi_final_modulated} by setting the proactive coefficients to zero. Expanding around one of the diagonal pinning directions, e.g., $\Phi=\pi/4$, and comparing the resulting expression with Eq.~\eqref{eq:pin-global}, one readily identifies
\begin{equation}\label{eq:r-glob-et-A}
r^{\mathrm{glob}} = 4S^{(4)}_{\mathrm{R}}
\quad\mbox{and}\quad
\mathcal{A}
=
-C^{(4)}_{\mathrm{R}}
\cos\left(4\delta\phi\right).
\end{equation}
Numerical measurements performed at $T=0.1$, $\epsilon=0.3$, and for a system size $L=200$ yield
$
r^{\mathrm{glob}}\simeq 0.40,
$
while $C^{(4)}_{\mathrm{R}}$ is compatible with zero within statistical fluctuations, as expected from the symmetry argument discussed in App.~\ref{app:modulated-mean}. Moreover, we find
$
\sigma_{\delta\phi}\simeq 3\times 10^{-3},
$
$
\sigma_{\mathcal A}\simeq 4\times 10^{-4},
$
and
$
\mathrm{Corr}(\mathcal A,\delta\phi)\simeq -0.17
$
\footnote{
The correlation coefficient $\mathrm{Corr}(\mathcal A,\delta\phi)$ is computed from the stationary time series of $\mathcal A(t)$ and $\delta\phi(t)$ as the normalized covariance,
\(
\mathrm{Corr}(\mathcal A,\delta\phi)
=
\overline{
\left(\mathcal A(t)-\overline{\mathcal A}\right)
\left(\delta\phi(t)-\overline{\delta\phi}\right)
}
/
{
\sigma_{\mathcal A}\sigma_{\delta\phi}
},
\)
where the overline denotes a time average in the stationary state. By construction,
$-1\le \mathrm{Corr}(\mathcal A,\delta\phi)\le 1$.
}.
Accordingly, the component of $\mathcal A$ that is correlated with the fluctuations of the mean orientation --- given by 
$
\sigma_{\mathcal A}
|\mathrm{Corr}(\mathcal A,\delta\phi)|
$, ---
is much smaller than the restoring contribution
$
r^{\mathrm{glob}}\sigma_{\delta\phi},
$
and the condition in Eq.~\eqref{eq:OU-validity} is 
satisfied. Thus, the dynamics of $\delta\phi$ is accurately described by an OU process with relaxation time $1/r^{\mathrm{glob}}$, so that the autocorrelation time $1/r^{\mathrm{OU}}$ determined numerically should coincide with the analytical prediction $1/r^{\mathrm{glob}}$.
We test this relationship in Fig.~\ref{fig:r-sq}(a) (with $T = 0.1$, $L = 200$,  $dt = 0.05$) where, as a function of the non-reciprocity parameter $\epsilon$,  we compare the values of $r^{\mathrm{glob}}$, obtained by numerically by averaging the corresponding analytical expressions (see Eqs.~\eqref{eq:r-glob-et-A} and \eqref{eq-app:SR1-glob-SK}), with the values of $r^{\mathrm{OU}}$, extracted by fitting the autocorrelation function of the mean orientation with an exponential decay. 
The two estimates are found to be in excellent agreement over the entire range of non-reciprocity parameters $\epsilon$ considered. 
In Fig.~\ref{fig:r-sq}(b) we compare $r^{\rm glob}$ and $r^{\rm OU}$
after including %
the proactive term %
in the dynamics %
($T = 0.3$, $L = 200$,  $dt = 0.05$), finding an equally good agreement; This confirms that the OU description remains valid. 
%

The situation changes qualitatively for the sinusoidal coupling with the sole reactive Langevin dynamics.
In this case, the analytical expression for the dynamics of the mean orientation $\phi(t)$ is obtained from Eq.~\eqref{eq:meanphi_sinusoidal} by setting the proactive coefficients to zero. Expanding around one of the pinned lattice directions, e.g., $\Phi=0$, and comparing the resulting expression with Eq.~\eqref{eq:pin-global}, one identifies
\begin{equation}
r^{\mathrm{glob}} = -S^{(1)}_{\mathrm{R}}
\qquad\mbox{and}\qquad
\mathcal{A}
=
C^{(1)}_{\mathrm{R}}
\cos\left(\delta\phi\right).
\end{equation}
For the parameters $T=0.2$, $L=200$, and $\epsilon=2.0$, we find numerically
\(
r^{\mathrm{glob}}\simeq 10^{-2}
\), 
\(
C_{\rm R}^{(1)}
\)
compatible with zero,
\(
\sigma_{\mathcal A}\simeq 10^{-3}
\),
\(
\sigma_{\delta\phi}\simeq 6\times 10^{-2}
\), and
\(
\mathrm{Corr}(\mathcal A,\delta\phi)\simeq 0.73.
\)
In this case one finds
\begin{equation}
\sigma_{\mathcal A}
\left|
\mathrm{Corr}(\mathcal A,\delta\phi)
\right|
\sim
r^{\mathrm{glob}}\sigma_{\delta\phi},
\end{equation}
so that the condition in Eq.~\eqref{eq:OU-validity} is no longer satisfied. 
Accordingly, the contribution of $\mathcal A$ cannot be regarded as a small perturbation 
or as an additional uncorrelated source of noise.
Instead, it contributes significantly to the dynamics of the mean orientation.
Consistently with this picture, the relaxation rate extracted from the autocorrelation function is found to be
\(
r^{\mathrm{OU}}\simeq 10^{-3},
\)
approximately one order of magnitude smaller than the value $r^{\mathrm{glob}}\simeq 10^{-2}$ predicted analytically. 
This discrepancy indicates that, in this regime, the dynamics of the mean orientation cannot be accurately described by a simple OU process with relaxation rate determined solely by the global pinning coefficient.

By systematically analyzing all the interaction kernels considered in this work (see Sec.~\ref{sec:kernels}), 
we find that 
the relaxation rate $r^{\rm OU}$ provides a reliable estimate of the global pinning coefficient $r^{\rm glob}$ only for the modulated coupling, both in the purely reactive dynamics (see Fig.~\ref{fig:r-sq}(a)) and  when the proactive term is included (see Fig.~\ref{fig:r-sq}(b)). 
For the remaining kernels, the additional terms contained in $\mathcal A(t)$ contribute significantly to the dynamics of the mean orientation and do not allow a direct identification of $r^{\rm OU}$ and $r^{\rm glob}$.

\subsection{Pinning extracted from the structure factor}\label{sec:correlation-length}
\begin{figure}
    \centering
    \includegraphics[width=0.99\linewidth]{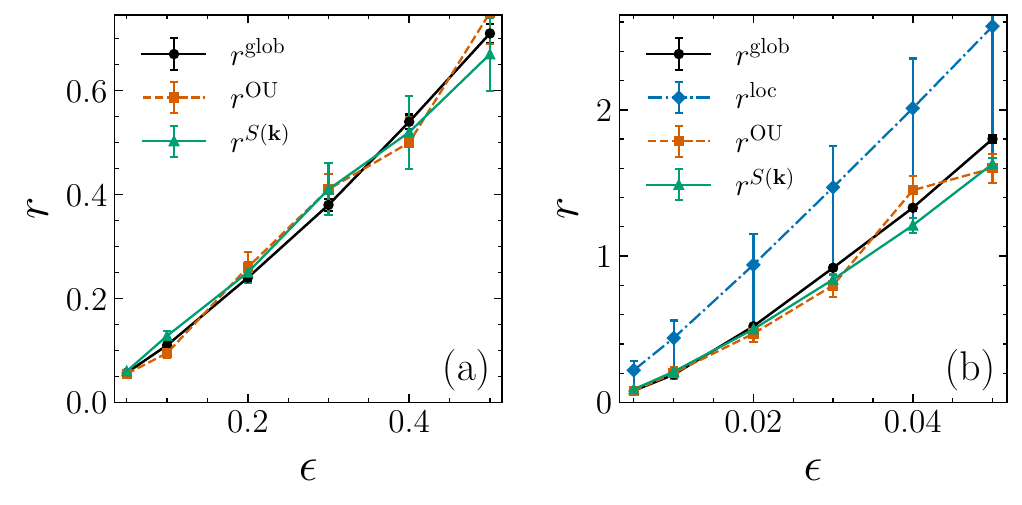}
    \caption{%
Pinning coefficients for the modulated-coupling kernel (see Sec.~\ref{sec:kernels}) obtained by using four different methods.
Panel (a) corresponds to the Langevin dynamics with only the reactive term. In this case, the global pinning coefficient $r^{\mathrm{glob}}$, the relaxation rate $r^{\mathrm{OU}}$ extracted from the OU autocorrelation time, and the pinning coefficient $r^{S(\mathbf{k})}$ obtained from the static structure factor are in excellent agreement. This indicates that, despite the absence of an explicit local pinning term in the continuum equations, the spatial fluctuations are accurately described by an effective pinning equal to $r^{\mathrm{glob}}$.
Panel (b) corresponds to the complete Langevin dynamics including both the reactive and proactive terms. Again, $r^{\mathrm{glob}}$, $r^{\mathrm{OU}}$, and $r^{S(\mathbf{k})}$ are found to be in very good agreement. By contrast, the mean local pinning coefficient $r^{\mathrm{loc}}$, which is generated by the proactive term and which is absent in panel (a), differs systematically from these quantities. %
The ranges of the non-reciprocity parameter $\epsilon$ explored in panels (a) and (b) are different, but were chosen so as to produce pinning coefficients of comparable magnitude.
}
\label{fig:r-sq}
\end{figure}

In this section, we compare the global pinning coefficient $r^{\rm glob}$ and the average local pinning coefficient $r^{\rm loc}$ with the effective  pinning term $r^{S(\mathbf{k})}$
extracted from the static structure factor of configurations in the stationary state.

We start from the evolution equation for the fluctuations $\delta\phi_i$ around a pinned direction $\Phi$, 
given in Eq.~\eqref{eq:local-continuum}.
If the modulation of the diffusion coefficient is weak, i.e., if $\kappa(\delta\phi_i)\simeq \kappa$, the fluctuations 
of the local pinning coefficient 
satisfy
$
\Delta_{r^{\mathrm{loc}}}
\ll
r^{\mathrm{loc}},
$
and the additional contributions $\mathcal A_i$  can be neglected, 
then Eq.~\eqref{eq:local-continuum} reduces to the simplest scalar theory characterized by a pinning coefficient $r^{\mathrm{loc}}$. 
In this regime, $r^{\mathrm{loc}}$ directly controls the long-wavelength spatial correlations.
However, once the fluctuations of the local pinning coefficient and the neglected terms $\mathcal A_i$ become non-negligible, the effective pinning governing the spatial correlations may 
be renormalized by fluctuations.
To investigate the 
magnitude 
of this effect, we compare the analytical estimates of the pinning coefficients with the pinning extracted from the static structure factor \(
S(\mathbf{k})
=
\left\langle
|\phi(\mathbf{k})|^2
\right\rangle
\)
of the fluctuations,
where $\phi(\mathbf{k})$ denotes the Fourier transform of the local field $\phi_i$.

For a scalar field obeying the equation
\begin{equation}
\partial_t \delta\phi_i
=
\kappa^{S(\mathbf k)}
\nabla_{\mathrm d}^2 \delta\phi_i
-r^{S(\mathbf{k})}\delta\phi_i
+\sqrt{2T}\,\eta_i,
\label{eq:effective-Sk}
\end{equation}
the stationary structure factor is given by the lattice Ornstein--Zernike form
\begin{equation}
S(\mathbf{k})
=
\frac{T}
     {r^{S(\mathbf{k})}
      +
      \kappa^{S(\mathbf k)}\,\hat{k}^2},
\label{eq:fit-Sk}
\end{equation}
where $\hat{k}^2=4\sin^2(k_x/2)+4\sin^2(k_y/2)$  is the squared lattice momentum.
In the fits discussed below, the temperature $T$ is fixed to the value used for the simulations, while both
$r^{S(\mathbf{k})}$ and $\kappa^{S(\mathbf k)}$ are treated as free parameters.
The central question is whether the pinning coefficient $r^{S(\mathbf k)}$ extracted from the spatial correlations can be identified with either the average local pinning coefficient $r^{\mathrm{loc}}$  or the global pinning coefficients $r^{\mathrm{glob}}$, and under which conditions such identification is valid.

We address this question for the modulated coupling kernel, considering both the purely reactive dynamics and the complete dynamics which includes also the proactive contribution.
In the purely reactive case, the evolution equation \eqref{eq:local-harmonics-txt} expressed in terms of the 
fluctuations $\delta\phi_{i}=\phi_{i}-\pi/4$ around the pinned direction $\Phi = \pi/4$ takes the form (see Eq.~\eqref{eq:modulated-reactive-local})
\begin{equation}
\partial_t \delta\phi_{i}
=
\left[1-\epsilon\cos{(4\delta\phi_i)}\right]
\nabla^2_{\mathrm{d}}\delta\phi_{i}
+
\sqrt{2T}\,\eta_{i}(t).
\end{equation}
Comparing with Eq.~\eqref{eq:local-continuum}, one can identify $\kappa(\delta\phi_{i})=1-\epsilon\cos{(4\delta\phi_{i})}$, 
while no local pinning term of the form $r^{\mathrm{loc}}_{i}\delta\phi_i$ is present.
This is consistent with the discussion in Sec.~\ref{sec:pinning-modulated}: the continuum equation does not contain a local pinning coefficient $r^{\mathrm{loc}}_{i}$ 
that can be directly compared with the effective pinning extracted from the structure factor.
In the absence of the local pinning term, a natural 
approximation 
is to use the global pinning coefficient
$r^{\mathrm{glob}}$ as a proxy for the 
pinning experienced by the local fluctuations 
and compare it with the effective 
pinning coefficient $r^{S(\mathbf{k})}$ extracted from the static structure factor. 
This is essentially the approach adopted in Ref.~\cite{dopierala2025}, where the local fluctuations are assumed to 
experience the same pinning strength as the global 
orientation.
In the present case, $r^{\mathrm{glob}}$ is obtained by evaluating the analytical expression in Eq.~\eqref{eq-app:SR4glob} and averaging it over the stationary fluctuations. 

In Fig.~\ref{fig:r-sq}(a), we compare the values of $r^{S(\mathbf{k})}$, obtained by fitting the static structure factor with Eq.~\eqref{eq:fit-Sk}, with the corresponding values  
for $r^{\mathrm{glob}}$ for stationary configurations generated at $T=0.1$ and $L=800$, as a function of the non-reciprocity parameter $\epsilon$. 
We find good agreement throughout the entire range of $\epsilon$ considered. This indicates that, for the modulated coupling, the global pinning coefficient provides an accurate effective description of the spatial fluctuations despite the absence of an explicit local pinning term in the evolution equations of the fluctuations.

In the case of the modulated-coupling kernel with the complete dynamics, 
the evolution equation \eqref{eq:local-harmonics-txt} expressed in terms of the 
fluctuations $\delta\phi_{i}=\phi_{i}$ around the pinned direction $\Phi = \pi/4$ takes the form (see Eqs.~\eqref{eq:modulated-reactive-local} and \eqref{eq:sp4-loc-mod})
\begin{equation}
\begin{aligned}
\partial_t \delta\phi_{i}
=&\;
\left[
1+\epsilon\cos\left(4\delta\phi_{i}\right)
\right]
\nabla_{\mathrm{d}}^2\delta\phi_{i}
\\
&+
\frac{\sin\left(4\delta\phi_{i}\right)}
{4\delta\phi_{i}}
S_{\mathrm{P},i}^{(4), \mathrm{loc}}\,
\delta\phi_{i}
+
\sqrt{2T}\,\eta_{i}(t).
\end{aligned}
\end{equation}
Comparing with Eq.~\eqref{eq:local-continuum}, one readily identifies
\begin{equation}
\kappa(\delta\phi_{i})
=
\left[
1+\epsilon\cos\left(4\delta\phi_{i}\right)
\right]
\ \ \mbox{and}\ \ 
r^{\mathrm{loc}}_{i}
=
\frac{\sin\left(4\delta\phi_{i}\right)}
{4\delta\phi_{i}}
S_{\mathrm{P},i}^{(4), \mathrm{loc}}.
\label{eq:diffusion-and-pinning}
\end{equation}
%
Unlike the purely reactive case, an explicit  pinning term is already present in the local equation of motion, since $r^{\mathrm{loc}}_{i}$ remains finite in the limit $\delta\phi_i \to 0$. 
This enables a direct comparison between the 
(spatial) average
$r^{\mathrm{loc}}$ of 
$r_i^{\mathrm{loc}}$ and the effective pinning coefficient
$r^{S(\mathbf{k})}$ extracted from the static structure factor.

In Fig.~\ref{fig:r-sq}(b), we report the comparison between 
$r^{\mathrm{loc}}$ and $r^{S(\mathbf{k})}$ as a function of $\epsilon$. We find that 
the two quantities
are of the same order of magnitude, although they agree less than
$r^{\mathrm{glob}}$ and $r^{\mathrm{OU}}$.
Interestingly, we find that the values of
$r^{S(\mathbf{k})}$ are actually closer to the global pinning coefficient
$r^{\mathrm{glob}}$ than to the  local coefficient
$r^{\mathrm{loc}}$.
This suggests that, 
in the regime where
$\Delta_{r^{\mathrm{loc}}}\lesssim r^{\mathrm{loc}}$ 
(see the errobars of $r^{\rm loc}$ in Fig.\ref{fig:r-sq}(b)\footnote{We emphasize that the error bars associated with $r^{\mathrm{loc}}$ are of a fundamentally different nature from those of $r^{\mathrm{OU}}$, $r^{\mathrm{glob}}$, and $r^{S(\mathbf{k})}$. While the fluctuations underlying the latter quantities vanish in the thermodynamic limit, the uncertainty associated with $r_i^{\mathrm{loc}}$ corresponds to the intrinsic standard deviation $\Delta_{r^{\mathrm{loc}}}$. It therefore reflects genuine spatial and temporal variability of the local pinning coefficient rather than finite-size statistical uncertainty.}), 
the local pinning coefficient undergoes a non-negligible renormalization due to the remaining terms in the local equation of motion. 
As a result, the effective pinning governing the spatial correlations appears to be closer to the global pinning coefficient
$r^{\mathrm{glob}}$ than to the local quantity
$r^{\mathrm{loc}}$ itself. 

The discussion above focused on the simple case of the modulated coupling. For interaction kernels with vision-induced anisotropy, additional complications arise.
First, the form of the evolution equations of the fluctuations in the continuum, analogous to Eq.~\eqref{eq:local-continuum}, become considerably more involved, containing several additional contributions that can renormalize the effective pinning. Second, the local pinning coefficient $r_{i}^{\mathrm{loc}}$ may exhibit large fluctuations. 
This is the case, for instance, of the sinusoidal coupling (see Sec.~\ref{sec:kernels}) for the parameters values considered in Sec.~\ref{sec:pinning-sinusoidal}, which correspond to those studied in Ref.~\cite{dopierala2025}. 
In fact, although the average value $r^{\mathrm{loc}}$ of $r_{i}^{\mathrm{loc}}$ remains positive, indicating an overall tendency toward pinning along lattice directions, strong local fluctuations ($\Delta_{r^{\rm loc}} \gg r^{\rm loc}$) can drive $r_{i}^{\mathrm{loc}}$ towards strongly negative values within finite spatial regions and finite time intervals. 
As a consequence, 
a quantitative comparison between 
$r^{\mathrm{loc}}$ and $r^{\mathrm{glob}}$ with the effective pinning coefficient $r^{S(\mathbf{k})}$ extracted from the static structure factor becomes considerably more subtle.
%
This scenario likely contributes to the discrepancy, of approximately a factor of three, reported in Ref.~\cite{dopierala2025} between the correlation length computed from the global pinning coefficient ($\xi = \sqrt{\kappa / r^{\rm glob}}$), and the value of the correlation length inferred directly from the structure factor.

To summarize, for the modulated-coupling kernel, the assumption that local fluctuations experience the same effective pinning strength as the global orientation 
provides a remarkably accurate description.
This is true both for the purely reactive dynamics, where no explicit local pinning term is present, and for the complete dynamics including the proactive contribution.

In the latter case, the results suggest that the local pinning coefficient is subject to significant renormalization due to the additional terms in the local dynamics, with $r^{\mathrm{glob}}$ playing the role of an effective renormalized pinning that controls the fluctuations of the spins.

The situation is considerably more subtle for the sinusoidal coupling. As already discussed in Sec.~\ref{sec:correlation-time}, even the comparison between $r^{\mathrm{glob}}$ and the relaxation rate $r^{\mathrm{OU}}$ fails in this case. Furthermore, the local pinning coefficient exhibits very large fluctuations, $\Delta_{r^{\mathrm{loc}}}\gg r^{\mathrm{loc}}$, making any effective mean-field description based on a single renormalized pinning coefficient highly questionable. We therefore expect the relation between the analytical pinning coefficients and the correlation length to be substantially more intricate in this case.

\section{Conclusions}
\label{sec:conclusions}

In this work, we carried out a systematic analysis of a single-species non-reciprocal XY model with anisotropic interactions.
We considered four different interaction kernels, three of which feature vision-induced anisotropy.
We showed that two distinct implementations of the spin dynamics naturally emerge. The first one is derived from a selfish-energy functional and leads to a Langevin equation containing an additional term  proportional to the derivative of the interaction kernel. The second consists of a Langevin dynamics in which this term is absent.
We interpreted the additional term as a \emph{proactive} component of the dynamics. While the standard alignment mechanism drives agents to align with their visible neighbors (to react to the sensory input, therefore, a \emph{reactive} term),
the proactive term captures the additional tendency to actively increase their sensory input.
In practice, this term favors orientations that maximize the effective number of visible and partially aligned neighbors,
thereby enhancing the agent's ability to gather information from its surroundings. 

We demonstrated that the pinning of the mean spin orientation induced by the underlying lattice depends sensitively on both the form of the interaction kernel and the microscopic dynamics.
This conclusion is supported by a combination of mean-field arguments, extensive numerical simulations, and explicit analytical calculations of the effective pinning terms acting on the global orientation.

A key result of our analysis is that the existence of a global pinning mechanism does not necessarily imply the presence of a corresponding local restoring force in the dynamic equations for the individual spins.
For the sinusoidal coupling kernel, the reactive dynamics generates a local pinning contribution already at the level of the single-spin equations of motion, whereas the proactive dynamics does not. The opposite situation occurs for the modulated kernel: the reactive contribution does not produce local pinning, while the proactive one does. 
By contrast, all the dynamical contributions considered in this work generate an effective pinning term for the global orientation. Accordingly, global pinning emerges as a more robust and generic feature of the dynamics than local pinning.

Our results also clarify several apparent discrepancies in the existing literature. In particular, for the von Mises coupling kernel with purely reactive dynamics, we show that local pinning originates already at order ${\cal O}(\sigma)$, corresponding to the sinusoidal component of the interaction. The global pinning term, however, receives contributions not only from the ${\cal O}(\sigma)$ term but also, crucially, from the ${\cal O}(\sigma^4)$ contribution, which is the first term to generate the $\mathbb{Z}_4$-symmetric modulation responsible for the selection of the diagonal directions.
As a result, the truncation of the von Mises coupling kernel performed at order ${\cal O}(\sigma^2)$ (such as that considered Ref.~\cite{popli2025}),
is insufficient to fully capture the pinning mechanism and the resulting orientations selection.

We also clarified the distinct physical significance of the local and global pinning coefficients and their respective estimators.
The global coefficient $r^{\mathrm{glob}}$ governs the fluctuations of the mean orientation and, whenever the dynamics is well described by an Ornstein--Uhlenbeck process, can be directly inferred from its relaxation time.
By contrast, the local coefficient $r^{\mathrm{loc}}$ enters the local equations governing the spatial fluctuations of the orientation field, but does not necessarily coincide with the pinning extracted from the static structure factor $r^{S(\mathbf k)}$.
For the modulated-coupling kernel, we find that $r^{S(\mathbf k)}$ is remarkably close to the global pinning coefficient $r^{\mathrm{glob}}$, even when the explicit local pinning term is present. For more complex interaction kernels, however, large fluctuations and additional renormalization effects can substantially modify the dynamics and therefore the relation between the analytical pinning coefficients and the effective pinning extracted from the static structure factor.
We further identify the conditions under which the different estimators provide a meaningful and quantitatively consistent characterization of the pinning mechanism.

An interesting avenue for future work concerns the extension of the present 
framework
to off-lattice systems of self-propelled particles.  The results obtained here suggest that the proactive term can qualitatively modify the effective alignment dynamics. Understanding how these mechanisms affect collective phenomena such as flocking, orientational ordering, and the emergence of long-range correlations in active matter therefore represents a promising direction for future investigations.

More broadly, one might wonder whether mechanisms analogous to those identified here play a role in biological collectives. 
In particular, gathering visual sensory information from the environment and from the neighboring individuals is crucial for coordinated motion and for the ability of the group to respond collectively to external perturbations \cite{strandburg2013visual}. 
While anisotropic interactions are known to be relevant in systems such as bird flocks, the underlying anisotropy may differ qualitatively from the forward-oriented vision kernels considered here. For instance, inference of the effective interactions from the experimental data of starling flocks suggests stronger interactions with lateral neighbors rather than with those located in front \cite{cavagna2015short}. Investigating how different forms of perceptual anisotropy shape collective ordering and whether proactive-like contributions play a role in biological groups remains an open question.

\bibliographystyle{apsrev4-2}
\providecommand{\noopsort}[1]{}\providecommand{\singleletter}[1]{#1}%
%

\appendix
\onecolumngrid

\section{Local equations}\label{app:local}

In order to identify the local pinning terms, we start from the Langevin equation reported in Eq.~\eqref{eq:langevin}, substitute the different interaction kernels discussed in Sec.~\ref{sec:kernels}, and rearrange the resulting expressions so as to highlight the various contributions to the expansion in Eq.~\eqref{eq:local-harmonics-txt} introduced in 
Sec.~\ref{sec:pinning-local-global}. 
The coefficients $S_{\alpha,i}^{(n),\mathrm{loc}}(t)$ and $C_{\alpha,i}^{(n),\mathrm{loc}}(t)$ of the expansion depend on the instantaneous configuration of the neighboring spins of spin $i$ and can be expressed as functions of the 
local angular differences, as in Eq.~\eqref{eq:local-coeffs}.
The mean values and fluctuations of these coefficients can be computed numerically in the stationary state, once the system is pinned along a given direction $\Phi$. Since, in the stationary state, the system is both time-translation invariant and, due to the periodic boundary conditions, space-translation invariant, averages and standard deviations can be computed either over time or over space. Here, we choose to perform spatial averages, as this makes it more transparent which terms vanish on average as a consequence of the periodic boundary conditions. %
Expanding around the pinned configuration in terms of the fluctuations $\delta\phi_i=\phi_i-\Phi$, one can investigate whether a local restoring term of the form $-r_i^{\mathrm{loc}}\delta\phi_i$ emerges in the (overdamped) evolution equation of $\phi_i$, with $r_i^{\mathrm{loc}}$ depending on the coefficients $S_{\alpha,i}^{(n),\mathrm{loc}}$ and $C_{\alpha,i}^{(n),\mathrm{loc}}$. In general, $r_i^{\mathrm{loc}}$ is itself a fluctuating quantity, characterized by a mean value ${r^{\mathrm{loc}}}$ and a standard deviation $\Delta_{r^{\mathrm{loc}}}$ (see Eq.~\eqref{eq:def-delta-rloc}), as discussed in Sec.~\ref{sec:pinning-local-global}. %

In order to formulate a description on the continuum, we formally replace $\delta\phi_i\to\delta\phi_{\mathbf{x}}$, $S_{\alpha,i}^{\cdots}\to S_{\alpha,\mathbf{x}}^{\cdots}$, $C_{\alpha,i}^{\cdots}\to C_{\alpha,\mathbf{x}}^{\cdots}$, 
and perform a gradient expansion, e.g., 
\begin{equation}
\delta\phi_{i\pm \hat{\mu}}
=
\delta\phi_{\mathbf{x}}
\pm
a\,\partial_\mu\delta\phi_{\mathbf{x}}
+
\frac12 a^2\,
\partial_\mu^2
\delta\phi_{\mathbf{x}}
+
\mathcal{\cal O}(a^3),
\label{eq-app:grexp}
\end{equation}
where $\hat{\mu}$ is the unit vector of the $\mu$-th spatial direction, while we eventually set the lattice spacing $a$ to unity. 
We emphasize that this construction on the continuum is mainly intended to facilitate comparison with coarse-grained descriptions which are reported in the literature, in particular that one of Refs.~\cite{popli2025}. %
For notational simplicity, throughout this appendix we consider only the deterministic part of the dynamics and omit the Gaussian noise term.

\subsection{Modulated coupling}\label{app:modulated-local}

We consider Eq.~\eqref{eq:langevin} for $\phi_i$ with the modulated coupling kernel (see Eq.~\eqref{eq:mod-coup}) and rewrite it in the form of Eq.~\eqref{eq:local-harmonics-txt}, in order to identify the effective pinning terms. We then introduce the coarse-grained field $\phi_{\mathbf{x}}$ and perform the continuum expansion described in Eq.~\eqref{eq-app:grexp}. 
\\
\paragraph*{Reactive term.}

The reactive contribution (indicated by the subscript ${\rm R}$) reads
\begin{equation}\label{eq:modulated-reactive-local}
\begin{aligned}
(\partial_t{\phi}_i)_{\rm R} &=
-\sum_{j\in\mathcal{N}_i} \left[1+\epsilon\cos(4\phi_i)\right]\sin(\phi_i-\phi_j)\\
&=C_{\mathrm{R},i}^{(0), \text{loc}} + C_{\mathrm{R},i}^{(4), \text{loc}}\cos{(4\phi_i)},
\end{aligned}
\end{equation}
where $\epsilon C_{{\rm R},i}^{(0), \text{loc}} =  C_{{\rm R},i}^{(4), \text{loc}} = \epsilon \sum_{j\in\mathcal{N}_i} \sin{(\phi_i-\phi_j)}$.
The coefficients $C_{{\rm R},i}^{(0), \text{loc}}$ and $C_{{\rm R},i}^{(4), \text{loc}}$ vanish upon averaging over the entire lattice.
Consequently, the reactive contribution does not generate any local pinning term, but only a modulation of the diffusion.
This conclusion is also confirmed in the continuum limit, in which the previous equation becomes
\begin{equation}
(\partial_t{\phi_{\mathbf{x}}})_{{\rm R}}
\simeq
\left[1+\epsilon\cos(4\phi_{\mathbf{x}})\right] \nabla^2\phi_{\mathbf{x}}.
\end{equation}
This expression explicitly shows the presence of a modulated diffusion term and the absence of a local restoring force.
\\

\paragraph*{Proactive term.}

We now consider the proactive contribution (indicated by the subscript ${\rm P}$),
\begin{equation}\label{eq:sp4-loc-mod}
(\partial_t{\phi}_i)_{{\rm P}}
=
-4\epsilon \sum_{j\in\mathcal{N}_i}
\sin(4\phi_i)
\cos(\phi_i-\phi_j)
=
S_{{\rm P},i}^{(4),\mathrm{loc}}
\sin(4\phi_i)\quad\mbox{with}\quad S_{{\rm P},i}^{(4),\mathrm{loc}} = -4\epsilon \sum_{j\in\mathcal{N}_i}
\cos(\phi_i-\phi_j).
\end{equation}
To identify the local pinning coefficient (see Eq.~\eqref{eq:local-continuum}), we expand this equation in terms of the fluctuations $\delta\phi_i$ around one of the pinned directions, e.g., $\Phi = 0$.
Isolating the term which is linear in $\delta\phi_i$ by writing 
$\sin(4\delta\phi_i) = \frac{\sin(4\delta\phi_i)}{4\delta\phi_i}4\delta\phi_i$,
one obtains the following expression for the coefficient $r_i^{\mathrm{loc}}$ (introduced in Eq.~\eqref{eq:local-continuum}) of such a linear term:
\begin{equation}
r_i^{\mathrm{loc}}
=
-4
\frac{\sin{4\delta\phi_i}}{4\delta\phi_i}
S_{\mathrm{P},i}^{(4),\mathrm{loc}}.
\end{equation}
Passing to the continuum limit, one finds
\begin{equation}
\sum_{j\in\mathcal{N}_i}
\cos(\phi_i-\phi_j)
\simeq
4
-
|\nabla\delta\phi_{\mathbf{x}}|^2,
\end{equation}
which yields
\begin{equation}
\label{eq:modulated-pro-continuum}
(\partial_t{\delta\phi_{\mathbf{x}}})_{{\rm P}}
\simeq
-r_{\mathbf{x}}^{\mathrm{loc}} \delta\phi_{\mathbf{x}} \simeq 
-16\epsilon
\frac{\sin{(4\delta\phi_{\mathbf{x}})}}{4\delta\phi_{\mathbf{x}}}
\left(
4
-
|\nabla\delta\phi_{\mathbf{x}}|^2
\right)\delta\phi_{\mathbf{x}}.
\end{equation}
Accordingly, the proactive interaction generates a
genuinely local pinning mechanism. By contrast, in the purely reactive dynamics, the selection of preferred
orientation emerges only at the collective level, as it is shown furtgher below in App.~\ref{app:modulated-mean}, where  the corresponding equations for the global orientation are derived.

\subsection{Sinusoidal coupling}\label{app:sin-local}

We rewrite Eq.~\eqref{eq:langevin} for $\phi_i$ in the case of the sinusoidal coupling (see Eq.~\eqref{eq:sin-kernel}) in the form of Eq.~\eqref{eq:local-harmonics-txt}, in order to identify the pinning terms. We then take the continuum limit by introducing the coarse-grained field $\phi_{\mathbf{x}}$ and perform the continuum expansion as in Eq.~\eqref{eq-app:grexp}.
\\

\paragraph*{Reactive term.}
We consider the reactive term
\begin{equation}
(\partial_t \phi_i)_{{\rm R}}
=
-\sum_{j\in \mathcal{N}_i}
\Bigl[1+\epsilon \cos(\phi_i-\vartheta_{ij})\Bigr]
\sin(\phi_i-\phi_j),
\label{eq:local_full}
\end{equation}
which can be rewritten in the form
\begin{align}
(\partial_t \phi_i)_{{\rm R}}
&=
-\sum_{j\in \mathcal{N}_i} \sin(\phi_i-\phi_j)
\nonumber\\
&\quad
-\epsilon \cos\phi_i
\Bigl[
\sin(\phi_i-\phi_{i+\hat x})
-
\sin(\phi_i-\phi_{i-\hat x})
\Bigr]
\nonumber\\
&\quad
-\epsilon \sin\phi_i
\Bigl[
\sin(\phi_i-\phi_{i+\hat y})
-
\sin(\phi_i-\phi_{i-\hat y})
\Bigr],
\label{eq:local_square_explicit}
\end{align}
where $\hat x$ and $\hat y$ are unit vectors along the $x$ and $y$ axis of the lattice.
Comparing this expression with Eq.~\eqref{eq:local-harmonics-txt}, we identify the coefficients 
\begin{align}
    C_{\text{R},i}^{(1), \textrm{loc}} = -\epsilon \Big[\sin(\phi_i-\phi_{i+\hat x})
-
\sin(\phi_i-\phi_{i-\hat x})
\Bigr],\label{eq-app:CR1loc-SK}  \\
S_{\text{R},i}^{(1), \textrm{loc}} = -\epsilon \Big[\sin(\phi_i-\phi_{i+\hat y})
-
\sin(\phi_i-\phi_{i-\hat y})
\label{eq-app:SR1loc-SK}
\Bigr].
\end{align}

The symmetry properties of these coefficients follow from the same symmetry arguments discussed for the corresponding global coefficients in App.~\ref{app:sin-global}. We therefore refer the reader to the symmetry analysis presented there. In particular, when the system is pinned along $\Phi=0$, one has $C_{\mathrm{R},i}^{(1),\mathrm{loc}}=0$ and $S_{\mathrm{R},i}^{(1),\mathrm{loc}}\neq0$, whereas for a state pinned along $\Phi=\pi/2$ the roles of the two coefficients are exchanged: $S_{\mathrm{R},i}^{(1),\mathrm{loc}}=0$ and $C_{\mathrm{R},i}^{(1),\mathrm{loc}}=-S_{\mathrm{R},i}^{(1),\mathrm{loc}}|_{\Phi=0}$.
These relations are the local counterparts of those derived in App.~\ref{app:sin-global} for the global coefficients.

Expanding Eq.~\eqref{eq:local_square_explicit} in spatial gradients,  we find the following expression on the continuum: 
\begin{equation}
    (\partial_t \phi_{\mathbf{x}})_{{\rm R}} = \nabla^2 \phi_{\mathbf{x}} + 2\epsilon(\sin{\phi_{\mathbf{x}}}\partial_y \phi_{\mathbf{x}} + \cos{\phi_{\mathbf{x}}}\partial_x\phi_{\mathbf{x}}) + \frac{\epsilon}{3}\left\{\sin{\phi_{\mathbf{x}}}\left[\partial_y^3 \phi_{\mathbf{x}} -(\partial_y \phi_{\mathbf{x}})^3\right]  + \cos{\phi_{\mathbf{x}}}\left[\partial_x^3 \phi_{\mathbf{x}} -(\partial_x \phi_{\mathbf{x}})^3\right]\right\},
\end{equation}
which corresponds to Eq.~(5) in  Ref.~\cite{popli2025}. 
When we search for the first potentially non-vanishing contributions to the local pinning --- i.e., linear in the fluctuations of $\phi_{\mathbf{x}}$ --- we see that the terms $\partial_x \phi_{\mathbf{x}}$, $\partial_y\phi_{\mathbf{x}}$, $\partial_x^3 \phi_{\mathbf{x}}$, $\partial_y^3 \phi_{\mathbf{x}}$ are total derivatives and therefore vanish upon spatial averaging under periodic boundary conditions. By contrast, the terms $(\partial_y \phi_{\mathbf{x}})^3$ and $(\partial_x \phi_{\mathbf{x}})^3$ are not total derivatives and thus provide the candidate for the leading (non-vanishing) local pinning contributions. %
\\

\paragraph*{Proactive term.}
We now consider the proactive contribution
\begin{equation}
(\partial_t \phi_i)_{\rm P}
=
-\epsilon \sum_{j\in \mathcal{N}_i}
\sin(\phi_i-\vartheta_{ij})\,\cos(\phi_i-\phi_j),
\label{eq:proactive_full}
\end{equation}
which can be written in the form
\begin{align}
(\partial_t \phi_i)_{\rm P}
=&
-\epsilon 
\Bigl[
\cos(\phi_i-\phi_{i+\hat x})
-
\cos(\phi_i-\phi_{i-\hat x})
\Bigr]\sin\phi_i
\nonumber\\
&
+\epsilon 
\Bigl[
\cos(\phi_i-\phi_{i+\hat y})
-
\cos(\phi_i-\phi_{i-\hat y})
\Bigr] \cos\phi_i\\
=& C_{{\rm P},i}^{(1),\mathrm{loc}}\cos{\phi_i} + S_{{\rm P},i}^{(1),\mathrm{loc}}\sin{\phi_i}.
\label{eq:proactive_square_explicit}
\end{align}
The coefficients \(C_{\mathrm{P},i}^{(1),\mathrm{loc}}\) and \(S_{\mathrm{P},i}^{(1),\mathrm{loc}}\) (the expressions of which are readily obtained by comparing the last equality with the previous two lines) vanish upon averaging over a system with periodic boundary conditions. This can be seen also on the continuum by expanding the equation in gradients, as one obtains 
\begin{equation}
    (\partial_t \phi_{\mathbf{x}})_{\mathrm P}
    =
    \frac{\epsilon}{2}
    \Big\{
        \sin\phi_{\mathbf{x}}\, \partial_x\big[(\partial_x \phi_{\mathbf{x}})^2\big]
        -
        \cos\phi_{\mathbf{x}}\, \partial_y\big[(\partial_y \phi_{\mathbf{x}})^2\big]
    \Big\}.
\end{equation}
Note that the terms $\partial_x\big[(\partial_x \phi_{\mathbf{x}})^2\big]$ and $\partial_y\big[(\partial_y \phi_{\mathbf{x}})^2\big]$ corresponding to $S_{\mathrm{P},i}^{(1),\mathrm{loc}}$ and $C_{\mathrm{P},i}^{(1),\mathrm{loc}}$, which could be responsible for pinning,  are actually total derivatives
and thus they have a vanishing spatial average in the presence of periodic boundary conditions.
Accordingly, for the sinusoidal coupling,  we conclude that 
the proactive term does not induce local pinning, and thus a possible local pinning must originate from the reactive contribution.%

\subsection{Von Mises coupling}\label{app:vonmises-local}

We now consider Eq.~\eqref{eq:langevin} for the evolution of $\phi_i$ in the case of the von Mises coupling (see Eq.~\eqref{eq:vonmises-kernel}), casting in the form of Eq.~\eqref{eq:local-harmonics-txt} in order to identify the pinning terms. 
As discussed in Sec.~\ref{sec:pinning-vonmises}, we restrict the analysis to the contributions of order ${\mathcal O}(\sigma^0)$, ${\mathcal O}(\sigma)$, and ${\mathcal O}(\sigma^4)$, which are sufficient to understand the emergence of  pinning in the model. 
Within this approximation, the resulting dynamics 
turns out to contain terms which also appear in the evolution with the sinusoidal and the modulated coupling, as discussed further below. 
\\

\paragraph*{Reactive term.} We write the local equation for the reactive term
\begin{equation}
(\partial_t \phi_i)_{{\rm R}}
=
-\sum_{j\in \mathcal{N}_i}
\left[1+\sigma \cos(\phi_i-\vartheta_{ij}) + \frac{\sigma^4}{24}\cos^4{(\phi_i-\vartheta_{ij})}\right]
\sin(\phi_i-\phi_j),
\label{eq:local_full}
\end{equation}
as
\begin{align}
(\partial_t \phi_i)_{\rm R} =
&-
\sum_{j\in\mathcal{N}_i}
\sin(\phi_i-\phi_j)
\nonumber\\
&-\sigma\cos\phi_i
\Big[
\sin(\phi_i-\phi_{i+\hat{x}})
-\sin(\phi_i-\phi_{i-\hat{x}})
\Big]
\nonumber\\
&-\sigma\sin\phi_i
\Big[
\sin(\phi_i-\phi_{i+\hat{y}})
-\sin(\phi_i-\phi_{i-\hat{y}})
\Big]
\nonumber\\
&-\frac{\sigma^4}{24}\cos^4\phi_i
\Big[
\sin(\phi_i-\phi_{i+\hat{x}})
+\sin(\phi_i-\phi_{i-\hat{x}})
\Big]
\nonumber\\
&-\frac{\sigma^4}{24}\sin^4\phi_i
\Big[
\sin(\phi_i-\phi_{i+\hat{y}})
+\sin(\phi_i-\phi_{i-\hat{y}})
\Big],
\end{align}
which, after using the following trigonometric identities, 
\begin{equation}
\cos^4x=\frac{3}{8}+\frac{1}{2}\cos(2x)+\frac{1}{8}\cos(4x) \quad \mbox{and} \quad \sin^4x=\frac{3}{8}-\frac{1}{2}\cos(2x)+\frac{1}{8}\cos(4x),
\label{eq-app:cos4sin4}
\end{equation}
can be rewritten as in Eq.~\eqref{eq:local-harmonics-txt}, i.e.,  
\begin{equation}
(\partial_t \phi_i)_{{\rm R}} = C_{\mathrm{R},i}^{(0),\mathrm{loc}} + S_{\mathrm{R},i}^{(1),\mathrm{loc}} \sin \phi_i + 
C_{\mathrm{R},i}^{(1),\mathrm{loc}}\cos \phi_i + C_{\mathrm{R},i}^{(2),\mathrm{loc}}\cos (2\phi_i) +  
C_{\mathrm{R},i}^{(4),\mathrm{loc}}\cos(4 \phi_i)
\end{equation}
with
\begin{align}
C_{\mathrm{R},i}^{(0),\mathrm{loc}}
&=
-\left(1+\frac{\sigma^4}{64}\right)
\sum_{j\in\mathcal{N}_i}
\sin(\phi_i-\phi_j),
\label{eq-app:CR0-loc-vM}
\\
S_{\mathrm{R},i}^{(1),\mathrm{loc}}
&=
-\sigma
\left[
\sin(\phi_i-\phi_{i+\hat y})
-
\sin(\phi_i-\phi_{i-\hat y})
\right],
\label{eq-app:SR1-loc-vM}
\\
C_{\mathrm{R},i}^{(1),\mathrm{loc}}
&=
-\sigma
\left[
\sin(\phi_i-\phi_{i+\hat x})
-
\sin(\phi_i-\phi_{i-\hat x})
\right],
\label{eq-app:CR1-loc-vM}
\\
C_{\mathrm{R},i}^{(2),\mathrm{loc}}
&=
\frac{\sigma^4}{48}
\left[
\sin(\phi_i-\phi_{i+\hat y})
+
\sin(\phi_i-\phi_{i-\hat y})
-
\sin(\phi_i-\phi_{i+\hat x})
-
\sin(\phi_i-\phi_{i-\hat x})
\right],
\label{eq-app:CR2-loc-vM}
\\
C_{\mathrm{R},i}^{(4),\mathrm{loc}}
&=
-\frac{\sigma^4}{192}
\sum_{j\in\mathcal{N}_i}
\sin(\phi_i-\phi_j).
\label{eq-app:CR4-loc-vM}
\end{align}
Note that apart from a prefactor,  $C_{\mathrm{R},i}^{(1),\mathrm{loc}}$ and $S_{\mathrm{R},i}^{(1),\mathrm{loc}}$ have the same expression as for the sinusoidal coupling, see Eqs.~\eqref{eq-app:CR1loc-SK} and \eqref{eq-app:SR1loc-SK}). 
Moreover, upon averaging over the system with periodic boundary conditions, only the coefficients $C_{\mathrm{R},i}^{(1),\mathrm{loc}}$ and $S_{\mathrm{R},i}^{(1),\mathrm{loc}}$ may have non-vanishing mean values and therefore they may contribute to the local pinning. By contrast, the coefficients $C_{\mathrm{R},i}^{(0),\mathrm{loc}}$, $C_{\mathrm{R},i}^{(2),\mathrm{loc}}$, and $C_{\mathrm{R},i}^{(4),\mathrm{loc}}$ have zero mean and thus they only contribute to the diffusive part of the dynamics.
Moreover, symmetry properties imply that $S_{\mathrm{R}, i}^{(1),\mathrm{loc}} = - C_{\mathrm{R}, i}^{(1),\mathrm{loc}}$ when the system is pinned along diagonal directions, as discussed further below in App.~\ref{app:vonmises-mean}.
\\

\paragraph*{Proactive term.}

We now focus on the contributions of the proactive terms to the equation of motion, keeping only the terms of order $\mathcal{O}(\sigma)$ and $\mathcal{O}(\sigma^4)$ in $\partial_{\phi_i}g(\phi, \vartheta_{ij})$ (the term of order $\mathcal{O}(\sigma^0)$ does not contribute to $\partial_{\phi_i}g(\phi, \vartheta_{ij})$).
Starting from Eq.~\eqref{eq:langevin} and \eqref{eq:vonmises-kernel}, 
one finds
\begin{equation}
\begin{aligned}
    (\partial_t \phi_i)_{{\rm P}} =\;& -\sum_{j\in\mathcal{N}_i} 
    \left[\sigma \sin(\phi_i -\vartheta_{ij}) 
    + \frac{\sigma^4}{6}\sin(\phi_i-\vartheta_{ij})\cos^3(\phi_i - \vartheta_{ij})\right] 
    \cos(\phi_i - \phi_j)
    \\
    =\;& -\sum_{j\in\mathcal{N}_i} 
    \Big[
    \sigma \sin(\phi_i -\vartheta_{ij})
    + \frac{\sigma^4}{24}\sin\big(2(\phi_i-\vartheta_{ij})\big)
    + \frac{\sigma^4}{48}\sin\big(4(\phi_i-\vartheta_{ij})\big)
    \Big]
    \cos(\phi_i - \phi_j)\,.
\end{aligned}
\end{equation}
This expression can eventually be rewritten
as in Eq.~\eqref{eq:local-harmonics-txt}, i.e.,  
\begin{equation}
(\partial_t \phi_i)_{{\rm P}} = S_{\mathrm{P},i}^{(1),\mathrm{loc}}\sin\phi_i 
+ C_{\mathrm{P},i}^{(1),\mathrm{loc}} \cos \phi_i
+ S_{\mathrm{P},i}^{(2),\mathrm{loc}} \sin (2\phi_i)
+ S_{\mathrm{P},i}^{(4),\mathrm{loc}} \sin (4\phi_i),
\label{eq:proactive_vm_square_explicit}
\end{equation}
where
\begin{align}
S_{\mathrm{P},i}^{(1),\mathrm{loc}}
&=
-\sigma
\Big[
\cos(\phi_i-\phi_{i+\hat x})
-
\cos(\phi_i-\phi_{i-\hat x})
\Big],
\\
C_{\mathrm{P},i}^{(1),\mathrm{loc}}
&=
\sigma
\Big[
\cos(\phi_i-\phi_{i+\hat y})
-
\cos(\phi_i-\phi_{i-\hat y})
\Big],
\\
S_{\mathrm{P},i}^{(2),\mathrm{loc}}
&=
-\frac{\sigma^4}{24}
\Big[
\cos(\phi_i-\phi_{i+\hat x})
+
\cos(\phi_i-\phi_{i-\hat x})
-
\cos(\phi_i-\phi_{i+\hat y})
-
\cos(\phi_i-\phi_{i-\hat y})
\Big],
\\
S_{\mathrm{P},i}^{(4),\mathrm{loc}}
&=
-\frac{\sigma^4}{48}
\Big[
\cos(\phi_i-\phi_{i+\hat x})
+
\cos(\phi_i-\phi_{i-\hat x})
+
\cos(\phi_i-\phi_{i+\hat y})
+
\cos(\phi_i-\phi_{i-\hat y})
\Big].
\end{align}
In particular, the coefficients $S_{\mathrm{P},i}^{(1),\mathrm{loc}}$ and $C_{\mathrm{P},i}^{(1),\mathrm{loc}}$, up to an overall prefactor, have the same expression as for the sinusoidal coupling, see Eq.~\eqref{eq:proactive_square_explicit}.
As discussed in App.~\ref{app:sin-local}, these terms vanish upon averaging over the whole system with periodic boundary conditions. They do not contribute to the average local pinning.

By contrast, the coefficients $S_{\mathrm{P},i}^{(2),\mathrm{loc}}$ and $S_{\mathrm{P},i}^{(4),\mathrm{loc}}$ are not constrained to vanish by the same argument and can contribute to pinning.
We verified numerically that they both contribute to pinning along lattice directions. However, the magnitude of the $S_{\mathrm{P},i}^{(2),\mathrm{loc}}$ term is smaller than that of $S_{\mathrm{P},i}^{(4),\mathrm{loc}}$ for the parameters considered. 

This difference in magnitude is already apparent from their continuum expansions,
\begin{align}
S_{\mathrm{P},\mathbf{x}}^{(2),\mathrm{loc}}
&\simeq
-\frac{\sigma^4}{24}
\Big[
(\partial_y\phi_{\mathbf{x}})^2
-
(\partial_x\phi_{\mathbf{x}})^2
\Big],
\\
S_{\mathrm{P},\mathbf{x}}^{(4),\mathrm{loc}}
&\simeq
-\frac{\sigma^4}{48}
\Big[
4
-
|\nabla\phi_{\mathbf{x}}|^2
\Big].
\end{align}
In fact, the coefficient $S_{\mathrm{P},\mathbf{x}}^{(2),\mathrm{loc}}$ starts only at second order in spatial gradients and is therefore subleading. Conversely, $S_{\mathrm{P},\mathbf{x}}^{(4),\mathrm{loc}}$ contains a zeroth-order contribution in the gradient expansion, exactly as the proactive contribution of the modulated-coupling kernel discussed in App.~\ref{app:modulated-local}.


\section{Global equations}
\label{app:global}

In this appendix, we derive the evolution equation for the mean (global) orientation $\phi(t)$, defined in Eq.~\eqref{eq:def-phi}. To this end, we rewrite $\partial_t\phi(t)=N^{-1}\sum_i \partial_t\phi_i$ as a sum over lattice bonds, obtaining an expression of the general form reported in Eq.~\eqref{eq:global-harmonics-txt}. For notational simplicity, throughout this appendix we consider only the deterministic part of the dynamics and omit the Gaussian noise term.


\subsection{Modulated coupling}
\label{app:modulated-mean}

We consider the reactive (first line) and the proactive (second line) contributions together, and obtain
\begin{equation}
\begin{aligned}
\partial_t\phi 
&= 
-\frac{\epsilon}{N} \sum_{\langle i,j\rangle}
\big[
    \cos(4(\phi+\delta\tilde\phi_i))
    -\cos(4(\phi+\delta\tilde\phi_j))
\big]
\sin(\delta\tilde\phi_i-\delta\tilde\phi_j)
\\
&\quad
-\frac{4\epsilon}{N} \sum_{\langle i,j\rangle} 
\big[
    \sin(4(\phi+\delta\tilde\phi_i))
    +\sin(4(\phi+\delta\tilde\phi_j))
\big]
\cos(\delta\tilde\phi_i-\delta\tilde\phi_j),
\end{aligned}
\label{eq:meanphi_full}
\end{equation}
where we expressed the local value $\phi_i$ of the phase in terms of its deviation $\delta \tilde\phi_i$ from the global orientation $\phi$, according to Eq.~\eqref{eq:def-tphii}. 
After using standard trigonometric identities,
Eq.~\eqref{eq:meanphi_full} can be rearranged into the form
\begin{equation}
\begin{aligned}
\partial_t\phi &=
-2\epsilon\cos(4\phi)
\Big\langle
    \left[\cos(4\delta\tilde\phi_i)-\cos(4\delta\tilde\phi_j)\right]
    \sin(\delta\tilde\phi_i-\delta\tilde\phi_j)
\Big\rangle
\\
&\quad
+2\epsilon\sin(4\phi)
\Big\langle
    \left[\sin(4\delta\tilde\phi_i)-\sin(4\delta\tilde\phi_j)\right]
    \sin(\delta\tilde\phi_i-\delta\tilde\phi_j)
\Big\rangle
\\
&\quad
-8\epsilon\cos(4\phi)
\Big\langle
    \left[\sin(4\delta\tilde\phi_i)+\sin(4\delta\tilde\phi_j)\right]
    \cos(\delta\tilde\phi_i-\delta\tilde\phi_j)
\Big\rangle
\\
&\quad
-8\epsilon\sin(4\phi)
\Big\langle
    \left[\cos(4\delta\tilde\phi_i)+\cos(4\delta\tilde\phi_j)\right]
    \cos(\delta\tilde\phi_i-\delta\tilde\phi_j)
\Big\rangle,
\end{aligned}
\label{eq:meanphi_decomposed}
\end{equation}
where the averages $\langle\cdots\rangle$ are taken over all the lattice bonds at a given time, i.e., $\langle\dots\rangle = \frac{1}{2N}\sum_{\langle ij\rangle} (\dots)$.
Note that the first two lines of the expression above originate from the \emph{reactive} part of the dynamics,
while the last two stem from the \emph{proactive} term.

By collecting the coefficients of $\sin(4\phi)$ and $\cos(4\phi)$,  
Eq.~\eqref{eq:meanphi_decomposed} can be written in the form of Eq.~\eqref{eq:global-harmonics-txt} as
\begin{equation}
\partial_t\phi 
= \big(
S^{(4), \mathrm{glob}}_{\text{R}}+S^{(4), \mathrm{glob}}_{\text{P}}\big)\sin(4\phi) + \big(C^{(4),\mathrm{glob}}_{\text{R}}+C^{(4),\mathrm{glob}}_{\text{P}}\big)\cos(4\phi)
\label{eq:meanphi_final}
\end{equation}
where the coefficients $S^{(4), \mathrm{glob}}_{\text{R}, {\rm P}}$ and $C^{(4), \mathrm{glob}}_{\text{R}, {\rm P}}$ are defined as
\begin{align}
C^{(4), \mathrm{glob}}_{\text{R}} &=
-2\epsilon\Big\langle
\left[\cos(4\delta\tilde\phi_i)-\cos(4\delta\tilde\phi_j)\right]
\sin(\delta\tilde\phi_i-\delta\tilde\phi_j)
\Big\rangle, \label{eq-app:CR4glob}\\
S^{(4), \mathrm{glob}}_{\text{R}} &=
+ 2\epsilon\Big\langle
\left[\sin(4\delta\tilde\phi_i)-\sin(4\delta\tilde\phi_j)\right]
\sin(\delta\tilde\phi_i-\delta\tilde\phi_j)
\Big\rangle,\label{eq-app:SR4glob}
\\
C^{(4), \mathrm{glob}}_{\text{P}} &=
-8\epsilon\Big\langle
\left[\sin(4\delta\tilde\phi_i)+\sin(4\delta\tilde\phi_j)\right]
\cos(\delta\tilde\phi_i-\delta\tilde\phi_j)
\Big\rangle, \label{eq-app:CP4glob}\\
S^{(4), \mathrm{glob}}_{\text{P}} &=
-8\epsilon\Big\langle
\left[\cos(4\delta\tilde\phi_i)+\cos(4\delta\tilde\phi_j)\right]
\cos(\delta\tilde\phi_i-\delta\tilde\phi_j)
\Big\rangle.\label{eq-app:SP4glob}
\end{align}

\paragraph*{Symmetry considerations.}
The dynamics of the system is generically invariant under a global $\mathbb{Z}_4$, i.e., under the simultaneous transformation $\phi_i \mapsto \phi_i + n \pi/4$ at all lattice sites $i$, with $n\in \{0,1,2,3\}$. Once the pinning direction $\Phi$ has been selected spontaneously by the system as discussed in Sec.~\ref{sec:pinning-modulated}, 
%
%
the dynamics remains invariant under the global reflection symmetry ${\cal R}: \delta\phi \to - \delta\phi, \delta\tilde\phi_i \to -\delta\tilde\phi_i$ done simultaneously at all lattice sites $i$.
As a consequence, the fluctuation-induced coefficients can be classified
according to their parity under ${\cal R}$.
In particular, $C^{(4), \mathrm{glob}}_{\mathrm{R}}$ in Eq.~\eqref{eq-app:CR4glob} and $C^{(4), \mathrm{glob}}_{\mathrm{P}}$ in Eq.~\eqref{eq-app:CP4glob} 
 are odd under
${\cal R}$ and therefore they vanish identically, whereas
$S_{\mathrm{R}}^{(4), \mathrm{glob}}$ in Eq.~\eqref{eq-app:SR4glob} and $S_{\mathrm{P}}^{(4), \mathrm{glob}}$ in Eq.~\eqref{eq-app:SP4glob} are even and thus they might be finite.
Accordingly, the only term in Eq.~\eqref{eq:meanphi_final} which survives is the one  proportional to $\sin(4\phi)$.
The overall sign of the corresponding prefactor (determined via numerical simulations, see Sec.~\ref{sec:numerical-mean-orienetation}) 
%
%
selects the pinned
orientations of the global direction and locks the system either to
$\Phi = 0 + n\pi/2$ or to $\Phi = \pi/4 + n\pi/2$, with $n\in\mathbb{Z}$.%
\\

\paragraph*{Small-fluctuation expansion.}
We now expand the coefficients $S_{\mathrm R}^{(4), \mathrm{glob}}$ and $S_{\mathrm P}^{(4), \mathrm{glob}}$ in Eqs.~\eqref{eq-app:SR4glob} and \eqref{eq-app:SP4glob}, respectively, assuming small fluctuations $\delta\tilde\phi_i$ and then consider formally to the continuum limit by introducing the field $\delta\tilde\phi_{\mathbf{x}}(t)$ instead of $\delta\tilde\phi_i$.
For the reactive contribution, we obtain
\begin{equation}
    (\partial_t \phi)_{\mathrm R}
    =
    4\epsilon \,
    \big\langle |\nabla\delta\tilde{\phi}_{\mathbf{x}}|^2 \big\rangle
    \,\sin(4\phi),
\end{equation}
which is actually the expression found in Ref.~\cite{dopierala2025}.
For the proactive contribution, one finds, instead,
\begin{equation}
    (\partial_t \phi)_{\mathrm P}
    =
    -4\epsilon \sin(4\phi)\Big[
    4
    -32\,\langle \delta\tilde\phi_{\mathbf{x}}^2 \rangle
    -\langle |\nabla \delta\tilde\phi_{\mathbf{x}}|^2 \rangle
    \Big].
\end{equation}
Note that, differently from the dynamics of the local phase fluctuation $\delta\phi_{\mathbf{x}}$ on the continuum reported in Eq.~\eqref{eq:modulated-pro-continuum}, the spatially averaged onsite fluctuations $\langle \delta\tilde\phi_{\mathbf{x}}^2 \rangle$ of the field explicitly appear in the evolution equation for the global phase. 
%

\subsection{Sinusoidal coupling}
\label{app:sin-global}

For the case of the sinusoidal coupling in Eq.~\eqref{eq:sin-kernel}, we find convenient to treat separately the reactive and proactive contributions to the equation of motion in Eq.~\eqref{eq:langevin}, which we then use to determine the equation of motion for the average orientation $\phi(t)$, as done above. 
\\

\paragraph*{Reactive term.}
The deterministic contribution to the Langevin dynamics originating from the reactive term is
\begin{equation}\label{eq:sinus_active_start}
\begin{split}
    (\partial_t \phi)_{\rm R}
    =  -\frac{\epsilon}{N} 
    \sum_{ i}&\Big\{
    \big[\cos(\phi + \delta\tilde{\phi}_{i})
        + \cos(\phi + \delta\tilde{\phi}_{i+\hat{x}})\big]
    \sin(\delta\tilde{\phi}_i - \delta\tilde{\phi}_{i+\hat{x}}) \\
   & + \big[\sin(\phi + \delta\tilde{\phi}_i)
        + \sin(\phi + \delta\tilde{\phi}_{i+\hat{y}})\big]
    \sin(\delta\tilde{\phi}_i - \delta\tilde{\phi}_{i+\hat{y}})\Big\},
\end{split}
\end{equation}
where, as done in the previous sections, the local value $\phi_i$ of the phase is expressed in terms of its deviation $\delta \tilde\phi_i$ from the global orientation $\phi$, according to Eq.~\eqref{eq:def-tphii}. 
Applying trigonometric identities, we obtain
\begin{equation}
\begin{aligned}
(\partial_t \phi)_{\rm R}
= -\frac{\epsilon}{N}
\sum_{ i}&\Big\{
\big[(\cos\delta\tilde{\phi}_i+\cos\delta\tilde{\phi}_{i+\hat{x}})\cos\phi
      -(\sin\delta\tilde{\phi}_i+\sin\delta\tilde{\phi}_{i+\hat{x}})\sin\phi \big]
      \sin(\delta\tilde{\phi}_i-\delta\tilde{\phi}_{i+\hat{x}})
\\
&
+\big[(\cos\delta\tilde{\phi}_i+\cos\delta\tilde{\phi}_{i+\hat{y}})\sin\phi
      +(\sin\delta\tilde{\phi}_i+\sin\delta\tilde{\phi}_{i+\hat{y}})\cos\phi\big]
      \sin(\delta\tilde{\phi}_i-\delta\tilde{\phi}_{i+\hat{y}})\Big\}.
\end{aligned}
\end{equation}
Collecting the coefficients of $\sin\phi$ and $\cos\phi$ and indicating with $\langle\cdots\rangle$ the average over the lattice sites, the result can be written as 
\begin{equation}
\begin{aligned}
(\partial_t \phi)_{\rm R}
&= -\epsilon\,
\Big[
\langle(\cos\delta\tilde{\phi}_i+\cos\delta\tilde{\phi}_{i+\hat{y}})
\sin(\delta\tilde{\phi}_i-\delta\tilde{\phi}_{i+\hat{y}}) \rangle
-\langle(\sin\delta\tilde{\phi}_i+\sin\delta\tilde{\phi}_{i+\hat{x}})
\sin(\delta\tilde{\phi}_i-\delta\tilde{\phi}_{i+\hat{x}}) \rangle
\Big] \sin\phi
\\
&\quad
-\epsilon\,
\Big[
\langle(\sin\delta\tilde{\phi}_i+\sin\delta\tilde{\phi}_{i+\hat{y}})
\sin(\delta\tilde{\phi}_i-\delta\tilde{\phi}_{i+\hat{y}}) \rangle
+\langle(\cos\delta\tilde{\phi}_i+\cos\delta\tilde{\phi}_{i+\hat{x}})
\sin(\delta\tilde{\phi}_i-\delta\tilde{\phi}_{i+\hat{x}}) \rangle
\Big]\cos\phi .
\end{aligned}
\label{eq:sinus_active_final}
\end{equation}
Defining the coefficients
\begin{align}
S_{\rm R}^{(1),\mathrm{glob}}(t) &= 
-\epsilon\left[\Big\langle(\cos\delta\tilde{\phi}_i+\cos\delta\tilde{\phi}_{i+\hat{y}})
\sin(\delta\tilde{\phi}_i-\delta\tilde{\phi}_{i+\hat{y}})\Big \rangle
-\Big\langle(\sin\delta\tilde{\phi}_i+\sin\delta\tilde{\phi}_{i+\hat{x}})
\sin(\delta\tilde{\phi}_i-\delta\tilde{\phi}_{i+\hat{x}})\Big \rangle\right],
\label{eq-app:SR1-glob-SK}
\\
C_{\rm R}^{(1),\mathrm{glob}}(t) &= 
-\epsilon\left[\Big\langle(\sin\delta\tilde{\phi}_i+\sin\delta\tilde{\phi}_{i+\hat{y}})
\sin(\delta\tilde{\phi}_i-\delta\tilde{\phi}_{i+\hat{y}})\Big \rangle
+\Big\langle(\cos\delta\tilde{\phi}_i+\cos\delta\tilde{\phi}_{i+\hat{x}})
\sin(\delta\tilde{\phi}_i-\delta\tilde{\phi}_{i+\hat{x}})\Big \rangle\right],
\label{eq-app:CR1-glob-SK}
\end{align}
the previous equation eventually can be written in the form of Eq.~\eqref{eq:global-harmonics-txt} as
\begin{equation}
    (\partial_t \phi)_{\rm R} = S_{\rm R}^{(1),\mathrm{glob}}\sin \phi  + C_{\rm R}^{(1),\mathrm{glob}}\cos\phi.
    \label{eq-app:SK-g-Rea}
\end{equation}

\textit{Proactive term.} The proactive contribution to the deterministic part of the Langevin dynamics, summed over all the lattice, reads
\begin{equation}
\begin{split}
    (\partial_t \phi)_{\rm P}
    = -\frac{\epsilon}{N}
    \sum_{i}&
    \Big\{\big[\sin(\phi + \delta\tilde{\phi}_i)
        - \sin(\phi + \delta\tilde{\phi}_{i+\hat{x}})\big]
    \cos(\delta\tilde{\phi}_i - \delta\tilde{\phi}_{i+\hat{x}}) \\
    & + \big[-\cos(\phi + \delta\tilde{\phi}_i)
         + \cos(\phi + \delta\tilde{\phi}_{i+\hat{y}})\big]
    \cos(\delta\tilde{\phi}_i - \delta\tilde{\phi}_{i+\hat{y}})\Big\}.
\end{split}
\label{eq:sinus_proactive_start}
\end{equation}
By using trigonometric identities this equation can be cast in the form
\begin{equation}
\begin{split}
(\partial_t \phi)_{\rm P}
= -\frac{\epsilon}{N}
\sum_i &
\Big\{\big[\sin\phi\,(\cos\delta\tilde{\phi}_i-\cos\delta\tilde{\phi}_{i+\hat{x}})
+\cos\phi\,(\sin\delta\tilde{\phi}_i-\sin\delta\tilde{\phi}_{i+\hat{x}})\big]
\cos(\delta\tilde{\phi}_i-\delta\tilde{\phi}_{i+\hat{x}})
\\
&
+ \big[-\cos\phi\,(\cos\delta\tilde{\phi}_i-\cos\delta\tilde{\phi}_{i+\hat{y}})
+\sin\phi\,(\sin\delta\tilde{\phi}_i-\sin\delta\tilde{\phi}_{i+\hat{y}})\big]
\cos(\delta\tilde{\phi}_i-\delta\tilde{\phi}_{i+\hat{y}})\Big\}.
\end{split}
\label{eq-app:SK-g-Pro}
\end{equation}
Introducing
\begin{align}
S_{\rm P}^{(1),\mathrm{glob}}(t) &= 
-\epsilon\left[\Big\langle(\cos\delta\tilde{\phi}_i-\cos\delta\tilde{\phi}_{i+\hat{x}})
\cos(\delta\tilde{\phi}_i-\delta\tilde{\phi}_{i+\hat{x}})\Big \rangle
+\Big\langle(\sin\delta\tilde{\phi}_i-\sin\delta\tilde{\phi}_{i+\hat{y}})
\cos(\delta\tilde{\phi}_i-\delta\tilde{\phi}_{i+\hat{y}})\Big\rangle\right],
\label{eq-app:SP1-glob-SK}
\\
C_{\rm P}^{(1),\mathrm{glob}}(t) &=
-\epsilon\left[\Big\langle(\sin\delta\tilde{\phi}_i-\sin\delta\tilde{\phi}_{i+\hat{x}})
\cos(\delta\tilde{\phi}_i-\delta\tilde{\phi}_{i+\hat{x}})\Big \rangle
-\Big\langle(\cos\delta\tilde{\phi}_i-\cos\delta\tilde{\phi}_{i+\hat{y}})
\cos(\delta\tilde{\phi}_i-\delta\tilde{\phi}_{i+\hat{y}})\Big\rangle\right],
\label{eq-app:CP1-glob-SK}
\end{align}
the proactive contribution in Eq.~\eqref{eq-app:SK-g-Pro} can eventually be written as
\begin{equation}
    (\partial_t \phi)_{\rm P} = 
    S_{\rm P}^{(1),\mathrm{glob}}\sin \phi  + C_{\rm P}^{(1),\mathrm{glob}}\cos \phi .
\end{equation}
Finally, combining this proactive contribution with the reactive one in Eq.~\eqref{eq-app:SK-g-Rea} one finds the complete equation for the mean orientation of the sinusoidal coupling in the general form of Eq.~\eqref{eq:global-harmonics-txt}:
\begin{equation}
    \partial_t \phi = (\partial_t \phi)_{\rm R} + (\partial_t \phi)_{\rm P}= 
    (S_{\rm R}^{(1),\mathrm{glob}}+S_{\rm P}^{(1),\mathrm{glob}})\sin\phi
    + (C_{\rm R}^{(1),\mathrm{glob}}+C_{\rm P}^{(1),\mathrm{glob}})\cos\phi.
    \label{eq-app:evphi-SK}
\end{equation}

\paragraph*{Symmetry considerations.}
We now analyze the coefficients obtained above from the perspective of the symmetries of the system, as done above for the modulated coupling, in order to show that the $\mathbb{Z}_4$ symmetry of the square lattice is preserved by the equations.

If the ordered state is pinned along $\Phi = 0$,
one has $\phi = \delta \phi$ (see Eq.~\eqref{eq:deltaphi}), 
%
and the symmetry transformation is given by the global reflection ${\mathcal R}_y:\delta \phi \to -\delta \phi$ and $\delta\tilde{\phi}_i \to -\delta\tilde{\phi}_i$, combined with a spatial reflection $y \to -y$. 
Under this transformation, the coefficients $S_R^{(1),\mathrm{glob}}$ and $S_P^{(1),\mathrm{glob}}$ in Eqs.~\eqref{eq-app:SR1-glob-SK} and \eqref{eq-app:SP1-glob-SK}, respectively, are even, and therefore their averages might be different from zero. 
By contrast, $C_R^{(1),\mathrm{glob}}$ and $C_P^{(1),\mathrm{glob}}$ in Eqs.~\eqref{eq-app:CR1-glob-SK} and \eqref{eq-app:CP1-glob-SK}, respectively, are odd and thus their spatial averages vanish. 
As a result, the evolution of the mean orientation prescribed by Eq.~\eqref{eq-app:evphi-SK} involves only the pinning term of the form $-r^{\text{glob}}\sin\delta\phi$, with $r^{\text{glob}} = S_R^{(1),\mathrm{glob}}$ in the purely reactive case, and $r^{\text{glob}} = S_R^{(1),\mathrm{glob}} + S_P^{(1),\mathrm{glob}}$ for the dynamics which involves both the reactive and the proactive terms. The numerical simulations discussed in Sec.~\ref{sec:stationary-state} indicate that $r^{\text{glob}}>0$ in both cases.

If, instead, the system is pinned along $\Phi = \pi/2$, 
one has $\delta\phi = \phi - \pi/2$ (see Eq.~\eqref{eq:deltaphi}). The symmetry transformation is given by the global reflection ${\mathcal R}_x: \delta\phi \to -\delta\phi$ and $\delta\tilde{\phi}_i \to -\delta\tilde{\phi}_i$, now combined with $x \to -x$.
Under ${\mathcal R}_x$, the non-vanishing contributions come from $C_R^{(1),\mathrm{glob}}$ and $C_P^{(1),\mathrm{glob}}$, because it is easy to check that $S_R^{(1),\mathrm{glob}}$ and $S_P^{(1),\mathrm{glob}}$ in 
Eqs.~\eqref{eq-app:SR1-glob-SK} and \eqref{eq-app:SP1-glob-SK} are now odd and therefore vanish. 
The pinning term for the total dynamics then takes the form $-r^{\text{glob}}\sin\delta\phi$, using $\cos(\pi/2 + \delta\phi) = -\sin\delta\phi$, with $r^{\text{glob}} = -(C_R^{(1),\mathrm{glob}} + C_P^{(1),\mathrm{glob}})$.
Moreover, a rotation of the lattice by $\pi/2$ (counterclockwise) maps $x \to y$ and $y \to -x$. Under this transformation, one finds that $S_R^{(1),\mathrm{glob}} \leftrightarrow -C_R^{(1),\mathrm{glob}}$ (compare Eqs.~\eqref{eq-app:SR1-glob-SK} and \eqref{eq-app:CR1-glob-SK}) and similarly $S_P^{(1),\mathrm{glob}} \leftrightarrow -C_P^{(1),\mathrm{glob}}$ (compare Eqs.~\eqref{eq-app:SP1-glob-SK} and \eqref{eq-app:CP1-glob-SK}). 
Accordingly, the pinning strength $r^{\rm glob}$ at $\Phi = \pi/2$ has the same sign as $r^{\rm glob}$ at $\Phi = 0$. Repeating the same argument for $\Phi = \pi$ and $\Phi = 3\pi/2$, one concludes that the pinning coefficient $r^{\rm glob}$ has the $\mathbb{Z}_4$ symmetry of the square lattice, and it leads always to a pinning along the lattice directions. The same symmetry argument hold for the local counterparts for these coefficients, as anticipated in App.~\ref{app:sin-local}. 
\\

\paragraph*{Equality of $S_{\mathrm R}^{(1),\mathrm{glob}}$ and $S_{\mathrm P}^{(1),\mathrm{glob}}$ in the presence of periodic boundary conditions.}

We show here that $S_{\mathrm R}^{(1),\mathrm{glob}}(t)$ in Eq.~\eqref{eq-app:SR1-glob-SK} and $S_{\mathrm P}^{(1),\mathrm{glob}}(t)$ in Eq.~\eqref{eq-app:SP1-glob-SK} are actually equal when periodic boundary conditions are assumed on the lattice.
We first consider the vertical contribution, i.e., the one which involves the sites $i$ and $i+\hat y$, corresponding to the first average in square brackets in Eq.~\eqref{eq-app:SR1-glob-SK} for $S_{\mathrm R}^{(1),\mathrm{glob}}(t)$ and to the second one in Eq.~\eqref{eq-app:SP1-glob-SK} for $S_{\mathrm P}^{(1),\mathrm{glob}}(t)$. Using elementary trigonometric identities, one finds that
the difference $(\cos \delta\tilde\phi_i+\cos \delta\tilde\phi_{i+\hat y})\sin(\delta\tilde\phi_i-\delta\tilde\phi_{i+\hat y})
-
(\sin \delta\tilde\phi_i-\sin \delta\tilde\phi_{i+\hat y})\cos(\delta\tilde\phi_i-\delta\tilde\phi_{i+\hat y})$ between the quantities which are then averaged in the two contributions is equal to
%
$\sin \delta\tilde\phi_i-\sin \delta\tilde\phi_{i+\hat y}$.
Similarly, for the remaining horizontal contribution, i.e., the one which involves  the sites $i$ and $i+\hat x$ (the second average in square brackets in Eq.~\eqref{eq-app:SR1-glob-SK} for $S_{\mathrm R}^{(1),\mathrm{glob}}(t)$ and to the first one in Eq.~\eqref{eq-app:SP1-glob-SK} for $S_{\mathrm P}^{(1),\mathrm{glob}}(t)$), one finds that the difference 
$-(\sin \delta\tilde\phi_i+\sin \delta\tilde\phi_{i+\hat x})\sin(\delta\tilde\phi_i-\delta\tilde\phi_{i+\hat x})
-
(\cos \delta\tilde\phi_i-\cos \delta\tilde\phi_{i+\hat x})\cos(\delta\tilde\phi_i-\delta\tilde\phi_{i+\hat x})$ 
is actually equal to $\cos \delta\tilde\phi_i-\cos \delta\tilde\phi_{i+\hat x}$.
Accordingly, the difference $S_R^{(1),{\rm glob}}(t)-S_P^{(1),{\rm glob}}(t)$ between the two coefficients can be written as
$S_R^{(1),{\rm glob}}(t)-S_P^{(1),{\rm glob}}(t)
=
- \epsilon\big[\langle \sin \delta\tilde\phi_i-\sin \delta\tilde\phi_{i+\hat y}\rangle
+
\langle \cos \delta\tilde\phi_i-\cos \delta\tilde\phi_{i+\hat x} \rangle\big]$.
Under periodic boundary conditions, the spatial average $\langle \cdots \rangle$ of a discrete gradient --- such as those on the r.h.s.~of the previous equation --- vanishes identically, i.e.,
$\sum_i \left[f_i-f_{i+\hat\mu}\right]=0$,
for $\mu\in\{x,y\}$.
Accordingly, we conclude that $S_R^{(1),\mathrm{glob}}(t)=S_P^{(1),\mathrm{glob}}(t)$
for periodic boundary conditions. Similar equality holds for $C_R^{(1),\mathrm{glob}}(t)$ and $C_P^{(1),\mathrm{glob}}(t)$.

\subsection{Von Mises coupling}
\label{app:vonmises-mean}

As done above, also for the case of the von Mises coupling in Eq.~\eqref{eq:vonmises-kernel}, we find convenient to treat separately the reactive and proactive contributions to the equation of motion in Eq.~\eqref{eq:langevin}, which we then use to determine the equation of motion for the average orientation $\phi(t)$. In particular, we consider the orders ${\mathcal O}(\sigma)$ and ${\mathcal O}(\sigma^4)$ of the expansion of the interaction kernel in increasing powers of $\sigma$. The order ${\mathcal O}(\sigma^0)$ does not appear in the equations for the mean orientation, because it is reciprocal. 
%

\textit{Reactive term.}
The deterministic contribution to the Langevin dynamics in Eq.~\eqref{eq:langevin} originating from the reactive part alone is
\begin{equation}
\begin{aligned}
(\partial_t \phi)_{\rm R}
&=
-\frac{1}{N}\sum_{i}
\Bigg\{
    \sigma[\cos(\phi+\delta\tilde\phi_i)
          +\cos(\phi+\delta\tilde\phi_{i+\hat{x}})]
    +\tfrac{1}{24}\sigma^4
     [\cos^4(\phi+\delta\tilde\phi_i)
      -\cos^4(\phi+\delta\tilde\phi_{i+\hat{x}})]
\Bigg\}
\sin(\delta\tilde\phi_i-\delta\tilde\phi_{i+\hat{x}})
\\
&\quad
-\frac{1}{N}\sum_{i}
\Bigg\{
    \sigma[\sin(\phi+\delta\tilde\phi_i)
          +\sin(\phi+\delta\tilde\phi_{i+\hat{y}})]
    +\tfrac{1}{24}\sigma^4
     [\sin^4(\phi+\delta\tilde\phi_i)
      -\sin^4(\phi+\delta\tilde\phi_{i+\hat{y}})]
\Bigg\}
\sin(\delta\tilde\phi_i-\delta\tilde\phi_{i+\hat{y}}),
\end{aligned}
\label{eq:vonmises_active}
\end{equation}
where, as done in the previous sections, the local value $\phi_i$ of the phase is expressed in terms of its deviation $\delta \tilde\phi_i$ from the global orientation $\phi$, according to Eq.~\eqref{eq:def-tphii}. 
Using the trigonometric identities in Eq.~\eqref{eq-app:cos4sin4},
we can rewrite this equation in the form of  Eq.~\eqref{eq:global-harmonics-txt}, i.e.,
\begin{equation}
\begin{aligned}
(\partial_t \phi)_{\rm R}
=\;&
S_{\rm R}^{(1),\mathrm{glob}}\sin\phi + C_{\rm R}^{(1),\mathrm{glob}}\cos\phi
+S_{\rm R}^{(2),\mathrm{glob}}\sin(2\phi)+C_{\rm R}^{(2),\mathrm{glob}}\cos(2\phi)
+S_{\rm R}^{(4),\mathrm{glob}}\sin(4\phi)+C_{\rm R}^{(4),\mathrm{glob}}\cos(4\phi),
\end{aligned}
\label{eq:vonmises_active_expanded_sigma4_compact}
\end{equation}
where $S_{\rm R}^{(1),\mathrm{glob}}$ and $C_{\rm R}^{(1),\mathrm{glob}}$ have, up to a factor, the same expressions as those for the sinusoidal interaction (see Eqs.~\eqref{eq-app:SR1-glob-SK} and \eqref{eq-app:CR1-glob-SK}, 
respectively), i.e.,
\begin{align}
S_{\rm R}^{(1),\mathrm{glob}} &= 
-\sigma\left[\Big\langle(\cos\delta\tilde{\phi}_i-\cos\delta\tilde{\phi}_{i+\hat{x}})
\cos(\delta\tilde{\phi}_i-\delta\tilde{\phi}_{i+\hat{x}})\Big \rangle
+\Big\langle(\sin\delta\tilde{\phi}_i-\sin\delta\tilde{\phi}_{i+\hat{y}})
\cos(\delta\tilde{\phi}_i-\delta\tilde{\phi}_{i+\hat{y}})\Big\rangle\right],  \label{eq-app:SR1-glo-vM}\\
C_{\rm R}^{(1),\mathrm{glob}} &=
-\sigma\left[\Big\langle(\sin\delta\tilde{\phi}_i-\sin\delta\tilde{\phi}_{i+\hat{x}})
\cos(\delta\tilde{\phi}_i-\delta\tilde{\phi}_{i+\hat{x}})\Big \rangle
-\Big\langle(\cos\delta\tilde{\phi}_i-\cos\delta\tilde{\phi}_{i+\hat{y}})
\cos(\delta\tilde{\phi}_i-\delta\tilde{\phi}_{i+\hat{y}})\Big\rangle\right], \label{eq-app:CR1-glo-vM}
\end{align}
while 
\begin{align}
S_{\rm R}^{(2),\mathrm{glob}}
&=
\frac{\sigma^4}{48}\left\langle \big[\sin{(2\delta\tilde\phi_i)}- \sin{(2\delta\tilde\phi_{i+\hat{x}})}\big]\sin{(\delta\tilde\phi_i-\delta\tilde\phi_{i+\hat{x}})} - \big[\sin{(2\delta\tilde\phi_i)}- \sin{(2\delta\tilde\phi_{i+\hat{y}})}\big]\sin{(\delta\tilde\phi_i-\delta\tilde\phi_{i+\hat{y}})} \right\rangle
,\label{eq-app:SR2-glo-vM}
\\
C_{\rm R}^{(2),\mathrm{glob}}
&=
\frac{\sigma^4}{48}\left\langle -\big[\cos{(2\delta\tilde\phi_i)}- \cos{(2\delta\tilde\phi_{i+\hat{x}})}\big]\sin{(\delta\tilde\phi_i-\delta\tilde\phi_{i+\hat{x}})} + \big[\cos{(2\delta\tilde\phi_i)}- \cos{(2\delta\tilde\phi_{i+\hat{y}})}\big]\sin{(\delta\tilde\phi_i-\delta\tilde\phi_{i+\hat{y}})} \right\rangle, \label{eq-app:CR2-glo-vM}
\\
S_{\rm R}^{(4),\mathrm{glob}}
&=
\frac{\sigma^4}{96}\Big\langle
\big[\sin(4\delta\tilde\phi_i)-\sin(4\delta\tilde\phi_j)\big]
\sin(\delta\tilde\phi_i-\delta\tilde\phi_j)
\Big\rangle,
\label{eq-app:SR4-glob-vM}
\\
C_{\rm R}^{(4),\mathrm{glob}}
&=
-\frac{\sigma^4}{96}\Big\langle
\big[\cos(4\delta\tilde\phi_i)-\cos(4\delta\tilde\phi_j)\big]
\sin(\delta\tilde\phi_i-\delta\tilde\phi_j)
\Big\rangle .
\label{eq-app:CR4-glob-vM}
\end{align}
the expression of $S_{\rm R}^{(4),\mathrm{glob}}$ and $C_{\rm P}^{(R),\mathrm{glob}}$ reported here are the same, up to a factor, as the corresponding ones for the modulated coupling, reported in Eqs.~\eqref{eq-app:SR4glob} and \eqref{eq-app:CR4glob}, respectively. 
\\

\paragraph*{Symmetry considerations.}
If the system is pinned along $\Phi=\pi/4$ (the reactive dynamics leads to pinning to diagonals as discussed in Sec.~\ref{sec:pinning-vonmises}), the stationary state is invariant under the diagonal reflection $\mathcal{R}_d$:
$\delta\phi \to
-\delta\phi$, $\delta\tilde{\phi}_i \to -\delta\tilde{\phi}_i$, combined with $x \leftrightarrow
y$.
Under this symmetry, one finds that
$
S_{\mathrm{R}}^{(1),\mathrm{glob}}
\leftrightarrow
-
C_{\mathrm{R}}^{(1),\mathrm{glob}},
$
(compare Eqs.~\eqref{eq-app:SR1-glo-vM} and \eqref{eq-app:CR1-glo-vM})
from which it follows that
$
S_{\mathrm{R}}^{(1),\mathrm{glob}}
=
-
C_{\mathrm{R}}^{(1),\mathrm{glob}}.
$
Numerically, we further find that
$
S_{\mathrm{R}}^{(1),\mathrm{glob}}
<
0
$
(see Tab.~\ref{tab:pinning_coeffs})
Accordingly,  the first two terms on the r.h.s.~of  Eq.~\eqref{eq:vonmises_active_expanded_sigma4_compact} can be written as 
$S_{\mathrm{R}}^{(1),\mathrm{glob}}(\sin{\phi}-\cos{\phi}) = 
\sqrt{2}\,
S_{\mathrm{R}}^{(1),\mathrm{glob}}
\sin(\phi-\pi/4),
$
and expanding in $\delta\phi$ around $\phi=\pi/4+\delta\phi$, one obtains
$
\sqrt{2}\,
S_{\mathrm{R}}^{(1),\mathrm{glob}}
\,\delta\phi,
$
which contributes to a restoring force toward $\delta\phi=0$, consistently with the stability of the diagonal pinned direction. 
A similar symmetry argument applies for the corresponding local coefficients $S_{\mathrm{R}, i}^{(1),\mathrm{loc}}$ and $C_{\mathrm{R}, i}^{(1),\mathrm{loc}}$ discussed in App.~\ref{app:vonmises-local}, which yields the relation $S_{\mathrm{R}, i}^{(1),\mathrm{loc}} = - C_{\mathrm{R}, i}^{(1),\mathrm{loc}}$ anticipated in the main text and thus to Eq.~\eqref{eq:vonmises-loc-pinning}. 

For the second harmonic, the same symmetry implies that the term $S_{\mathrm R}^{(2),\mathrm{glob}}$ in Eq.~\eqref{eq:vonmises_active_expanded_sigma4_compact} is odd under $\mathcal R_d$ and therefore vanishes in the diagonal pinned state. By contrast, $C_{\mathrm R}^{(2),\mathrm{glob}}$ is even under $\mathcal R_d$ and is not constrained to vanish by symmetry. Expanding the corresponding contribution around $\phi=\pi/4+\delta\phi$, one finds $C_{\mathrm R}^{(2),\mathrm{glob}}\cos(2\phi)=C_{\mathrm R}^{(2),\mathrm{glob}}\cos(\pi/2+2\delta\phi)\simeq -2C_{\mathrm R}^{(2),\mathrm{glob}}\delta\phi$, which therefore contributes to the global relaxation rate. We find numerically that $C_{\mathrm R}^{(2),\mathrm{glob}}>0$ but we not report this contribution in Tab.~\ref{tab:pinning_coeffs}, since we find numerically that it is much smaller than the one arising from the fourth harmonic.

We now turn to the fourth-harmonic terms. In this case, the diagonal reflection $\mathcal R_d$ reduces to the same symmetry argument used for the modulated-coupling kernel in App.~\ref{app:modulated-mean}, since the two spatial directions enter on the same footing. As a consequence, $C_{\mathrm R}^{(4),\mathrm{glob}}$ is odd under the symmetry and vanishes, whereas $S_{\mathrm R}^{(4),\mathrm{glob}}$ is even and can be nonzero. Expanding around $\phi=\pi/4+\delta\phi$, one has $S_{\mathrm R}^{(4),\mathrm{glob}}\sin(4\phi)=S_{\mathrm R}^{(4),\mathrm{glob}}\sin(\pi+4\delta\phi)\simeq -4S_{\mathrm R}^{(4),\mathrm{glob}}\delta\phi$, corresponding to a contribution $4S_{\mathrm R}^{(4),\mathrm{glob}}$ to the global relaxation rate when $S_{\mathrm R}^{(4),\mathrm{glob}}>0$.

Altogether, the reactive von Mises dynamics pinned along the diagonal direction is therefore characterized, to linear order in $\delta\phi$, by a restoring term with relaxation rate $r^{\mathrm{glob}}=-\sqrt{2}\,S_{\mathrm R}^{(1),\mathrm{glob}}+2C_{\mathrm R}^{(2),\mathrm{glob}}+4S_{\mathrm R}^{(4),\mathrm{glob}}$, where the second-harmonic contribution is allowed by symmetry but is numerically subleading for the parameters considered.

We also find that, for the parameters reported in Tab.~\ref{tab:pinning_coeffs}, the contribution proportional to $S_{\mathrm R}^{(1),\mathrm{glob}}$ is smaller than the one proportional to $S_{\mathrm R}^{(4),\mathrm{glob}}$. However, this hierarchy is parameter dependent: $S_{\mathrm R}^{(1),\mathrm{glob}}$ originates from the ${\cal O}(\sigma)$ part of the kernel, whereas $S_{\mathrm R}^{(4),\mathrm{glob}}$ originates from the ${\cal O}(\sigma^4)$ contribution. Therefore, in the small-$\sigma$ regime, the first-harmonic term is expected to become the leading contribution to the global pinning. We did not explore this regime systematically, since in the purely reactive case the pinning becomes too weak to be clearly resolved with the system sizes accessible in our simulations. 
\\
%

\textit{Proactive term.}
The deterministic contribution of the proactive term to the evolution equation for $\phi(t)$ is given by
\begin{equation}
\begin{aligned}
(\partial_t\phi)_{\rm P}
=&
-\frac{1}{N}\sum_i
\Bigg\{
\sigma
\Big[
\sin(\phi+\delta\tilde\phi_i)
-
\sin(\phi+\delta\tilde\phi_{i+\hat x})
\Big]
\\
&\qquad
-\frac{\sigma^4}{6}
\Big[
\sin(\phi+\delta\tilde\phi_i)\cos^3(\phi+\delta\tilde\phi_i)
+
\sin(\phi+\delta\tilde\phi_{i+\hat x})
\cos^3(\phi+\delta\tilde\phi_{i+\hat x})
\Big]
\Bigg\}
\cos(\delta\tilde\phi_i-\delta\tilde\phi_{i+\hat x})
\\
&+\frac{1}{N}\sum_i
\Bigg\{
\sigma
\Big[
\cos(\phi+\delta\tilde\phi_i)
-
\cos(\phi+\delta\tilde\phi_{i+\hat y})
\Big]
\\
&\qquad
+\frac{\sigma^4}{6}
\Big[ 
\cos(\phi+\delta\tilde\phi_i)\sin^3(\phi+\delta\tilde\phi_i)
+
\cos(\phi+\delta\tilde\phi_{i+\hat y})
\sin^3(\phi+\delta\tilde\phi_{i+\hat y})
\Big]
\Bigg\}
\cos(\delta\tilde\phi_i-\delta\tilde\phi_{i+\hat y}),
\end{aligned}
\end{equation}
where, as done above, $\phi_i$ which enters in the definition of $\phi(t)$ is expressed in terms of its deviation $\delta \tilde\phi_i$ from the global orientation $\phi$, according to Eq.~\eqref{eq:def-tphii}. 
Using the trigonometric identities $
\sin x\cos^3x
=
\frac{1}{4}\sin(2x)
+
\frac{1}{8}\sin(4x)$ and
$
\cos x \sin^3 x 
=
\frac{1}{4}\sin(2x)
-
\frac{1}{8}\sin(4x)
$ 
this equation can be cast in the form of Eq.~\eqref{eq:global-harmonics-txt}, with 
\begin{equation}\label{eq:vonmises_proactive_expanded_sigma4_compact}
(\partial_t\phi)_{\rm P}
=
S_{\rm P}^{(1),\mathrm{glob}}\sin\phi+C_{\rm P}^{(1),\mathrm{glob}}\cos\phi+
S_{\rm P}^{(2),\mathrm{glob}}\sin(2\phi)+C_{\rm P}^{(2),\mathrm{glob}}\cos(2\phi)+
S_{\rm P}^{(4),\mathrm{glob}}\sin(4\phi)+C_{\rm P}^{(4),\mathrm{glob}}\cos(4\phi),
\end{equation}
where the first harmonics $S_{\rm P}^{(1),\mathrm{glob}}$ and $C_{\rm P}^{(1),\mathrm{glob}}$ have, up to a factor, the same expression as those for the sinusoidal interaction reported in Eqs.~\eqref{eq-app:SP1-glob-SK} and \eqref{eq-app:CP1-glob-SK}, respectively, i.e.,
\begin{align}
S_{\rm P}^{(1),\mathrm{glob}}(t) &= 
-\sigma\left[\Big\langle(\cos\delta\tilde{\phi}_i-\cos\delta\tilde{\phi}_{i+\hat{x}})
\cos(\delta\tilde{\phi}_i-\delta\tilde{\phi}_{i+\hat{x}})\Big \rangle
+\Big\langle(\sin\delta\tilde{\phi}_i-\sin\delta\tilde{\phi}_{i+\hat{y}})
\cos(\delta\tilde{\phi}_i-\delta\tilde{\phi}_{i+\hat{y}})\Big\rangle\right],  \label{eq-app:SP1-glo-vM}\\
C_{\rm P}^{(1),\mathrm{glob}}(t) &=
-\sigma\left[\Big\langle(\sin\delta\tilde{\phi}_i-\sin\delta\tilde{\phi}_{i+\hat{x}})
\cos(\delta\tilde{\phi}_i-\delta\tilde{\phi}_{i+\hat{x}})\Big \rangle
-\Big\langle(\cos\delta\tilde{\phi}_i-\cos\delta\tilde{\phi}_{i+\hat{y}})
\cos(\delta\tilde{\phi}_i-\delta\tilde{\phi}_{i+\hat{y}})\Big\rangle\right],  \label{eq-app:CP1-glo-vM}
\end{align}
while the second harmonics are given by
\begin{align}
S_{\rm P}^{(2),\mathrm{glob}}
&=
\frac{\sigma^4}{24}
\Big\langle
\big[
\cos(2\delta\tilde\phi_i)+\cos(2\delta\tilde\phi_{i+\hat{y}})
\big]
\cos(\delta\tilde\phi_i-\delta\tilde\phi_{i+\hat{y}})
\Big\rangle
\nonumber\\
&\quad-
\frac{\sigma^4}{24}
\Big\langle
\big[
\cos(2\delta\tilde\phi_i)+\cos(2\delta\tilde\phi_{i+\hat{x}})
\big]
\cos(\delta\tilde\phi_i-\delta\tilde\phi_{i+\hat{x}})
\Big\rangle,\label{eq-app:SP2-glo-vM}
\\
C_{\rm P}^{(2),\mathrm{glob}}
&=
\frac{\sigma^4}{24}
\Big\langle
\big[
\sin(2\delta\tilde\phi_i)+\sin(2\delta\tilde\phi_{i+\hat{y}})
\big]
\cos(\delta\tilde\phi_i-\delta\tilde\phi_{i+\hat{y}})
\Big\rangle
\nonumber\\
&\quad-
\frac{\sigma^4}{24}
\Big\langle
\big[
\sin(2\delta\tilde\phi_i)+\sin(2\delta\tilde\phi_{i+\hat{x}})
\big]
\cos(\delta\tilde\phi_i-\delta\tilde\phi_{i+\hat{x}})
\Big\rangle.\label{eq-app:CP2-glo-vM}
\end{align}
Finally, for the fourth harmonic,
\begin{align}
S_{\rm P}^{(4),\mathrm{glob}}
&=
-\frac{\sigma^4}{24}
\Big\langle
\big[
\cos(4\delta\tilde\phi_i)+\cos(4\delta\tilde\phi_j)
\big]
\cos(\delta\tilde\phi_i-\delta\tilde\phi_j)
\Big\rangle,
\label{eq-app:SP4-glob-vM}
\\
C_{\rm P}^{(4),\mathrm{glob}}
&=
-\frac{\sigma^4}{24}
\Big\langle
\big[
\sin(4\delta\tilde\phi_i)+\sin(4\delta\tilde\phi_j)
\big]
\cos(\delta\tilde\phi_i-\delta\tilde\phi_j)
\Big\rangle.
\label{eq-app:CP4-glob-vM}
\end{align}
As expected, the expression of $S_{\rm P}^{(4),\mathrm{glob}}$ and $C_{\rm P}^{(4),\mathrm{glob}}$ reported here are the same, up to a factor, as the corresponding ones for the modulated coupling, reported in Eqs.~\eqref{eq-app:SP4glob} and \eqref{eq-app:CP4glob}, respectively. 
\\

\paragraph*{Symmetry considerations.}

If the system is pinned along the lattice direction $\Phi=0$ (the case realized by the full von Mises dynamics, see Sec.~\ref{sec:pinning-vonmises}), the relevant symmetry is the reflection $\mathcal R_y$ introduced in App.~\ref{app:sin-global}. Under this transformation, the proactive coefficients $S_{\mathrm P}^{(1),\mathrm{glob}}$, $S_{\mathrm P}^{(2),\mathrm{glob}}$, and $S_{\mathrm P}^{(4),\mathrm{glob}}$ are even and therefore they may be nonzero. 
For the fourth harmonic, the argument reduces to the same one discussed in App.~\ref{app:modulated-mean}, since the two spatial directions enter symmetrically in the corresponding expressions.

For the reactive contribution, the coefficients $S_{\mathrm R}^{(1),\mathrm{glob}}$, $S_{\mathrm R}^{(2),\mathrm{glob}}$, and $S_{\mathrm R}^{(4),\mathrm{glob}}$ are likewise even under $\mathcal R_y$. Furthermore, as shown in App.~\ref{app:sin-global}, one has $S_{\mathrm R}^{(1),\mathrm{glob}}=S_{\mathrm P}^{(1),\mathrm{glob}}$. Therefore, after expanding the sum of Eqs.~\eqref{eq:vonmises_active_expanded_sigma4_compact} and ~\eqref{eq:vonmises_proactive_expanded_sigma4_compact} around $\Phi = 0$ with $\phi=\delta\phi$, the global relaxation rate takes the form
$
r^{\mathrm{glob}}
=
-2S_{\mathrm R}^{(1),\mathrm{glob}}
-
2S_{\mathrm R}^{(2),\mathrm{glob}}
-
2S_{\mathrm P}^{(2),\mathrm{glob}}
-
4S_{\mathrm R}^{(4),\mathrm{glob}}
-
4S_{\mathrm P}^{(4),\mathrm{glob}}.
$
In Eq.~\eqref{eq:pin-glob-vonmises} of the main text, we reported only the first- and fourth-harmonic contributions. The first harmonic is of order ${\cal O}(\sigma)$ and is therefore expected to dominate in the small-$\sigma$ regime, even though, as already discussed in this section, we did not explore this regime numerically. The fourth harmonic is of order ${\cal O}(\sigma^4)$ and provides the leading higher-order correction. The second harmonic is also of order ${\cal O}(\sigma^4)$; however, we find numerically that its contribution to the pinning is significantly smaller than that of the fourth harmonic for the parameter values considered in this work. For this reason, it has been omitted from the discussion in the main text.

Finally, we comment on the equivalence among all lattice directions. Under a rotation by $\pi/2$, one has $x\to y$ and $y\to-x$. As discussed in App.~\ref{app:sin-global}, this implies $S_{\mathrm R}^{(1),\mathrm{glob}}\leftrightarrow -C_{\mathrm R}^{(1),\mathrm{glob}}$ (compare Eqs.~\eqref{eq-app:SR1-glo-vM} and \eqref{eq-app:CR1-glo-vM}) and, similarly, $S_{\mathrm P}^{(1),\mathrm{glob}}\leftrightarrow -C_{\mathrm P}^{(1),\mathrm{glob}}$ (compare Eqs.~\eqref{eq-app:SP1-glo-vM} and \eqref{eq-app:CP1-glo-vM}). 
The fourth-harmonic terms (Eqs.~\eqref{eq-app:SR4-glob-vM}, \eqref{eq-app:CR4-glob-vM},
\eqref{eq-app:SP4-glob-vM}, and \eqref{eq-app:CP4-glob-vM}) are unchanged by this rotation, since their analytical expression does not distinguish between the lattice directions. The second-harmonic coefficients (Eqs.~\eqref{eq-app:SR2-glo-vM}, \eqref{eq-app:CR2-glo-vM}, \eqref{eq-app:SP2-glo-vM}, and \eqref{eq-app:CP2-glo-vM}) instead, change sign, i.e., $S_{\mathrm R}^{(2),\mathrm{glob}}\to -S_{\mathrm R}^{(2),\mathrm{glob}}$, $C_{\mathrm R}^{(2),\mathrm{glob}}\to -C_{\mathrm R}^{(2),\mathrm{glob}}$, $S_{\mathrm P}^{(2),\mathrm{glob}}\to -S_{\mathrm P}^{(2),\mathrm{glob}}$, and $C_{\mathrm P}^{(2),\mathrm{glob}}\to -C_{\mathrm P}^{(2),\mathrm{glob}}$.
This sign change is consistent with the equivalence of the pinned lattice directions. For instance, a second-harmonic contribution of the form $(S_{\mathrm R}^{(2),\mathrm{glob}}+S_{\mathrm P}^{(2),\mathrm{glob}})\sin(2\phi)$ becomes $-(S_{\mathrm R}^{(2),\mathrm{glob}}|_{\Phi = 0}+S_{\mathrm P}^{(2),\mathrm{glob}}|_{\Phi = 0})\sin(2\phi)$ after a rotation by $\pi/2$. Expanding around $\Phi=\pi/2$, i.e., $\phi=\pi/2+\delta\phi$, one has $\sin(2\phi)=\sin(\pi+2\delta\phi)\simeq -2\delta\phi$. Accordingly, the additional minus sign in the coefficient is exactly compensated by the expansion of $\sin(2\phi)$ around the rotated pinned direction, yielding the same restoring contribution as the one around $\Phi=0$. This implies that the pinning strength is the same along all lattice directions, as required by the underlying $\mathbb Z_4$ symmetry.

\end{document}